\documentclass{jpp}

\usepackage{comment}
\usepackage{graphicx, color}
\usepackage{mathtools}          %.. extends amsmath
\usepackage{amssymb,amsmath}
\usepackage{bm,bbm}
\usepackage{float}
\usepackage[usenames,dvipsnames]{xcolor}
\usepackage{appendix}
\usepackage{url}
\usepackage[normalem]{ulem}  % in preamble
   \newcommand{\vct}[1]  {\ensuremath{\boldsymbol{#1}}}    %.. bold italic
 \newcommand{\vb} {\vct{b}}

 \newcommand{\vf} {\vct{f}}
 \newcommand{\vg} {\vct{g}}

 \newcommand{\vj} {\vct{j}}

 \def\vr{\vct{r}}               %.. redefines without error

 \newcommand{\vu} {\vct{u}}

 \newcommand{\vx} {\vct{x}}

 \newcommand{\vF} {\vct{F}}

 \newcommand{\vJ} {\vct{J}}

 \newcommand{\vS} {\vct{S}}

 \newcommand{\vOmega}{\vct{\Omega}}
 \newcommand{\vSigma}{\vct{\Sigma}}
 \renewcommand{\d} {\mathrm{d}}
   \newcommand{\PD}    [2]    {\frac{\partial   {#1}} {\partial{#2}} }

\newcommand{\squishlist}{
   \begin{list}{$\bullet$}
    { \setlength{\itemsep}{0pt}      \setlength{\parsep}{3pt}
      \setlength{\topsep}{3pt}       \setlength{\partopsep}{0pt}
      \setlength{\leftmargin}{1.5em} \setlength{\labelwidth}{1em}
      \setlength{\labelsep}{0.5em} } }

\newcommand{\squishlisttwo}{
   \begin{list}{$\bullet$}
    { \setlength{\itemsep}{0pt}    \setlength{\parsep}{0pt}
      \setlength{\topsep}{0pt}     \setlength{\partopsep}{0pt}
      \setlength{\leftmargin}{2em} \setlength{\labelwidth}{1.5em}
      \setlength{\labelsep}{0.5em} } }

\newcommand{\squishend}{
    \end{list}  }

\newcommand{\beq}{\begin{equation}}
\newcommand{\eeq}{\end{equation}}

\newcommand{\lan} {\langle}
\newcommand{\ran} {\rangle}

\newcommand{\eps}{\varepsilon}

\newcommand{\ol}  [1] {\overline{#1}}     %.. or use \bar

\newcommand{\oll} [1] {\overline{#1}^\ell}

\newcommand{\ou}{\overline{\vu}^\ell}
\newcommand{\ob}{\overline{\vb}^\ell}

\newcommand{\ovS}{\overline{\bm S}}
\newcommand{\ovJ}{\overline{\bm J}}
\newcommand{\ovOm}{\overline{\bm \Omega}}
\newcommand{\ovSS}{\overline{\bm \Sigma}}

\newcommand{\Sbarij}{ \overline{S}^\ell_{ij}} 

\newcommand{\Sbarji}{ \overline{S}^\ell_{ji}}

\newcommand{\tr}[1] 
  {\mathrm{Tr} \left\{ {#1} \right\}}
\newcommand{\red}  [1]{\textcolor{red}{#1}}
\newcommand{\blue} [1]{\textcolor{blue}{#1}}

\newcommand{\remark}[3] {\textcolor{#1} { ... {#2}: #3}\newline}

\newcommand{\damiano}[1]  {\remark{blue}{DC}{#1} }

\newcommand{\vSbl}  {\overline{\bm \Sigma}^{\ell} }

\newcommand{\vSul}  {\overline{\bm S}^{\ell} }

\newcommand{\vSFb} {\overline{\bm \Sigma}^{\sqrt{\theta}} }

\newcommand{\vSFu} {\overline{\bm S}^{\sqrt{\theta}} }

\newcommand{\vSFFb}{\overline{\overline{\bm \Sigma}^{\sqrt{\theta}}}^\phi }

\newcommand{\vSFFu}{\overline{\overline{\bm S}^{\sqrt{\theta}}}^\phi }

\newcommand{\vOb}  {\overline{\bm J}^{\ell} }

\newcommand{\vOFb} {\overline{\bm J}^{\sqrt{\theta}} }

\newcommand{\vOFu} {\overline{\bm \Omega}^{\sqrt{\theta}} }

\newcommand{\vOFFb}{\overline{\overline{\bm J}^{\sqrt{\theta}}}^\phi }

\newcommand{\vOFFu}{\overline{\overline{\bm \Omega}^{\sqrt{\theta}}}^\phi }

\shorttitle{Guidelines for authors}
\shortauthor{A. N. Other, H.-C. Smith and J. Q. Public}

\title{Energy transfer and conversion in Strongly Anisotropic Magnetohydrodynamic Turbulence}

\shortauthor{D. Capocci, 
S. Oughton, 
M. Linkmann}

\author{Damiano Capocci\aff{1}
\corresp{\email{dcapocci@ed.ac.uk}},
Sean Oughton\aff{2}, 
 \and Moritz Linkmann\aff{3}\corresp{\email{moritz.linkmann@ed.ac.uk}}}

\affiliation{
\aff{1}School of Physics and Astronomy and Higgs Centre, The University of Edinburgh, Edinburgh EH9 3FD, United Kingdom
\aff{2} Department of Mathematics, University of Waikato, Hamilton, New Zealand
\aff{3} School of Mathematics and Maxwell Institute for Mathematical Sciences, University of Edinburgh, Edinburgh, EH9 3FD, United Kingdom
}

\begin{document}

\maketitle

\begin{abstract}
In homogeneous magnetohydrodynamic (MHD) turbulence without a background magnetic field driven by mechanical forces, an exact decomposition of the energy fluxes (D. Capocci et al., \emph{Journal
of Plasma Physics}, 91(1), E11 (2025)) has shown that current-sheet thinning is the dominant physical mechanism responsible for transferring energy from large to small scales. In contrast, mechanisms that are characteristic of hydrodynamic turbulence, such as vortex stretching and strain self-amplification, are strongly suppressed. Here, we extend this analysis to MHD turbulence in the presence of weak and strong imposed magnetic field, as previously driven by mechanical forces, %. Relative to the isotropic case, our results 
and confirm that current-sheet thinning remains the leading process driving the energy cascade toward smaller scales in these more realistic configurations, and %Here, the anisotropy manifests itself through 
find enhanced scale invariance in the subfluxes. % and pronounced two-dimensionalisation of the dynamics. 
In addition to that, a decomposition of the contributions from the fluctuating and the background magnetic field to the conversion between kinetic and magnetic energies 
%nonlinear dynamo 
shows that the background-field-dependent contribution results in a nonlinear dynamo, that is an effective kinetic-to-magnetic conversion at large and intermediate scales. However, at small scales, it has the opposite effect, resulting in a net conversion of magnetic to kinetic energy. % magnetic-to-kinetic conversion at smaller scales. 

%Finally, we show that the background-field-dependent contribution to the resolved-scale conversion term induces an effective kinetic-to-magnetic conversion at large/intermediate scales and a magnetic-to-kinetic conversion at smaller scales.
\end{abstract}

\section{Introduction}

A wide range of astrophysical and laboratory plasmas are permeated by a strong large-scale magnetic field that introduces a preferred direction and, consequently, strong anisotropy in the turbulent dynamics. Canonical examples include the solar wind (e.g. \cite{goldstein1995}) and planetary magnetospheres \citep{leamon1998}, where turbulence develops in the presence of a mean magnetic field, and magnetically confined fusion plasmas, where a guide field is essential to confinement. In such systems, Alfv\'en-wave propagation and the tendency toward two-dimensionalisation motivate reduced descriptions such as reduced MHD (RMHD) in the strong-guide-field limit \citep{Strauss76,OughtonEA17-rmhd}. Moreover, the connection between anisotropy and dynamo processes, reflecting in the energy conversion between kinetic and magnetic energies, is still unclear. The above physical scenarios motivate a deeper characterisation of turbulent energy transfer and kinetic-magnetic energies exchanges in (strongly) anisotropic MHD configurations.

In the homogeneous case with no mean magnetic field, an exact decomposition of the MHD energy fluxes across scales was recently derived and applied to data pertaining to the saturated dynamo regime obtained by direct numerical simulation (DNS) \citep{capocci2025}. That analysis identified \emph{current-sheet thinning} as the dominant physical mechanism responsible for transferring energy from large to small scale, while processes characteristic of hydrodynamic turbulence like vortex stretching and strain self-amplification were found to be strongly suppressed in mean.
%ML cite alexakis, offermans
The purpose of the present work is to extend this approach to MHD turbulence in the presence of an imposed background magnetic field $\bm{B}_0$ and to quantify how strong anisotropy modifies (i) the interscale energy transfer mechanisms and (ii) the scale-by-scale kinetic-to-magnetic energy conversion. In particular, we seek to establish which aspects of the isotropic \emph{mechanistic} picture holds in the strong-guide-field regime.

%ML removed to comment briefly on RMHD in the conclusions
%In the strong-guide-field limit, anisotropic MHD turbulence is often described within the reduced-MHD (RMHD) framework introduced by \citet{Strauss76}. In that asymptotic regime, the dominant nonlinear dynamics are associated with fluctuations perpendicular to the imposed field, whereas parallel variations are weak and enter only at subleading order resulting in a quasi  two-dimensional approximation. From this perspective, the present analysis provides a useful complement to RMHD: rather than introducing a further asymptotic reduction, it interprets the quantified energy fluxes through the lens of RMHD. More specifically, our decomposition can be read as a manifestation of the RMHD ordering: if current-sheet thinning remains the dominant forward-transfer mechanism as the guide field is increased, then this identifies the leading cascade process retained in the RMHD limit, while the remaining much smaller subfluxes may be interpreted as processes that are progressively relegated to subleading orders from the RMHD expansion. In this sense, the analysis does not merely show which mechanism dominates in the anisotropic case; it also quantifies how the full MHD cascade reorganises itself into a hierarchy of leading and subleading processes that underlies the RMHD description, including effects dependent on the plasma $\beta$. %\blue{probably but I dont think to be honest, here we should mention something about the $\beta$.}

The structure of the paper is as follows: In section sec.~\ref{sec:theory} we recall the methodology that leads to the coarse-grained energy equations and MHD energy fluxes \cite{capocci2025}.  Section \ref{sec:numerics} contains a description of the numerical method and the DNS data. 
In sec.~\ref{sec:conversion}
we analyse how the presence of anisotropy modifies the expression of the conversion term describing the exchange between kinetic and magnetic energies. In \ref{sec:fluxes} we apply Johnsons's decomposition to study how and to what extent the physical mechanism behind the energy cascade is altered in (strongly) anisotropic configurations compared to the previously studied case without a background magnetic field. We conclude with a discussion in sec.~\ref{sec:conclusions}.

\section{Theory}
\label{sec:theory}

In this section we derive the coarse-grained energy
equations for MHD and giving the definitions of the scale-space energy fluxes and the energy conversion that appear in them. We then employ the Gaussian-filter methodology of Johnson \citep{Johnson21,Johnson2022}, which yields an \emph{exact} representation of the SGS stresses (and hence of the SGS fluxes) in terms of multiscale field gradients, and enables an exact decomposition of each MHD energy subflux into physically interpretable observables. 
%The structure of the paper follows the scheme outlined after the abstract: we first recall the methodology that leads to the coarse-grained energy equations and MHD energy fluxes; then we analyse how the presence of anisotropy impacts modifies the expression of the conversion term describing the exchange between kinetic and magnetic energies. Finally, we apply Johnsons's decomposition to study how and to what extent the the physical mechanism behind the energy cascade is altered in (strongly) anisotropic configurations compared to the previously studied isotropic counterpart.

The governing equations for these systems are:
\begin{align}
    &\frac{\partial\bm{u}}{\partial t} + \bm{u} \cdot \nabla \bm{u} = -\nabla \left(p + \frac{B^2}{2} \right) + \bm{B} \cdot \nabla \bm{B} + \nu_{\alpha} (-1)^{\alpha+1} \Delta^{\alpha} \bm{u} + \bm{F} \label{eq:mom_eq} \\
    &\frac{\partial \bm{B}}{\partial t} + \bm{u} \cdot \nabla \bm{B} =  \bm{B} \cdot \nabla \bm{u} + \mu_{\alpha} (-1)^{\alpha+1} \Delta^{\alpha} \bm{B} \label{eq:ind_eq} \\
    &\nabla \cdot \bm{u} =   0 \label{eq:divzeros_vel} \\
    &\nabla \cdot \bm{B} =   0  \label{eq:divzeros}
\end{align}
where $\bm{u}$ are the velocity fluctuations, $\bm{B}$ is the magnetic field, $p$ is the pressure divided by the constant fluid density, $\bm{F}$ a mechanical forcing term, $\nu_{\alpha}$ and $\mu_{\alpha}$ the kinematic hyper-viscosity and magnetic hyper-diffusivity, respectively, and $\alpha$ is the exponent of the Laplacian for hyper-diffusion. Using $\alpha=1$ refers to standard viscosity and diffusivity.  Here, we focus on anisotropic MHD configurations, realised by imposing a constant background magnetic field. %leading to $\langle \bm{B}\rangle \propto {B_0} $. 
Although a Lorenz transformation can always remove the anisotropy in the velocity field by enforcing $\langle \bm{u} \rangle=0$, this does not apply to the magnetic field. Hence, we write the magnetic field in terms of a fluctuation field and a superimposed time-independent and spatially uniform background field 
\begin{equation}
    \bm{B} = \bm{b} + B_0 \bm{\hat{e}_z} \ ,
    \label{eq:full}
\end{equation}
where, without any loss of generality, we set the background component to be aligned with the $z-$axis versor $\bm{\hat{e}_z}$. 
%Clearly, the presence of a constant non-zero background magnetic field breaks isotropy while preserving homogeneity. 
In consequence, eq.~\eqref{eq:ind_eq} is effectively the evolution equation of the magnetic field fluctuations $\bm{b}$ only, where the contribution of the background field is confined to the field-line stretching term $\bm{B} \cdot \nabla \bm{u} = B_0 \partial_z \bm{u} + \bm{b} \cdot \nabla \bm{u}$. Similarly, in the momentum equation the background magnetic field only contributes to the magnetic tension term $\bm{B} \cdot \nabla \bm{B} = B_0 \partial_z \bm{b} + \bm{b} \cdot \nabla \bm{b}$. 

The MHD variables and equations may be spatially coarse-grained using a suitable filtering kernel,
         $ G^\ell (\vr ) $
        \citep{Germano92,Aluie17},
whose role is the suppression of features with a length-scale smaller than the filtering scale $\ell$. The filtered (velocity) field is given by: 
\begin{equation}
      \ol{\vu}^\ell(\vx)
         =
      \int \d^3 r  \,  G^\ell (\vr ) \, \vu ( \vx + \vr) .
  \label{eq:defn-ubar}
\end{equation}
which, being a convolution product, can be interpreted as a weighted average of $\vu$ centered on the
position $\vx$. Moreover, filtering is a linear operation that commutes with differentiation. As for the hypotheses of the filter kernel function we require rapid decay for $\|\bm{r}\| \to \infty$, evenness and having a volume integral of unity. 
%ML remove?
%In this section, we do not specify the functional form of the filter while from sec~\ref{sec:exact} onwards we choose the Gaussian filter.
Coarse-graining each term in  eqs.~\eqref{eq:mom_eq}--\eqref{eq:ind_eq},  using the linearity of the filtering operation and re-writing the advective-type nonlinearities in terms of filtered fields,  requires introducing four sub-gridscale (SGS) stress tensors which have the following form
\begin{align}
      \tau^\ell (f_i, g_j)
       & =
        \ol{f_i g_j}^\ell
      - \ol{f_i}^\ell \, \ol{g_j}^\ell ,
 \label{eq:tau-fg}
\end{align}
where $\vf$ and $\vg$ are the solenoidal vectors relative to the advective-type non-linearity $ g_j \partial_j f_i $. In what follows, the advected field is the first argument in $ \tau^\ell(\cdot,\cdot) $ while the advecting field is the second argument. 

Evolution equations of the coarse-grained kinetic energy $ E^\ell_u (\vx,t)  = \frac{1}{2} \ou \cdot \ou $ and magnetic energy $ E^\ell_b (\vx,t)  = \frac{1}{2} \ob \cdot \ob $ can be obtained by filtering eqs.~\eqref{eq:mom_eq}--\eqref{eq:ind_eq} and scalar multiplication of each term with $\ou$ and $\ob$, respectively \citep[e.g.,][]{ZhouVahala91,KessarEA16,Aluie17,OffermansEA18,AlexakisChibbaro22}
%. The result can be rearranged as:
\begin{align}
%sign changed 13/02/24
  \partial_t {E}^\ell_u
        + \nabla \cdot \vct{{\cal J}}^\ell_u
      & =
        - \Pi^{I,\ell} - \Pi^{M,\ell}
        - {\cal W}^\ell
        - {\cal D}^\ell_{u} 
        + \ou \cdot \ol{\vF}^\ell,
  \label{eq:Eu-ls}
 \\
  \partial_t {E}^\ell_b
        + \nabla \cdot \vct{{\cal J}}^\ell_b
      & =
        - \Pi^{A,\ell} - \Pi^{D,\ell}
        + {\cal W}^\ell
        - {\cal D}^\ell_{b} ,
  \label{eq:Eb-ls}
\end{align}
where the $\vct{{\cal J}}$ terms correspond to the 
spatial transport of energy and the $\Pi$ terms describe the energy fluxes i.e. the energy transfer from scale $\ell$ to the sub-filter scales below $\ell$. 
We work with a sign convention that encodes forward transfer of energy, to scales smaller than $\ell$ by $\Pi > 0 $.
%According to our sign convention for any energy fluxes term, like eqs.~\eqref{eq:Pi-I}-\eqref{eq:Pi-D} defined below, a positive energy-flux, namely $\Pi > 0 $, corresponds to a forward transfer of energy, viz  to scales smaller than $\ell$.
The terms ${\cal D}^\ell$ account for (hyper)dissipative effects.
Finally we define $ {\cal W}^\ell = \ol{b}^\ell_i \ol{B}^\ell_j \partial_j \ol{u}^\ell_i $ as the
 \emph{resolved-scale conversion} (RSC) term, which, unlike the expression used by \cite{Aluie17} and \cite{capocci2025}, here depends on the full magnetic field $\bm{B}$ defined in eq.~\eqref{eq:full}. Since this term appears with opposite sign in each equation and depends only on filtered fields, it represents the energy \emph{exchange} between resolved-scale kinetic and magnetic energies. It is not a flux term. 
 %The RSC term describes energy conversion, and we note that it does not include scale smaller than $\ell$. It is not a flux term.
 %
 The expressions of spatial transport currents $ \vct{{\cal J}}_u, \vct{{\cal J}}_b $ depend on the form used for ${\cal W}^\ell$ and the value of $B_0$, other choices are discussed by \citet[e.g.,][]{KessarEA16,Aluie17,OffermansEA18,AlexakisChibbaro22}. 
Within this variety of expressions, the problem of Galilean invariance has been addressed by \cite{OffermansEA18} for MHD. The sign convention dictated by the form of the RSC term implies that ${\mathcal{W}}^\ell>0$ indicates kinetic-to-magnetic energy conversion while ${\mathcal{W}}^\ell<0$ the opposite i.e. kinetic-to-magnetic energy conversion. 
% ML: move to end of section and  point out that this term and some of the Pis are usually considered together in the fourier representation, the present approach cleanly separates transfer and conversion 

Having introduced all the terms appearing in the resolved-scale kinetic and magnetic energy equations, we now provide explicit expressions of the RSC term and the subfluxes, specifically taking contributions from the background magnetic field into account. That is, the RSC term is split into two components 
\begin{equation}
\mathcal{W}^\ell = 
\mathcal{W}^\ell_{iso} + \mathcal{W}^\ell_{anis} \ , 
\end{equation}
where
\begin{align}
\mathcal{W}^\ell_{iso} & =
 \overline{b}^\ell_i \overline{b}^\ell_j \,  \overline{S}^\ell_{ij} \\ 
 \mathcal{W}^\ell_{anis}  & = \overline{b}^\ell_i  B_0 \, \partial_z \overline{u}^\ell_i \ , 
\label{eq:W_def}
\end{align}
where $\Sbarij$ is the filtered strain-rate tensor. The term $\mathcal{W}^\ell_{iso}$ is that discussed by \cite{Aluie17, OffermansEA18, capocci2025}. 
%The second term contains the contribution from the background magnetic field magnitude with a symmetry breaking as $z-$axis is a preferred direction. 
%
%\red{ML: the B0 terms should be related to the derivative of $\mathcal{W}^\ell_{anis}$, we should calculate and point this out. Then point out that the Fourier representation does not include spatial transport (I think).}
%\damiano{connected to this see the comments above}
The expressions of the four energy subfluxes read
\begin{align}
%sign changed 13/02/24
  \Pi^{I,\ell} & =  - \PD{\ol{u}_i^\ell}{x_j} \, \tau^\ell(u_i,u_j) = -\Sbarji\tau^\ell(u_i,u_j) ,
 \label{eq:Pi-I} 
 \\
  \Pi^{M,\ell} & =  \;\; \PD{\ol{u}_i^\ell}{x_j} \, \tau^\ell(b_i,b_j) = \Sbarji \tau^\ell(b_i,b_j),
 \label{eq:Pi-M} 
 \\
  \Pi^{A,\ell} & =  - \PD{\ol{b}_i^\ell}{x_j} \, \tau^\ell(b_i,u_j) ,
 \label{eq:Pi-A} 
 \\
  \Pi^{D,\ell} & =  \;\; \PD{\ol{b}_i^\ell}{x_j} \, \tau^\ell(u_i,b_j) ,
 \label{eq:Pi-D}
\end{align}
%.. \inhouse{ub order implies u.grad b}
%where $\Jbarji$ is the filtered current tensor. \damiano{I think PiA PiA expressed with J as conctacting field, a la Aluie, would require to write the relative taus via u x b.} ML: correct
All terms defined in eqs.~\eqref{eq:Pi-I}-\eqref{eq:Pi-D} are
 functions of space and scale $\ell$ and, as they tend to zero for $\ell \to 0$, are proper flux terms. Clearly,
        $ \Pi^{I,\ell} + \Pi^{M,\ell} $
is the net flux of $ E^\ell_u $,
and
        $ \Pi^{A,\ell} + \Pi^{D,\ell} $
that for        $ E^\ell_b $, with $\Pi^{I,\ell}$ being the  kinetic energy flux from velocity-field interactions only, $\Pi^{M,\ell}$ the kinetic energy flux arising from interactions of the flow and the magnetic field,  $\Pi^{A,\ell}$ the magnetic energy flux due to the advection of magnetic field by the flow and $\Pi^{D,\ell}$ the magnetic energy flux that arises from changes in magnetic-field line geometry induced by the flow. The latter two, 
   $ \Pi^{A,\ell} $ and  $ \Pi^{D,\ell} $, 
have a common origin and may be readily combined to obtain the magnetic energy flux associated with the curl of the induced electric field, this is reflected in the corresponding SGS stress tensors being related by transposition.  In summary, through the definition of 
%For any choice of filter, these can be expressed in terms ofcontractions of filtered gradient tensors and SGS stress tensors. 
%In consequence of the nature of 
$\tau^\ell ( \cdot, \cdot) $,
there is a clear connection between the fluxes and the four advection type
nonlinearities in
        \eqref{eq:mom_eq}--\eqref{eq:ind_eq},
that we refer to as the Inertial, Maxwell (meaning from the Lorentz force), Advection, and Dynamo terms. 
Notably, eqs~\eqref{eq:Pi-I}--\eqref{eq:Pi-D} are invariant under a constant shift of the field.
%
%Note the capitalization.
 Taken together with eqns.~\eqref{eq:Eu-ls} and \eqref{eq:Eb-ls} 
 these definitions of the $\Pi$'s
 mean than the interpretation of the direction of an energy flux does not depend on which flux it is.
 This is why~\eqref{eq:Pi-M} and~\eqref{eq:Pi-D}
 lack a leading minus sign. 
Specifically, a positive value for 
any one of these fluxes
 % $\Pi^{I,\ell}$ and $\Pi^{A,\ell} $ 
corresponds to transfer of energy from scales greater than $\ell$ to scales smaller than $\ell$.

Before proceeding to a summary of the derivation of exact expressions for the SGS stresses and corresponding energy fluxes, we briefly compare the configuration-space approach used here with Fourier-based formulations of the energy flux commonly used in the literature. Therein, $\Pi^{M,\ell}$, $\mathcal{W}^\ell$ and the contribution to $\mathcal{J}_u$ from the Maxwell stresses are combined into one term, which precludes a clear distinction between interscale kinetic energy transfer, energy conversion and spatial transport of kinetic energy.  Similarly, $\Pi^{I,\ell}$ and the contribution to $\mathcal{J}_u$ from the Reynolds stress, which vanishes under spatial averaging, are combined. Consequences of the latter for fluctuation measurements in hydrodynamics are discussed by \citet{EyinkAluie09} and \citet{AluieEyink09}. A similar situation arises in the magnetic energy budget, where the Fourier-based formulation combines $\Pi^{D,\ell}$, $\mathcal{W}^\ell$ and a contribution to $\mathcal{J}_b$, and similarly, it combines $\Pi^{A,\ell}$ and the contribution to $\mathcal{J}_b$ that vanishes under spatial averaging.
The present formulation clearly distinguishes interscale magnetic energy fluxes from conversion terms and resolved-scale fluctuations that originate from spatial transport of resolved-scale energy. Again, we note that eqs.~\eqref{eq:Pi-I}-\eqref{eq:Pi-D} are indeed fluxes in the sense that they vanish for $\ell \to 0$.

\subsection{Exact expressions for $\tau^\ell$ and $\Pi^\ell$}
\label{sec:exact}

What has been derived so far holds for any generic type of filter kernel. Since we are interested in understanding the fundamental mechanisms governing the energy cascades in (anistropic) MHD turbulence we seek physically interpretable expressions of the energy fluxes, that is, of eqs.~\eqref{eq:Pi-I}--\eqref{eq:Pi-D}. As such, we summarise the methodology introduced in \cite{capocci2025} providing exact expressions of the MHD energy fluxes in terms of field gradients, which is an extension of that originally derived by \cite{Johnson20,Johnson21} for turbulence in non-conducting fluids. Since the nature of the following methodology is purely kinematic, we use two arbitrary fields $\bm{f}$ and $\bm{g}$ that, at the end, they will be replace by either the velocity and/or the magnetic-field fluctuations.

The key observation underlying the method is that for a Gaussian filter%\purple{isotropic} Gaussian filter 
\begin{equation}
    G^\ell (\vr )
        =
    \frac{1}{(2\pi\ell^2)^{3/2}}
       \exp \left( -\frac{|\vr|^2}{2 \ell^2} \right) \ ,       
 \label{eq:Gell}
\end{equation}
 any filtered field  is the solution of a diffusion equation in scale and space with initial data given by the un-filtered field
   \citep{Johnson20,Johnson21} 
\begin{align}
   \PD{ \ol{f}^\ell_j}{(\ell^2)}
  & =
   \frac{1}{2} \nabla^2 \ol{f}^\ell_j,
  \qquad 
    \ol{f}^\ell_j \Bigr\vert_{\ell=0}
   =
     f_j( \vx, t ),
 % j = 1,2,3 ,
 \label{eq:ubar-diffusion}
\end{align}
as the Gaussian is the Green's function for a diffusion equation. This approach can be used to show that the SGS stresses can be written as the solution of a diffusion equation forced by a contraction of field gradients $ \oll{X}_{ik} = \partial_k \oll{g}_i $ and $ \oll{Y}_{jk} = \partial_k \oll{f}_j $
\begin{align}
   \left(
     \PD{}{\ell^2} - \frac{1}{2} \nabla^2
   \right)
   \tau^\ell (g_i, f_j)
  & =
     \oll{X}_{ik} 
     \oll{Y}_{jk}
 ,
  \qquad
     \tau^{\ell=0} (g_i, f_j) = 0 .
 \label{eq:bu-diffusion}
\end{align} 
 %Similarly to \eqref{eq:ubar-diffusion}, is still a diffusion equation with an forcing term which is given in terms of filtered velocity gradients. 
 Eq.~\eqref{eq:bu-diffusion} admits an analytical solution 
  \citep{Johnson21}
\begin{align}
  \tau^\ell (g_i, f_j) 
   =  \int_0^{\ell^2} \d\theta \;
         \ol{ \ol{X}_{ik}^{\sqrt{\theta}} 
              \ol{Y}_{jk}^{\sqrt{\theta}} \,}^\phi 
     = 
  \ell^2 \; \oll{X}_{ik} \oll{Y}_{jk} 
    \; + \;
    \int_0^{\ell^2} \d\theta \; \tau^\phi \left( \ol{X}_{ik}^{\sqrt{\theta}}, \ol{Y}_{jk}^{\sqrt{\theta}} \right) ,
       %\left(
        % \ol{ \ol{X}_{ik}^{\sqrt{\theta}} 
         %     \ol{Y}_{jk}^{\sqrt{\theta}} \,}^\phi 
          % -
         %\ol{\ol{X}_{ik}^{\sqrt{\theta}}\,}^\phi \;
         %\ol{\ol{Y}_{jk}^{\sqrt{\theta}}\,}^\phi  
       %\right),
 \label{eq:tau-ell-bu}
\end{align}
where  $\phi = \sqrt{\ell^2 - \theta} $ and the solution has been split into a term that depends only on resolved-scale field gradients and a remainder that is multi-scale in nature and can be written as an integral over SGS stresses filtered at different scales $\phi \leqslant \ell$ .   
%On the RHS, the first term is opposite in sign to the second addendum in the integrand. This add/removal of the same term ensures that we can safely interpret the first terms a single-scale term, as everything depends on the scale $\ell$, while the second term can be viewed as a \emph{multi-scale} term because more scaled are involved in the integral. 
The first, single-scale, term corresponds to  the first-order term in the SGS stress expansion of \cite{Eyink06-multiscale} or equivalently to the non-linear gradient model used in \cite{leonard1975}, \cite{BorueOrszag98}, and \cite{MeneveauKatz00} in the context of subgrid-scale modelling. In a more general way, the single-scale term is also the leading-order term of the power law expansion in the filter limit going to zero relative of any filter kernel with bounded moments \citep[cf.\ sec 13.4.4 of][]{Pope}. 
%
%Considering the multi-scale term, we observe that it can be expressed as an SGS stress based on field gradients rather than fields viz $ \tau^\phi \left( \ol{X}_{ik}^{\sqrt{\theta}}, \ol{Y}_{jk}^{\sqrt{\theta}} \right) $. 
%
It is of key importance to underline that the solution \eqref{eq:tau-ell-bu} is a purely kinematic result that requires weak restrictions like the differentiability of the fields and the use of a Gaussian filter. %\purple{Although the above methodology can be extended to anisotropic Gaussian kernels, we use the isotropic form in eq.~\eqref{eq:Gell} throughout.}

Having summarised the underlying mathematical approach, we now consider the  gradient-based expansion of the energy fluxes of eqs.~\eqref{eq:Pi-I}--\eqref{eq:Pi-D}. To clarify the structure of the terms that arise from the present gradient-based decomposition of the energy fluxes we focus on the decomposition of the Advection flux as an example, see eq.~\eqref{eq:Pi-A}. This is obtained by setting $\bm{f}=\bm{u}$ and $\bm{g}=\bm{b}$ as entries in the SGS stress expansion on the LHS of eq.~\eqref{eq:tau-ell-bu}, implying that the corresponding gradient tensors become ${X}_{ik} = \partial_k {b}_i$ and ${Y}_{jk} = \partial_k {u}_j$ which have to be filtered accordingly. To obtain the corresponding magnetic energy flux we need to contract the present SGS expansion with the coarse-grained magnetic field gradient tensor i.e. $\overline{X}^\ell_{ij}=\p_j \ol{b}^\ell_i$ and respect the sign convention. 
%where we remind that the sign choice is to associate the positivity of each flux to forward energy transfer. 
This leads to the following expression for the Advection flux
\begin{align}
\label{eq:Pi-A-exact} 
  \Pi^{A,\ell} 
     &= 
	- \ell^2 \, \ol{X}_{ij}^\ell \ol{X}_{ik}^\ell \ol{Y}_{jk}^\ell 
    \; - \;
	\ol{X}_{ij}^\ell
      \int_0^{\ell^2} \d\theta 
       \left(
	 \ol{ \ol{X}_{ik}^{\sqrt{\theta}} 
	      \ol{Y}_{jk}^{\sqrt{\theta}} \,}^\phi 
           -
	 \ol{\ol{X}_{ik}^{\sqrt{\theta}}\,}^\phi \;
	 \ol{\ol{Y}_{jk}^{\sqrt{\theta}}\,}^\phi  
       \right)
 \\
    & =
    \Pi^{A,\ell}_s  +  \Pi^{A,\ell}_m 
 , 
\end{align} 
where the subscripts $s$ and $m$ denote the single- and multi-scale contributions, respectively. 
The tensor contraction in~\eqref{eq:Pi-A-exact} can be expressed as trace of the involved matrix products, after appropriate use of the transpose operation (superscript $t$),
for instance,
    $ \oll{X}_{ij} \oll{X}_{ik} \oll{Y}_{jk}
      = \tr{ \left(\oll{\bm X}\right)^t \, \oll{\bm X} \, \left(\oll{\bm Y}\right)^t}
    $.
    The trace formulation is, apart from notational simplificy, useful in proofs of kinematic identities and relations between different terms.   
As mentioned at the beginning, our aim is to express the energy flux via interpretable observables. For this purpose, we express field gradients as the sum of their symmetric and antisymmetric parts, that is we decompose the the velocity gradient $Y_{ij}$ in terms of strain-rate $ S_{ij} = ( X_{ij} + X_{ji} ) /2 $ and vorticity (or rotation-rate) $ {\Omega}_{ij} = ( X_{ij} - X_{ji} ) / 2 $, tensors, with $ {\Omega}_{ij} = -\epsilon_{ijk} {\omega}_k /2 $
in terms of the vorticity $ {\omega}_k = \epsilon_{ijk} \partial_i {u}_j $.
Similarly, the magnetic-field gradient is decomposed into magnetic strain-rate and current density tensors, $ Y_{ij} = \Sigma_{ij} + {J}_{ij} $, where the non-zero elements of $ J_{ij} = ( Y_{ij} - Y_{ji} ) / 2 = -\epsilon_{ijk} j_k/2 $ are the components of the electric current density $ \vj = \nabla \times \vb $, and  $\Sigma_{ij} = (Y_{ij} + Y_{ji})/2$ is the magnetic strain-rate (or shear) tensor. %All these tensors are clearly required to be filtered in the above discussion.

The present decomposition provides, a priori, eight single-scale and and eight multi-scale sub-fluxes. However, some of the resulting terms  cancel, vanish or coincide as a consequence of properties of the trace. An example of this is given by the single-scale contribution to the Advection flux
\begin{align}
  \tr{ \left(\oll{\bm X}\right)^t \, \oll{\bm X} \, \left(\oll{\bm Y}\right)^t}
    &=
  \tr{ 
     \left( \ovSS^\ell - \ovJ^\ell   \right)
     \left( \ovSS^\ell + \ovJ^\ell  \right)
     \left( \ovS^\ell - \ovOm^\ell \right)
    } 
  \label{eq:BBA-sym-anti-prelim}
 \\
     &=
   \tr{   
	  \ovSS^\ell   \ovSS^\ell   \ovS^\ell 
       -  \ovJ^\ell  \ovJ^\ell \, \ovS^\ell 
       + 2 \ovSS^\ell  \ovJ^\ell  \ovS^\ell
     },
  \label{eq:BBA-sym-anti}
\end{align}
which comprises only three distinct subfluxes. We write the single-scale flux  
for the Advection term by using the following symbols for each term as
\begin{align}
  \Pi^{A,\ell}_s 
    & = 
        \Pi^{A, \ell}_{s, \Sigma \Sigma S}
    +   \Pi^{A, \ell}_{s, J J S}
    +  2\Pi^{A, \ell}_{s, \Sigma J S}
,
 \label{eq:Pi-A-ss}
\end{align}
where
$ \Pi^{A,\ell}_{s, PQR} = -\ell^2 \tr{\left(\oll{\bm P}\right)^t \oll{ \bm Q} \left(\oll{ \bm R}\right)^t} $ and each of $ \bm P, \bm Q,\bm R$ are either symmetric or antisymmetric tensors. 
The multi-scale flux contributions, can be similarly decomposed
\begin{align}
   \Pi^{A,\ell}_m &=  \tr{ \left(\oll{\bm X}\right)^t \,  \tau^{\phi} \left( 
        \ol{\bm X}^{\sqrt{\theta}},
        \left(\ol{\bm Y}^{\sqrt{\theta}}\right)^t  \right) } \notag \\
        &= \tr{ \left( \ovSS^\ell - \ovJ^\ell   \right) \,  \tau^{\phi} \left( 
        \left( \ovSS^\ell + \ovJ^\ell   \right),
        \left( \ovS^\ell - \ovOm^\ell \right)  \right) } \ .
\label{eq:Pi-A-ms}
\end{align}
\begin{comment}
    \begin{align}
   \Pi^{A,\ell}_m &=  \tr{ \left(\oll{\bm X}\right)^t \,  \tau^{\phi} \left( 
        \ol{\bm X}^{\sqrt{\theta}},
        \left(\ol{\bm Y}^{\sqrt{\theta}}\right)^t  \right) } \\
        &= \tr{ \left( \ovSS^\ell - \ovJ^\ell   \right) \,  \tau^{\phi} \left( 
        \left( \ovSS^\ell + \ovJ^\ell   \right),
        \left( \ovS^\ell - \ovOm^\ell \right)  \right) } = \\       
  & +    \Pi^{A,\ell}_{m, \Sigma \Sigma S}
   +    \Pi^{A,\ell}_{m, \Sigma \Sigma \Omega} 
   +    \Pi^{A,\ell}_{m, \Sigma J S}
   +    \Pi^{A,\ell}_{m, \Sigma J \Omega}
   +    \Pi^{A,\ell}_{m, J \Sigma S}  
   +    \Pi^{A,\ell}_{m, J \Sigma \Omega}
   +    \Pi^{A,\ell}_{m, J J S}
   +    \Pi^{A,\ell}_{m, J J\Omega} 
\end{align}
\end{comment}
%in which, similarly to what was done above, each tensor is then written as the sum of the symmetric and antisymmetric parts. 
%DC: noted
A more generalised approach to provide expressions for the SGS stresses in eq.~\eqref{eq:tau-ell-bu} to decompose fluxes of quantities such as kinetic, magnetic and cross helicities is discussed in Appendix 1 of \cite{capocci2025}.
%ML -- list the interpretations here?
Since the SGS stress tensor and the field gradients are invariant under constant shifts of the underlying fields, the number and type of terms appearing in the exact flux decomposition is unchanged by the presence of a background magnetic field. That is, the decompositions of each flux given in eqs.~\eqref{eq:Pi-I}--\eqref{eq:Pi-D} in terms of field gradients and the physical interpretation of the resulting expressions provided by \cite{capocci2025} carry over to the anisotropic case. In what follows, we briefly describe these decompositions, exact expressions for each subflux term are given in Appendix~\ref{app:defintions}.
A numerical evaluation of all subflux terms from DNS data is provided in sec.~\ref{sec:fluxes} for each strong, weak and zero mean magnetic field. 

\subsection{Advection and Dynamo fluxes decomposition}
\label{sec:Pi-A_intro}
\begin{figure}
	\begin{center}
         \includegraphics[width=\columnwidth]{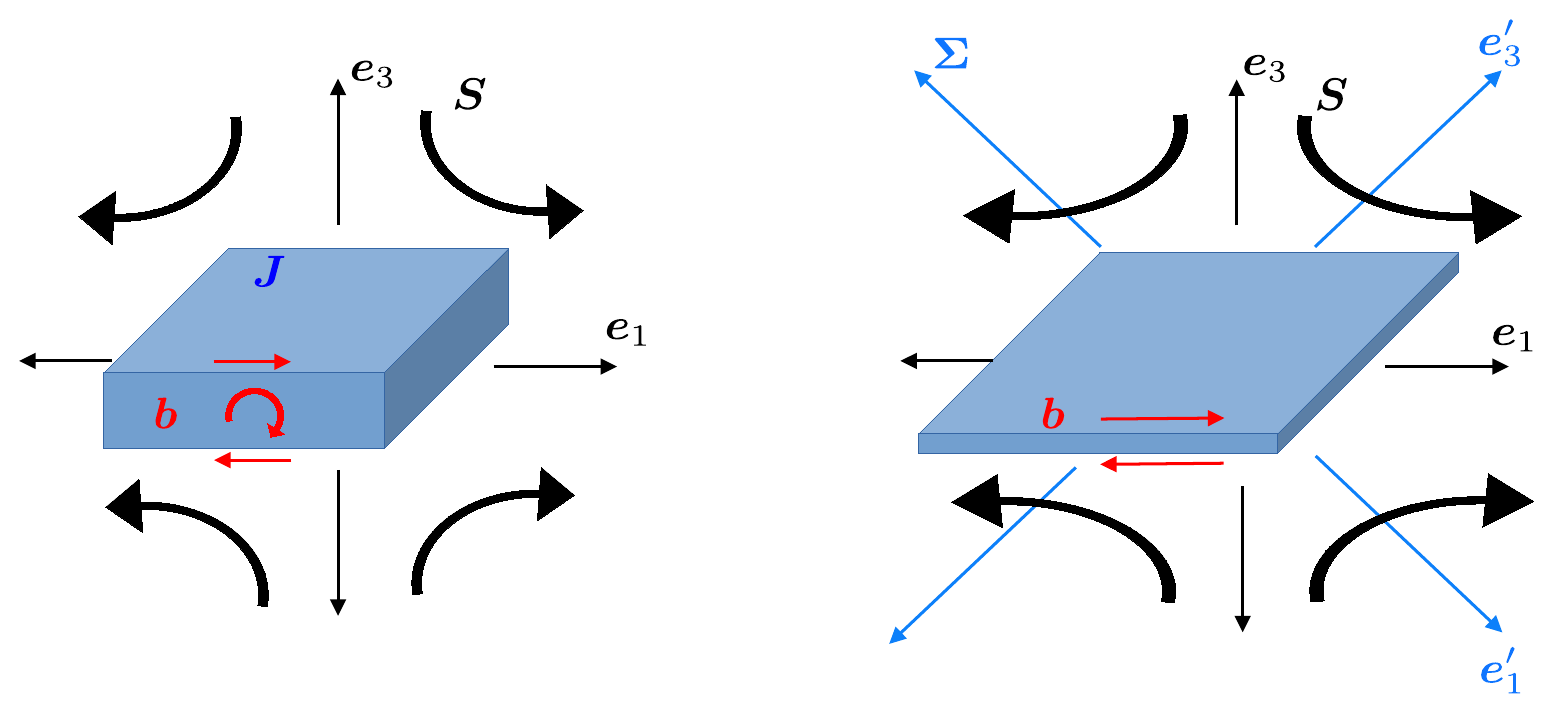} 
    \end{center}
	%\vspace{-5em}
	 \caption{
Two-dimensional sketch of current-sheet thinning. A current sheet, $\bm{J}$, is stretched by the large-scale strain-rate tensor $\bm{S}$ (left), producing a thinner magnetic shear layer $\bm{b}$ (right, red arrows). This process stretches the magnetic flux tubes within the sheet and, by conservation of magnetic flux, increases the magnetic-field strength at the newly generated smaller scales. Magnetic energy is therefore transferred from large to small scales.
The resulting magnetic shear layer is associated with a magnetic strain-rate field $\bm{\Sigma}$, whose principal axes (solid blue arrows) are oriented at $45^\circ$ with respect to those of the large-scale velocity strain-rate tensor (solid black arrows). Since the magnetic shear tends to align with the extensional direction of $\bm{S}$, the Lorentz force accelerates the fluid along the extensional directions and decelerates it along the compressional directions. The corresponding back-reaction on the velocity field arises by amplification of small-scale strain aligning with $\Sigma$.
%This feedback generates an enhanced small-scale strain-rate field, $\bm{S}'$, indicated by dashed arrows. 
The principal axes of the large-scale strain-rate tensor are denoted by $\bm{e}_1$ for an extensional direction and $\bm{e}_3$ for a contractile direction; the corresponding small-scale axes are labelled $\bm{e}_1'$ and $\bm{e}_3'$.
}
\label{fig:sketch-cst}
\end{figure}

We begin with the Advection flux, as it has been used above as an example to explain the SGS stress decomposition procedure. Hence, for the decomposition of eq.~\eqref{eq:Pi-A}, we refer directly to eqs.~\eqref{eq:Pi-A-ss} for the single-scale terms while to eq.~\eqref{eq:Pi-A-ms} for the multi-scale counterpart. Taken these terms together, the decomposition of the Advection flux reads
\begin{align}
   \Pi^{A,\ell} =\ &  
        \Pi^{A,\ell}_{s, \Sigma \Sigma S} 
   +    \Pi^{A,\ell}_{m, \Sigma \Sigma S}
   +    \Pi^{A,\ell}_{m, \Sigma \Sigma \Omega}
   +    \Pi^{A,\ell}_{s, \Sigma J S}
   +    \Pi^{A,\ell}_{m, \Sigma J S}
   +    \Pi^{A,\ell}_{s, \Sigma J \Omega}
  \\
   +&    \Pi^{A,\ell}_{m, \Sigma J \Omega}
   +   \Pi^{A,\ell}_{s, J \Sigma S}
   +    \Pi^{A,\ell}_{m, J \Sigma S}
   +    \Pi^{A,\ell}_{s, J \Sigma \Omega}
   +    \Pi^{A,\ell}_{m, J \Sigma \Omega}
   +    \Pi^{A,\ell}_{s, J J S}
   +    \Pi^{A,\ell}_{m, J J S}
   +    \Pi^{A,\ell}_{m, J J\Omega} %zero in s-sc
 ,
  \label{eq:adv_total} 
\end{align}
where we do not list terms that vanish identically, see  Appendix~\ref{app:defintions}. As shown in \citet{capocci2025}, several of these terms are also related, at the level of their mean values, by Betchov-type identities that link different Advection contributions to one another.
For a physical interpretation, terms of type $J \Sigma S$ are associated with current-sheet thinning, where large-scale straining motion first stretches a current sheet into a magnetic shear layer, in a manner analogous to vortex thinning in hydrodynamics. In doing so, it stretches the magnetic flux tubes embedded in the sheet and, by conservation of magnetic flux, enhances the magnetic-field strength at the newly generated smaller scales. This corresponds to a transfer of magnetic energy from large to small scales. This process also feeds back on the velocity field, amplifying smaller-scale strain by alignment with the magnetic strain associated with the newly formed magnetic shear layer. A schematic drawing of this process is provided in fig.~\ref{fig:sketch-cst}. We note that the single scale contribution to this process occurs twice, since the cyclic property of the trace gives $\Pi^{A,\ell}_{s,J \Sigma S}=\Pi^{A,\ell}_{s,\Sigma J S}$. 

\begin{figure}
	\begin{center}
         \includegraphics[width=\columnwidth]{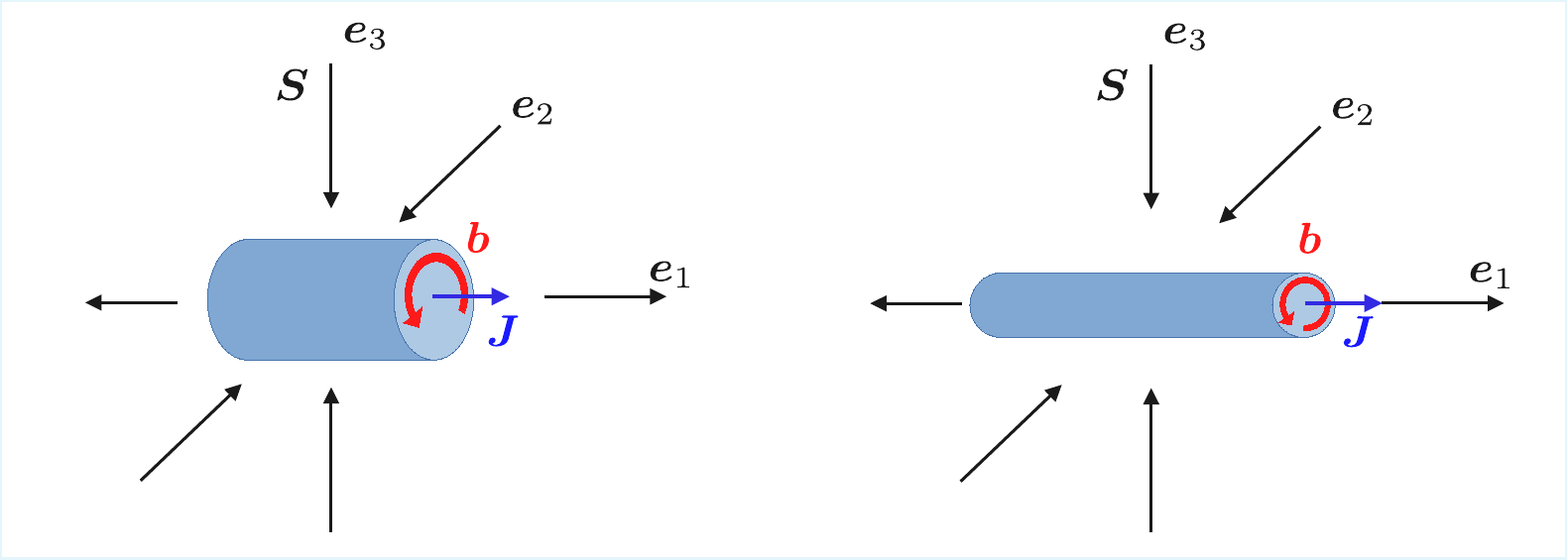} 
    \end{center}
	%\vspace{-5em}
	 \caption{
{Sketch of current-filament stretching. A current filament, $\bm{J}$, relative to the magnetic field $\bm{b}$, is stretched by the large-scale strain-rate tensor $\bm{S}$, which has one extensional and two compressional directions. The initially compact filament (a) is thereby elongated and thinned along the extensional direction (b). This deformation stretches the magnetic flux tubes within the filament and, by conservation of magnetic flux, increases the magnetic-field strength at the newly generated smaller scales (red arrows). Magnetic energy is therefore transferred from large to small scales. This mechanism is analogous to vortex stretching in hydrodynamic turbulence. The principal axes of the large-scale strain-rate tensor are denoted by $\bm{e}_1$ along the extensional direction and by $\bm{e}_2$ and $\bm{e}_3$ along the compressional directions.}
}
\label{fig:sketch-cfs}
\end{figure}

The $JJS$ subfluxes can be interpreted as a current filament stretching processes in which the large-scale strain motion stretches an electric current filament in a process analogous to vortex stretching in hydrodynamics, see fig.~\ref{fig:sketch-cfs} for a schematic representation. As regards the other terms, the $\Sigma \Sigma S$ contributions describe the
amplification of magnetic shear by strain motions whereas $\Sigma J \Omega$ terms can be interpreted as the bending of magnetic field lines into current filaments by rotational
flow, i.e., production of magnetic-field gradient induced by vortical motion, and terms of type $J J \Omega$ quantify current amplification by rotational motion.

Finally, to derive the exact decomposition of the Dynamo term $\Pi^{D,\ell} $ from eq.~\eqref{eq:Pi-D}, substitute $\bm{g}=\bm{u}$ and $\bm{f}=\bm{b}$ in eq.~\eqref{eq:tau-ell-bu} and contract the resulting expression with the magnetic field gradient tensor, resulting in
\begin{align}
    \Pi^{D,\ell} =\ &  
    \Pi^{D,\ell}_{s,\Sigma S \Sigma} 
+   \Pi^{D,\ell}_{m,\Sigma S \Sigma}   
+   \Pi^{D,\ell}_{s,\Sigma S J} 
+   \Pi^{D,\ell}_{m,\Sigma S J} 
+   \Pi^{D,\ell}_{m,\Sigma \Omega \Sigma}
+   \Pi^{D,\ell}_{s,\Sigma \Omega J}         %%same
+   \Pi^{D,\ell}_{m,\Sigma \Omega J}\\
+&   \Pi^{D,\ell}_{s,J S \Sigma}
+    \Pi^{D,\ell}_{m,J S \Sigma}
+  \Pi^{D,\ell}_{s, J S J} 
+   \Pi^{D,\ell}_{m, J S J}
+   \Pi^{D,\ell}_{s, J \Omega \Sigma}        %%same
+   \Pi^{D,\ell}_{m, J \Omega \Sigma}
+   \Pi^{D,\ell}_{m,J \Omega J} 
,
 \label{eq:dyn_total}
\end{align}
where terms that vanish individually are omitted; see Appendix~\ref{app:defintions}. Moreover, by using the cyclic property of the trace, one finds that $ \Pi^{D,\ell}_{s,\Sigma S J} =-\Pi^{D,\ell}_{s, J S \Sigma }$ so that these two single-scale contributions cancel exactly and therefore do not contribute to the net flux. As a result, the Dynamo subflux is predominantly multiscale. With reference to the Advection subflux decomposed and interpreted above, the following subflux pairs are opposite in sign, $ \Pi^{D,\ell}_{s,\Sigma S \Sigma} $ and $ \Pi^{A,\ell}_{s,\Sigma \Sigma S} $ as well as $ \Pi^{D,\ell}_{s, \Sigma \Omega J} $ and
$ \Pi^{A,\ell}_{s, \Sigma J \Omega} $ 
alongside their multi-scale counterparts.
Therefore, these terms cancel pairwise and do not contribute to the net magnetic energy flux where Advection and Dynamo fluxes, i.e. eqs.~\eqref{eq:Pi-A}--\eqref{eq:Pi-D}, are summed. On the other hand, 
$ \Pi^{D,\ell}_{s,J S J} ,$ and $ \Pi^{A,\ell}_{s,J J S} $ together with the relative multi-scale terms, are equal and thus contribute to the total flux. These redundancies and cancellations reflect the common physical origin of the Advection and Dynamo fluxes, both of which arise from the electric field.

\subsection{Maxwell flux}
\label{sec:Pi-M_intro}
The decomposition of the Maxwell flux, i.e. eq.~\eqref{eq:Pi-M}, is obtained by setting $\bm{f}=\bm{g}=\bm{b}$ in the general gradient decomposition of the SGS stress tensor, eq.~\eqref{eq:tau-ell-bu}, and subsequently contracting with the filtered strain-rate tensor, resulting in 
\begin{equation}
\label{eq:dec_maxwell}
    \Pi^{M,\ell} =  \Pi^{M,\ell}_{s,S  \Sigma  \Sigma}  +  \Pi^{M,\ell}_{m,S  \Sigma  \Sigma}  +   \Pi^{M,\ell}_{s, S J J}  +  \Pi^{M,\ell}_{m, S J J}  +  \Pi^{M,\ell}_{s, S J  \Sigma}  + \Pi^{M,\ell}_{m, S J  \Sigma} .
\end{equation}
 The definitions of each subflux are listed in Appendix.~\ref{app:defintions}. Terms of type $S\Sigma\Sigma$ are interpreted as strain amplification by magnetic shear, similar to strain self-amplification in hydrodynamics, whereas those of type $SJJ$ describe the back-reaction of current-filament stretching, the magnetic analogue of vortex stretching,  on the flow. More precisely, this process corresponds to the production of velocity gradients induced by the elongation of current filaments.  
Finally the last pair of terms, of type $SJ\Sigma$, quantify the feedback of the magnetic field onto the flow through amplification of velocity strain at smaller scales by the aforementioned current-sheet-thinning process, where we observe that the single-scale term $\Pi^{M,\ell}_{s,SJ\Sigma}$, see the definition in Appendix~\ref{app:defintions}, is identical to the Advection flux term $\Pi^{A,\ell}_{s,J \Sigma S}$, introduced in 
in sec.~\ref{sec:Pi-A_intro}, by virtue of the cyclic property of the trace. This term has been previously identified as the stretching, and by incompressibility the thinning, of current sheets into magnetic shear layers. 
%As shown in below this is the dominant mechanism contributing to the Maxwell subflux. 
The resulting small-scale magnetic shear layer has a magnetic rate-of-strain field that accelerates the fluid along its extensional directions and decelerates it along its compressional ones, thereby amplifying a suitably aligned velocity strain-rate field at smaller scales. 
The connection between the Maxwell and Advection subfluxes is not limited to identities following from the cyclic property of the trace. As shown in \citet{capocci2025}, several of these terms are also related, at the level of their mean values, by Betchov-type identities that link different Advection contributions to one another and to the Maxwell flux.
For the case of zero background magnetic field studied by \cite{capocci2025}, the single- and multi-scale  current-sheet thinning terms are the dominant processes in the mean energy transfer.

\subsection{Inertial flux}
\label{sec:Pi-I_intro}
 The exact decomposition of the Inertial term $\Pi^{I,\ell} $, eq.~\eqref{eq:Pi-I}, is obtained by by applying the general identity eq.~\eqref{eq:tau-ell-bu} with the choice $\bm{f}=\bm{g}=\bm{u}$ and then contracting the resulting SGS expansion with the filtered as prescribed by eq.~\eqref{eq:Pi-I}, resulting in \citep{capocci2025}
\begin{equation}
 \label{eq:inertial_dec}
    \Pi^{I,\ell} =  \Pi^{I,\ell}_{s, SSS}  +  \Pi^{I,\ell}_{m,SSS}  +   \Pi^{I,\ell}_{s, S \Omega \Omega}  +  \Pi^{I,\ell}_{m,S \Omega \Omega}  +  \Pi^{I,\ell}_{m, S \Omega S} 
\end{equation}
where the individual expressions of each subflux are reported in Appendix~\ref{app:defintions}. The above expression is the inertial-flux decomposition already obtained for hydrodynamic turbulence \citep{Johnson20,Johnson21}, now applied to the MHD Inertial subflux. The two terms of type $SSS$ are interpreted as strain self-amplification, the two terms of type $S\Omega\Omega$ are associated with vortex stretching, and the term of type $S\Omega S$ is related to vortex-thinning \cite{Johnson21,chen2006,kraichnan1976}. The latter appears only as a multi-scale contribution, since its single-scale counterpart vanishes identically \citep{Johnson20}, it its the only process that remains active in two-dimensional turbulence, where it governs the inverse energy cascade. Further discussions and schematic drawings of these processes have been provided in \cite{Johnson21,johnson2024}.

\section{Numerical methods and data}
\label{sec:numerics}

The results discussed so far are analytical: they provide exact expressions for the energy fluxes, their decompositions, and the resolved-scale conversion term. Quantifying these observables requires data. In this work, we use DNS data from three MHD simulations that differ in the strength of the background magnetic field introduced in eq.~\eqref{eq:full}. This strength is measured relative to the magnetic-field fluctuations through the ratio $B_0/B_{\rm rms}$, allowing us to distinguish zero-, weak-, and strong-BMF cases, corresponding respectively to $B_0/B_{\rm rms}=0$, $1.2$, and $12.7$. The datasets are publicly available through the SMART-Turb portal at \url{http://smart-turb.roma2.infn.it} and are documented in detail in \citet{capocciMHDdata}. Here, we provide only a brief overview of the nuemrical similations setup relevant to the present analysis.

In this regard, eqs.~\eqref{eq:mom_eq}--\eqref{eq:divzeros} are solved in a three-dimensional periodic domain using a standard pseudospectral method, where the time-stepping is performed using a second-order Runge--Kutta scheme with dealiasing performed by the two-thirds rule \citep{PattersonOrszag71,CanutoEA}. To obtain a satisfactory separation of scales, we focus on the hyperdiffusive datasets, for which the Laplacian power is $\alpha=5$, rather than on standard diffusive MHD, corresponding to $\alpha=1$. In all cases considered here, the magnetic Prandtl number is unity, $\mu_\alpha=\nu_\alpha$. The mechanical forcing $\bm{F}$, see eq.~\eqref{eq:mom_eq}, applied to the system is a drag-free Ornstein–Uhlenbeck process, active in the wavenumber band $k \in [2.5,5.0]$ for each configuration. This forcing scheme presents minimal injection of cross helicity as well. Key parameters and observables are summarised in table~\ref{tab:datasets}.

\begin{figure}
        \begin{center}
        %\hspace{0.2cm}
         \noindent\makebox[\textwidth]{
         \hspace{0.18cm}
         \includegraphics[width=.38\columnwidth]{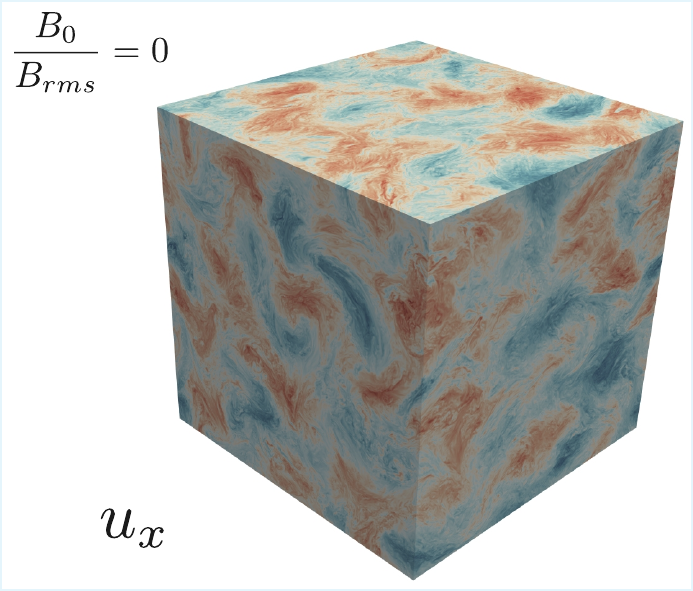}
         \hspace{0.2cm}
         \includegraphics[width=.35\columnwidth]{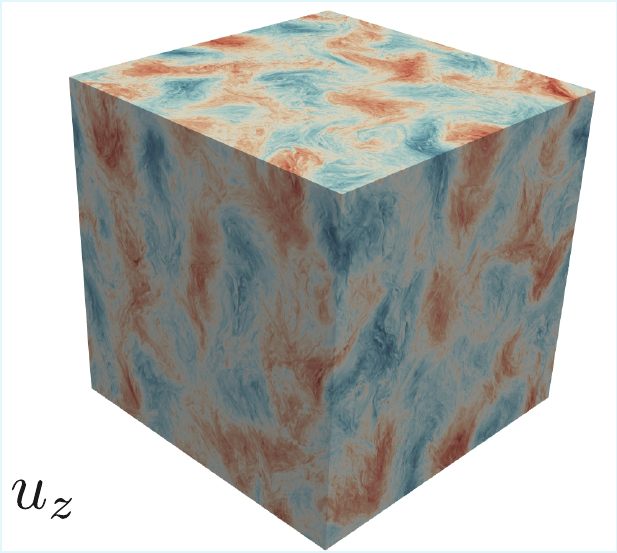}
         \hspace{-0.4cm}
         \includegraphics[width=.46\columnwidth]{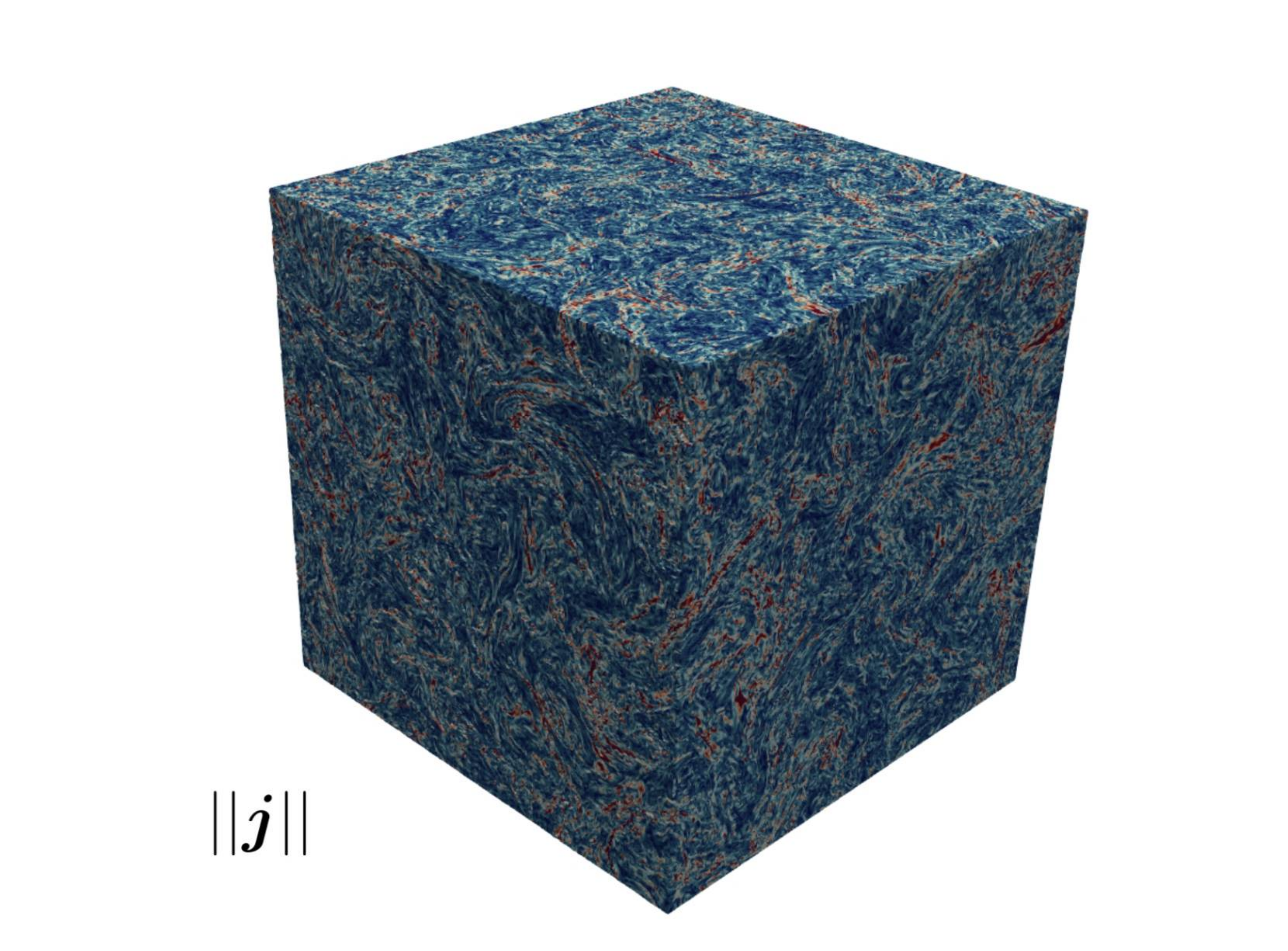}
         }
         \noindent\makebox[\textwidth]{
         \vspace{1cm}
         \hspace{0.2cm}
         \includegraphics[width=.38\columnwidth]{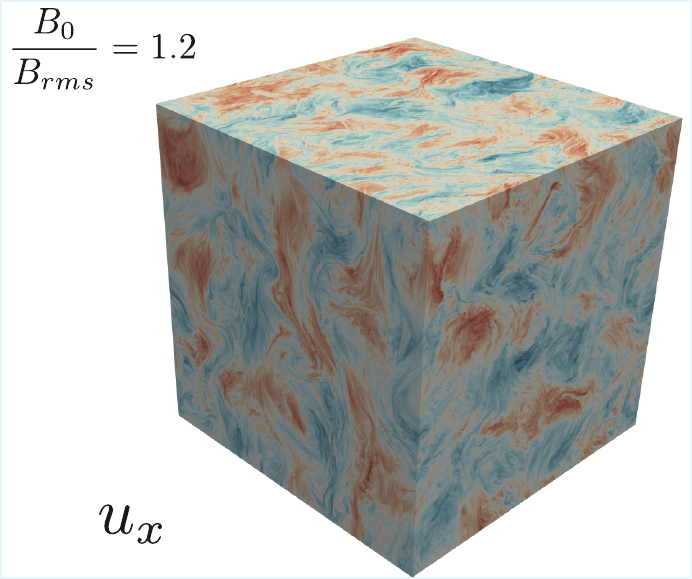}
         \includegraphics[width=.36\columnwidth]{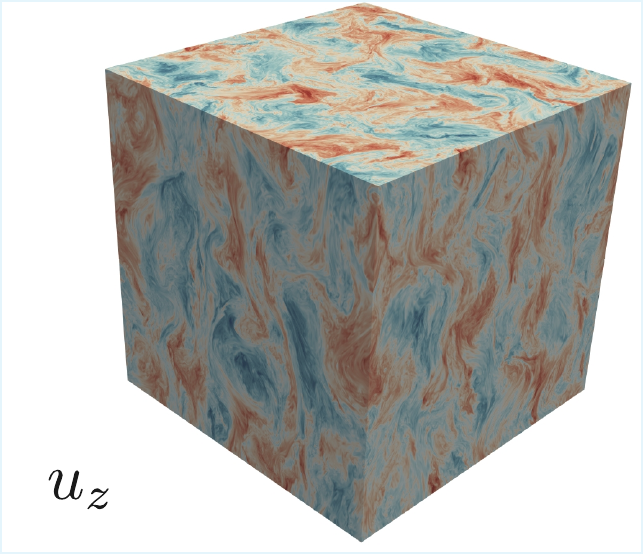}
         \hspace{-0.3cm}
         \includegraphics[width=.47\columnwidth]{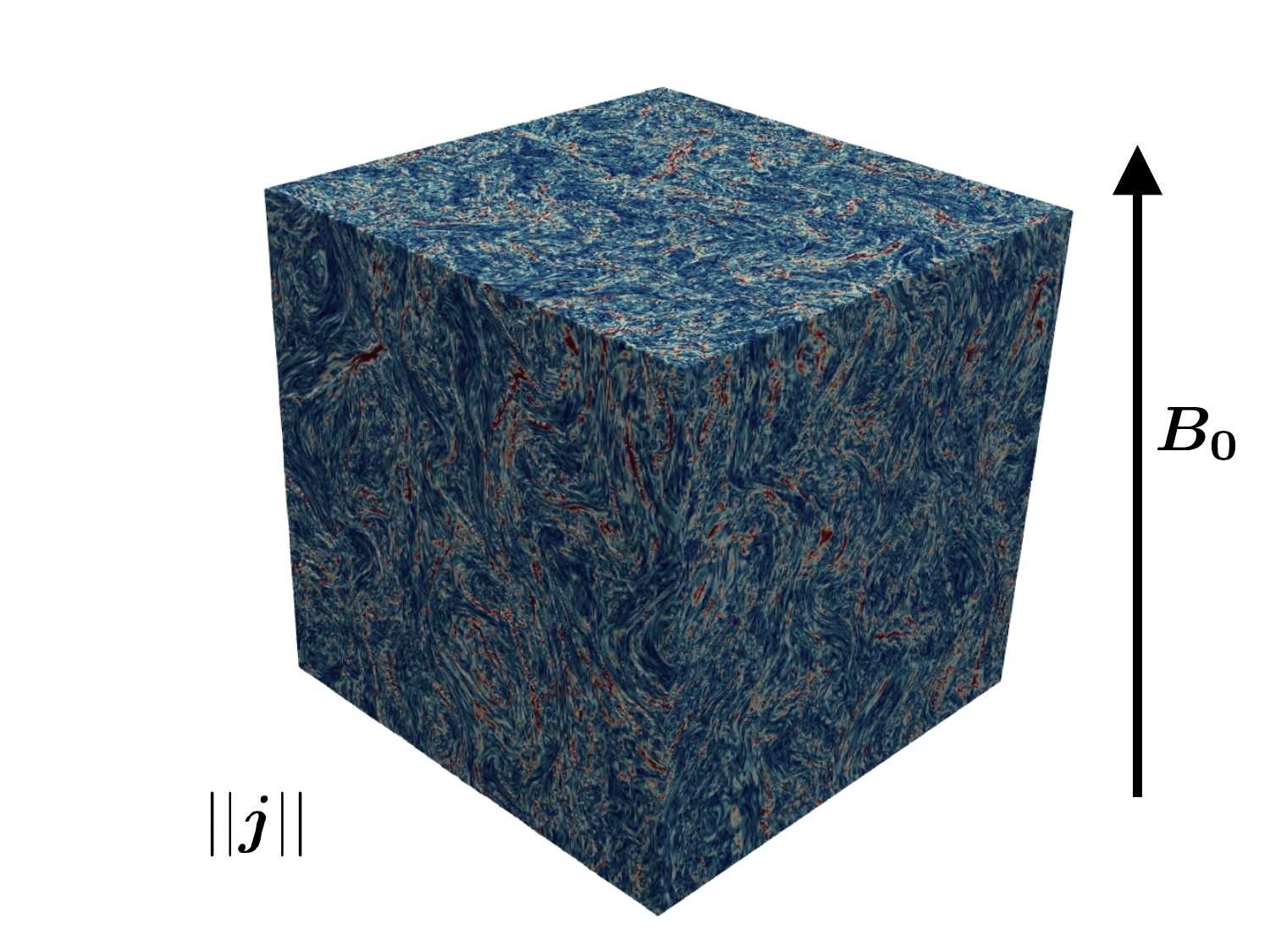}
         }
         \noindent\makebox[\textwidth]{
         \vspace{1cm}
         \hspace{0.3cm}
         \includegraphics[width=.38\columnwidth]{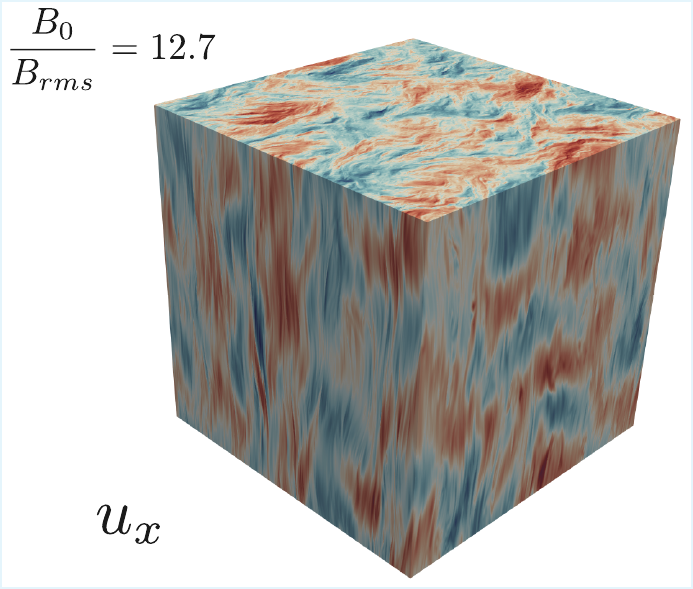} 
         \hspace{0.18cm}
         \includegraphics[width=.34\columnwidth]{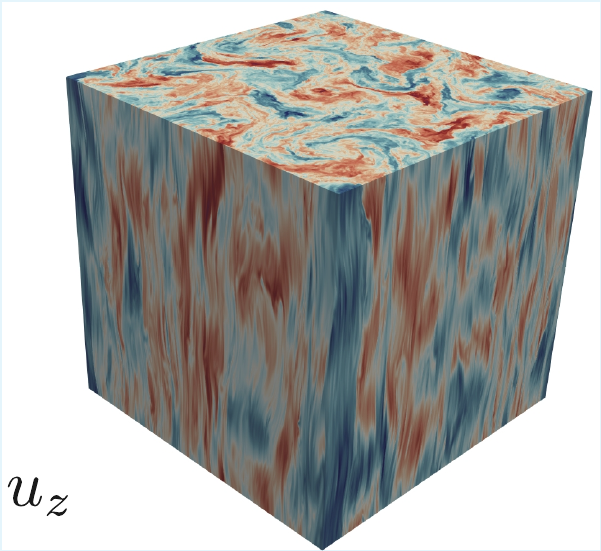} 
         \vspace{0.5cm}
         \hspace{-0.3cm}
         \includegraphics[width=.47\columnwidth]{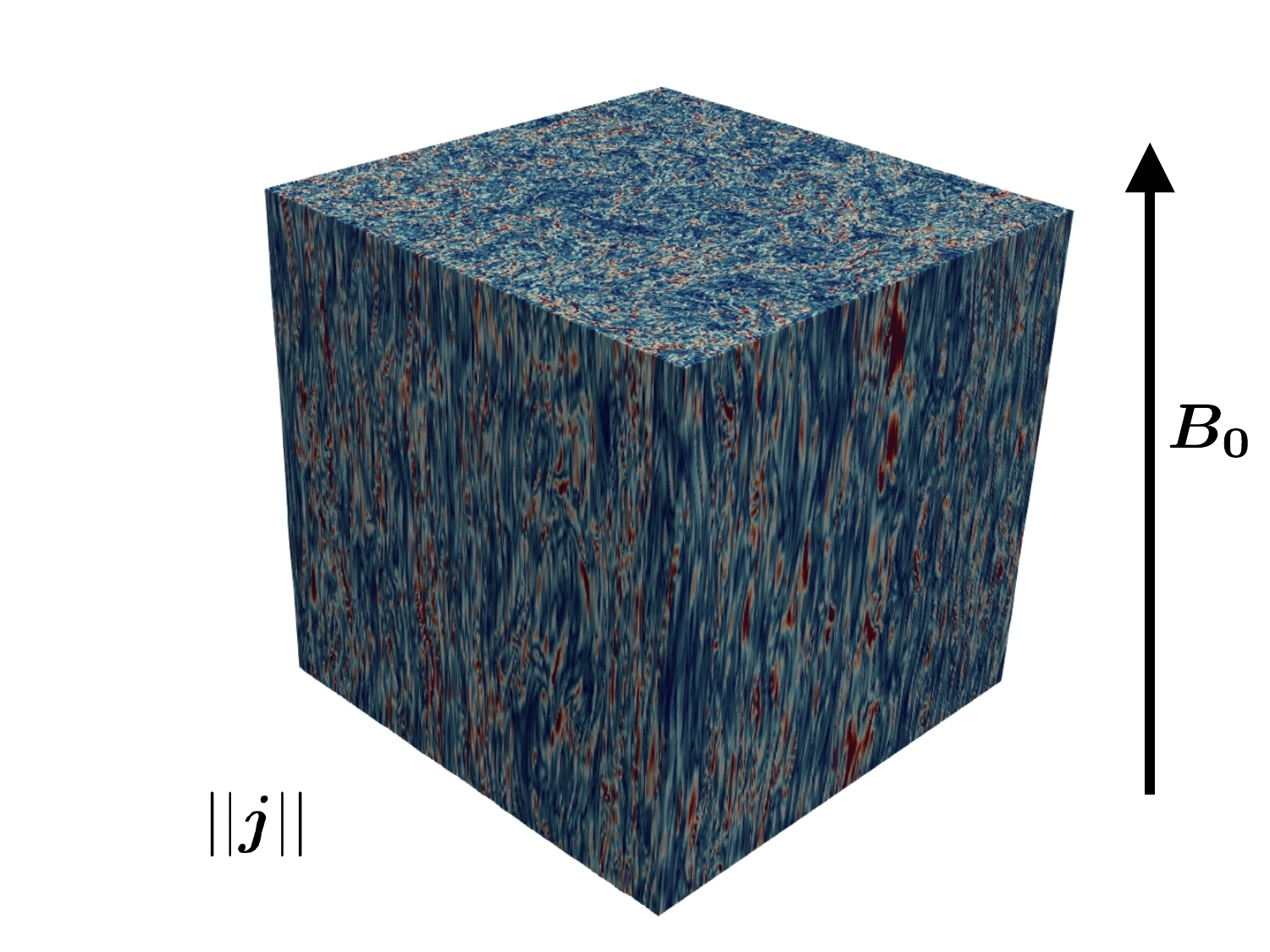} 
         %CAVE: the choice of spacing and size parameters was the result of a long trial&error. 
         }
    \end{center}
	 \caption{3D visualisation of fields. Each horizontal panel is formed by three visualisation cubes showing $u_x$ and the magnitude of $\bm{j} = \nabla \times \bm{b}$ as a function of the position $(\bm{x},\bm{y}$, $\bm{z})$. Top panel $B_0=0$, middle panel $B_0=1.2 \, B_{rms}$ and bottom panel $B_0=12.7 \, B_{rms}$. Visuaisations of the same field share the colour bar range. The arrays indicate the direction of the BMF $\bm{B}_0 = B_0 \,\hat{e}_z$. %\purple{placeholder: just say BMF.} %\orange{NOTE: that the pictures have been reduced in (memory) size to make the compilation time quicker. In the figures folder there are the higher resolution versions as well.}
     }
\label{fig:visuals}
\end{figure}

Figure \ref{fig:visuals} presents visualisations pertaining to single snapshots of $u_x$, $u_z$ and the magnitude of the electric current $j=\bm{\nabla} \times \bm{B}$ for each dataset. These visualisations provides a first qualitative indication of the increasing anisotropy induced by the BMF. in particular, the strong BMF case displays clear signatures of a two-dimensionalisation at the bottom row of panels of fig.~\ref{fig:visuals}. Here, the fields show extended coherent structures aligned with $\bm{B}_0$. In planes perpendicular to $\bm{B}_0$, by contrast, the velocity fluctuations are organised in smaller and more fragmented patches than in the zero and weak BMF cases. The weak BMF configuration displays shows an intermediate geometry i.e. with more broken and filamented patches than the $B_0=0$ case but without the strong anisotropy relative to the $B_0/B_{rms}=12.7$ case. Moreover, the stronger colouring in the strong-BMF visualisations reflects the higher kinetic-energy level reported in table~\ref{tab:datasets}, associated with the ongoing two-dimensionalisation.

%A similar difference is visible in the current field $\bm{j}$, with the $B_0/B_{rms}=1$ case appearing slightly more disordered than the $B_0/B_{rms}=0$ case. 
%Similar considerations apply to the visualisation of the electric current magnitude.

\begin{table}
  \begin{center}
\def~{\hphantom{0}}
   \begin{tabular}{ccccccccccccccccc}
        \hline
        \hline
		 id & $\frac{B_0}{B_{\rm rms}}$ & $E_u$ & $E_b$ & $\nu_\alpha$ & $\eps_u$ & $\eps_b$ & $L_u$ & $\tau$  & $\mbox{Re}$ & $\dfrac{k_\text{max}}{k_\text{diss}^u}$ & $\dfrac{k_\text{max}}{k_\text{diss}^b}$ & $\Delta t / \tau$ & \# \\
        \hline        
      A3 & 0  & 0.70 & 0.48 & $5 \times 10^{-23}$ & 0.31 & 0.43 & 0.53  & 0.78  & 4272  & 1.45 & 1.43 & 1.5  & 26  \\
      C1 &  1.2 & 0.73 & 0.76 & $5 \times 10^{-23}$ & 0.32 & 0.42 & 0.44  & 1.05   & 2742   & 1.45 & 1.43 & 1.1  & 26  \\
      C10 &  12.7 & 3.52 & 0.31 & $5 \times 10^{-23}$ & 0.32 & 0.40 & 1.56  & 1.08   & 7501   & 1.45 & 1.44 & 0.8  & 19  \\

        \hline
      \hline
        \end{tabular}
        \caption{Simulation parameters and key observables, where
        $E_u$ and $E_b$ are the mean kinetic and magnetic energies,
        $\nu_\alpha$ the kinematic (hyper)viscosity,
        $\eps_u$ and $\eps_b$ are the kinetic and Joule energy dissipation rates,
        $L_u = (3 \pi/4 E_u) \int_0^{k_\text{max}} dk \ E_u(k)/k$ the
        %longitudinal
        integral scale,
        $\tau = L_u/\sqrt{2E_u/3}$ the large-scale eddy-turnover time, $B_{\rm rms} = \sqrt{2E_b}$ the root-mean-square value of the magnetic field fluctuations. Each dataset employs a Laplacian power $\alpha=5$ used in the hyperdiffusion. $\eps_u$ and $\eps_b$ are the kinetic and magnetic energy dissipation rates. $\mbox{Re}= C \left(L_u/I_d \right)^{4/3}$ is the (effective) Reynolds number \citep{buzzicotti2018} calculated in terms of the ratio of the integral scale and the scale with maximum dissipation $I_d=\pi/\mathrm{argmax}\left(k^2\, E_u(k)\right)$ with $C=40$ being a fit constant determined in \cite{capocci2025}.
	  Conversely, $k_\text{diss}^u$ and $k_\text{diss}^b$ are the wavenumbers associated with the hyperdiffusive Kolmogorov scales $\eta^u_{\alpha} = (\nu_\alpha^3 / \varepsilon_u)^{1 / (6  \alpha -2)}$ and $\eta^b_{\alpha} = (\mu_\alpha^3 / \varepsilon_b)^{1 / (6  \alpha -2)}$ respectively calculated in terms of the hyperviscous and Joule hyperdissipation rates,
        $k_\text{max}$ the largest retained wavenumber component after dealiasing, 
        $\Delta t$ the mean of the snapshots sampling intervals,
        and \# indicates the number of snapshots used in the statistics. {Each configuration employs $N=1024^3$ grid points.}
        The magnetic Prandtl number,  
        $Pm = \nu_\alpha / \mu_\alpha $ defines the ratio between the hyperviscosity and magnetic hyperdiffusivity, equals unity for each dataset. Regarding C10, the parameters $E_u$, $E_b$, $L_u$, $\tau$ and $Re$ , are related to the last stationarity interval of the run corresponding to $t/\tau \in [95, 98] $ (see fig.~\ref{fig:B0_10}). 
        %\red{ML: Please include a very brief discussion of the definition of Re}
        }
   \label{tab:datasets}
  \end{center}
\end{table}

\begin{figure}
	\begin{center}
         \includegraphics[width=0.85\columnwidth]{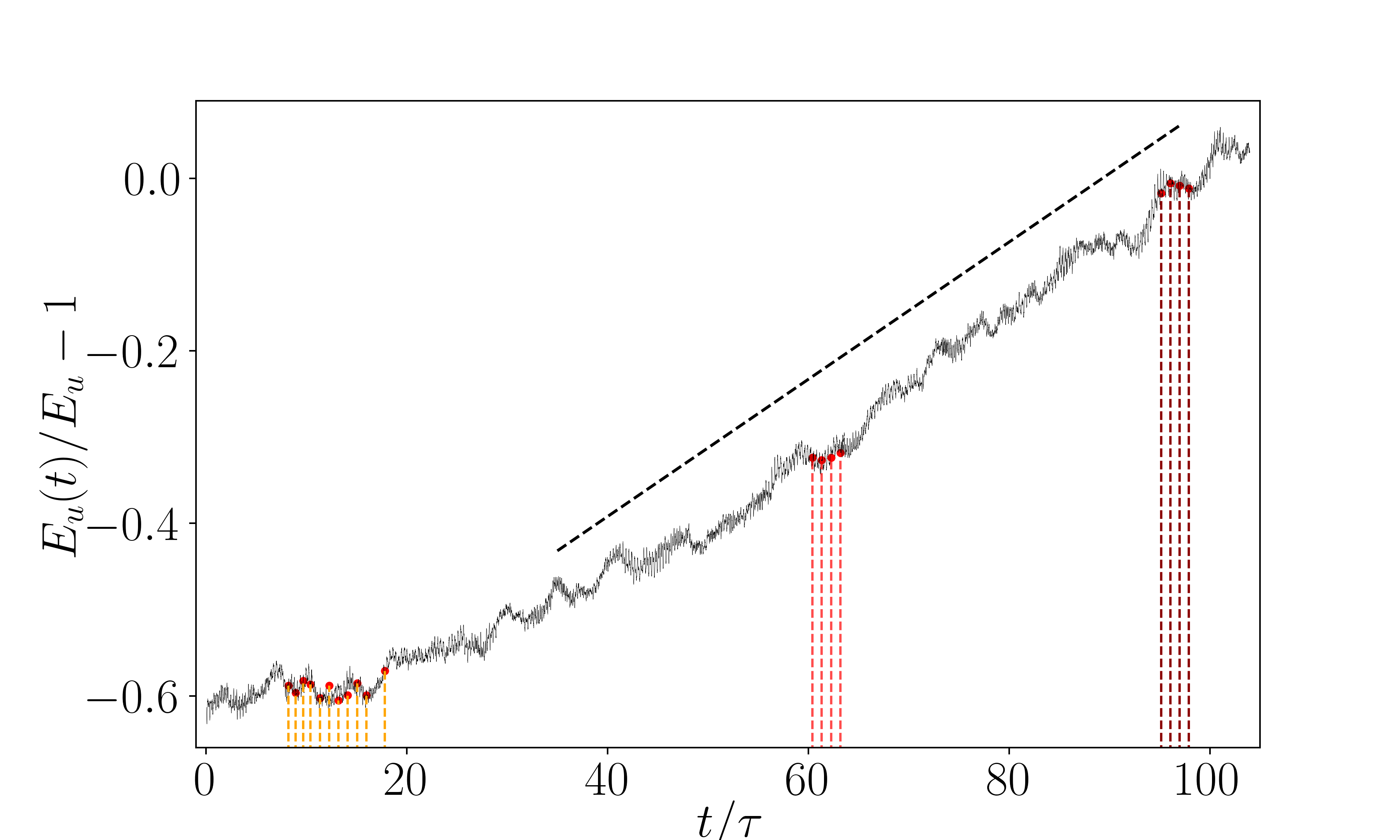}
    \end{center}
    \caption{Time evolution of mean kinetic normalised by the mean total energy corresponding to the last quasi-stationary time interval. The markers correspond to the sampled snapshots belonging to the three different sampling intervals.}
	 \label{fig:B0_10}
\end{figure}

The configurations with zero and weak BMF, datasets A3 and C1, are statistically stationary. By contrast, the strong BMF case (dataset C10) has not yet reached a stationary state: the kinetic energy continues to grow, {as shown in} fig.~\ref{fig:B0_10}, whereas the magnetic energy remains approximately statistically stationary in {all three} configurations, see \cite{capocciMHDdata}. {As a consequence of the non-stationarity, the snapshots of dataset C10} were sampled over time intervals in which the kinetic energy is quasi stationary, as indicated in fig.~\ref{fig:B0_10}. The growth of the total energy is associated with a process of 2D-dimensionalisation and partial inverse transfer of kinetic energy. In addition, for each dataset, the mean kinetic and magnetic dissipation rates, $\eps_u$ and $\eps_b$, remain stationary, as reported \cite{capocciMHDdata}. For dataset C10, this indicates that the system {does not} dissipate the entire injected energy, leading to the approximately linear growth of total energy visible in fig.~\ref{fig:B0_10}.

\begin{figure}
	\begin{center}
         \includegraphics[width=0.49\columnwidth]{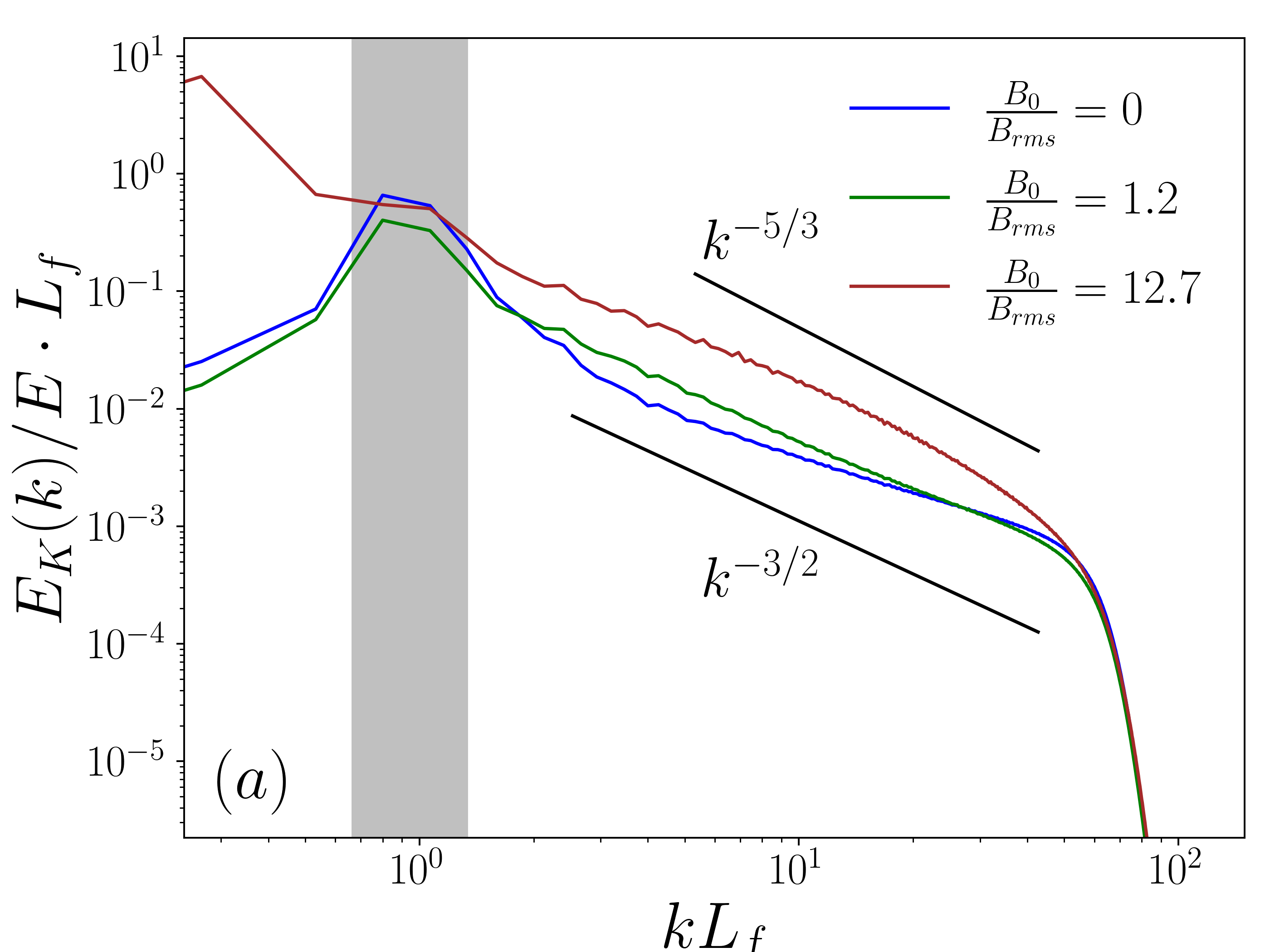}
         \includegraphics[width=0.49\columnwidth]{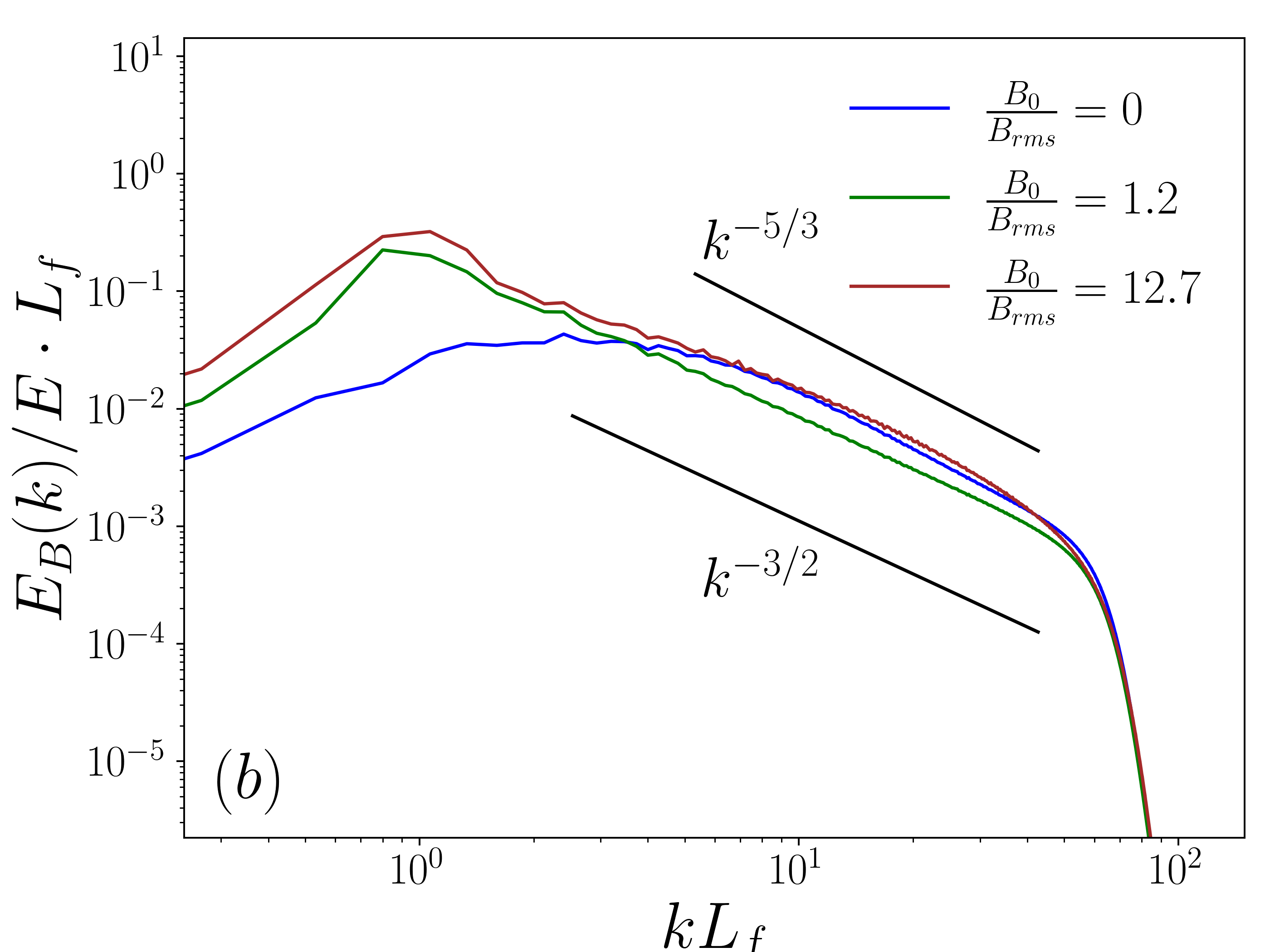}
    \end{center}
    \caption{Time-averaged omnidirectional spectra of velocity, panel (a), and magnetic field fluctuations, panel (b), corresponding to datasets A3, C1 and
C10. The grey region indicates the wavenumber band $k \in [2.5, 5.0]$ in which the velocity field is forced. The forcing scale $L_f$ is defined via the midpoint of the forcing band. }
	 \label{fig:spectra}
\end{figure}

{To assess the characteristic length-scales of the dynamics, fig.~\ref{fig:spectra} compares the time-averaged kinetic (panel a) and magnetic (panel b) energy spectra, where in C10, the spectra refer to the late-stage quasi-stationarity interval indicated in fig.~\ref{fig:B0_10}}. %\purple{Before discussing the spectra, we note that we consider omnidirectional spectra for all three configurations. This choice provides a common characterisation of the scale distribution of energy across datasets with different degrees of anisotropy, allowing the zero-, weak-, and strong BMF cases to be compared using the same scale variable. For the strong BMF case, separate perpendicular and parallel spectra would provide a more detailed description of the anisotropy, but would not substantially alter the present discussion, since the fields visualisations in fig.~\ref{fig:visuals} already show that the fluctuations along the BMF direction are strongly suppressed.}
In panel (a) the spectra of C10 indicate that modes {belonging} to the wavenumber {band} $0.2<k\, L_f<0.6$ are more excited than those relative to other two {BMF} configurations. This low-wavenumber modes population already indicates that an inverse cascade is enhanced when $B_0$ becomes \emph{large} enough. Moreover the only difference between C1 and C10 can be found in the power-law scaling where the latter appears to be closer to $k^{-3/2}$ than the former. Panel (b), {on the other hand}, shows a stronger similarity between the configurations $B_0 \neq 0$. {In both cases,} their peaks of magnetic spectra, approximately in the kinetic spectrum forcing band, resemble the effect of an energy injection {by external} forcing. {However, the $B_0/B_{\rm rms}=0$ case does not exhibit a comparable peak. Instead, its magnetic spectrum remains comparatively flatter over the same wavenumber band. This difference is easily explained by the presence of a nonzero BMF  acting as a zero-wavenumber mode coupling, which couples with the mechanically forced velocity modes and in turn excites magnetic-energy modes in the same wavenumber band as the velocity forcing itself.}

As concerns the scaling, we notice that for the three magnetic spectra it appears to be a power-law close to $k^{-5/3}$. This appears to be in agreement with spacecraft measurement of solar wind \citep{cho2003} and interplanetary magnetic field \citep{leamon1998} showing a $-5/3$ scaling. For $k\, L_f > 50$, the magnetic spectra are essentially coincident which is a consequence of the same value of (hyper)viscosity and (hyper)diffusivity employed in the runs. Consistently, in the strongly anisotropic regime we also observe, near the end of the inertial range ($k\,L_f\gtrsim 30$), an approximate equipartition between the kinetic and magnetic spectra, similar to what is expected in decaying MHD turbulence \citep{MullerEA03,HaugenBrandenburg04}.

\begin{figure}
        \begin{center}
         \includegraphics[width=.5\columnwidth]{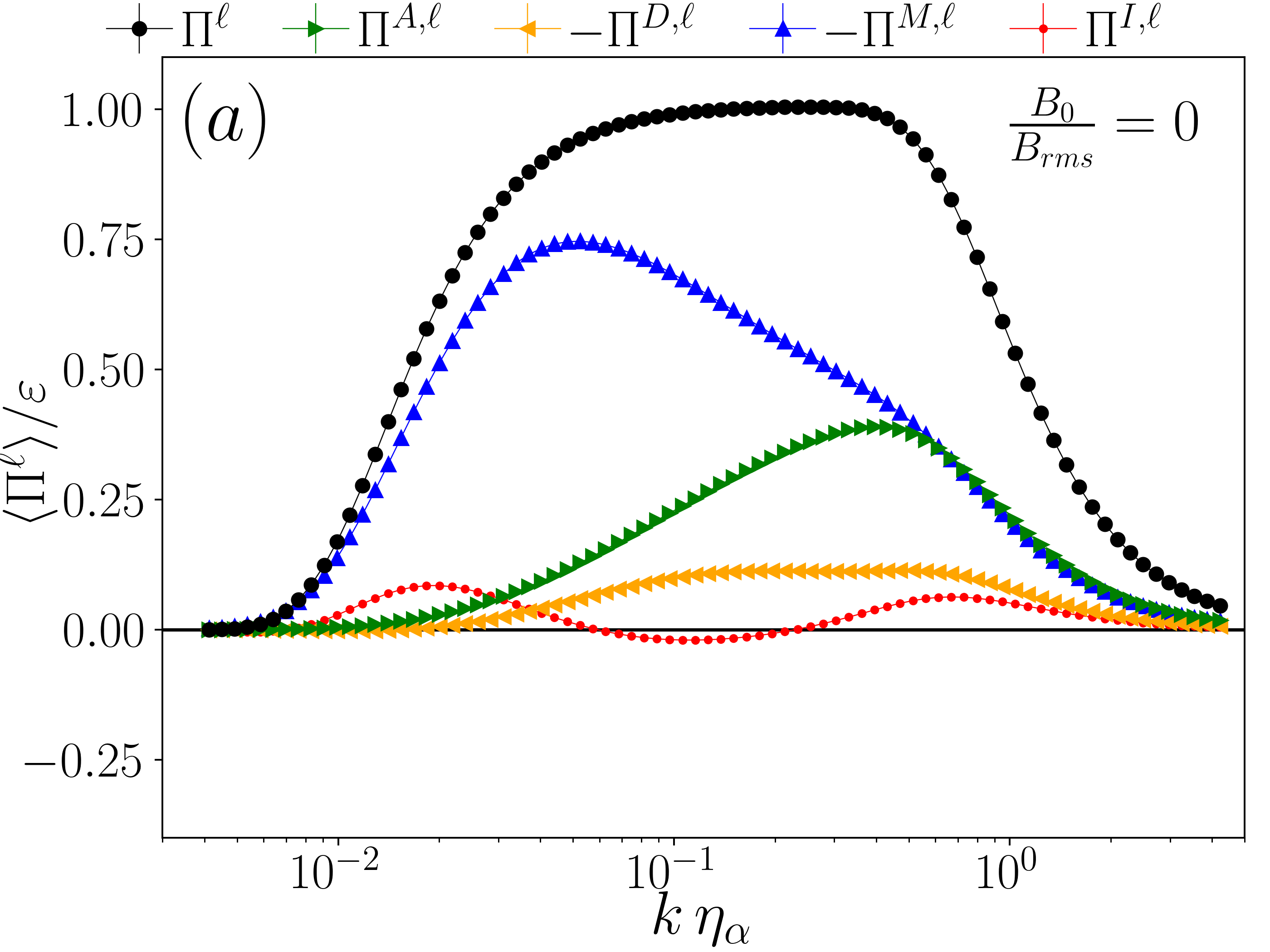}
          \noindent\makebox[\textwidth]{
         \includegraphics[width=.5\columnwidth]{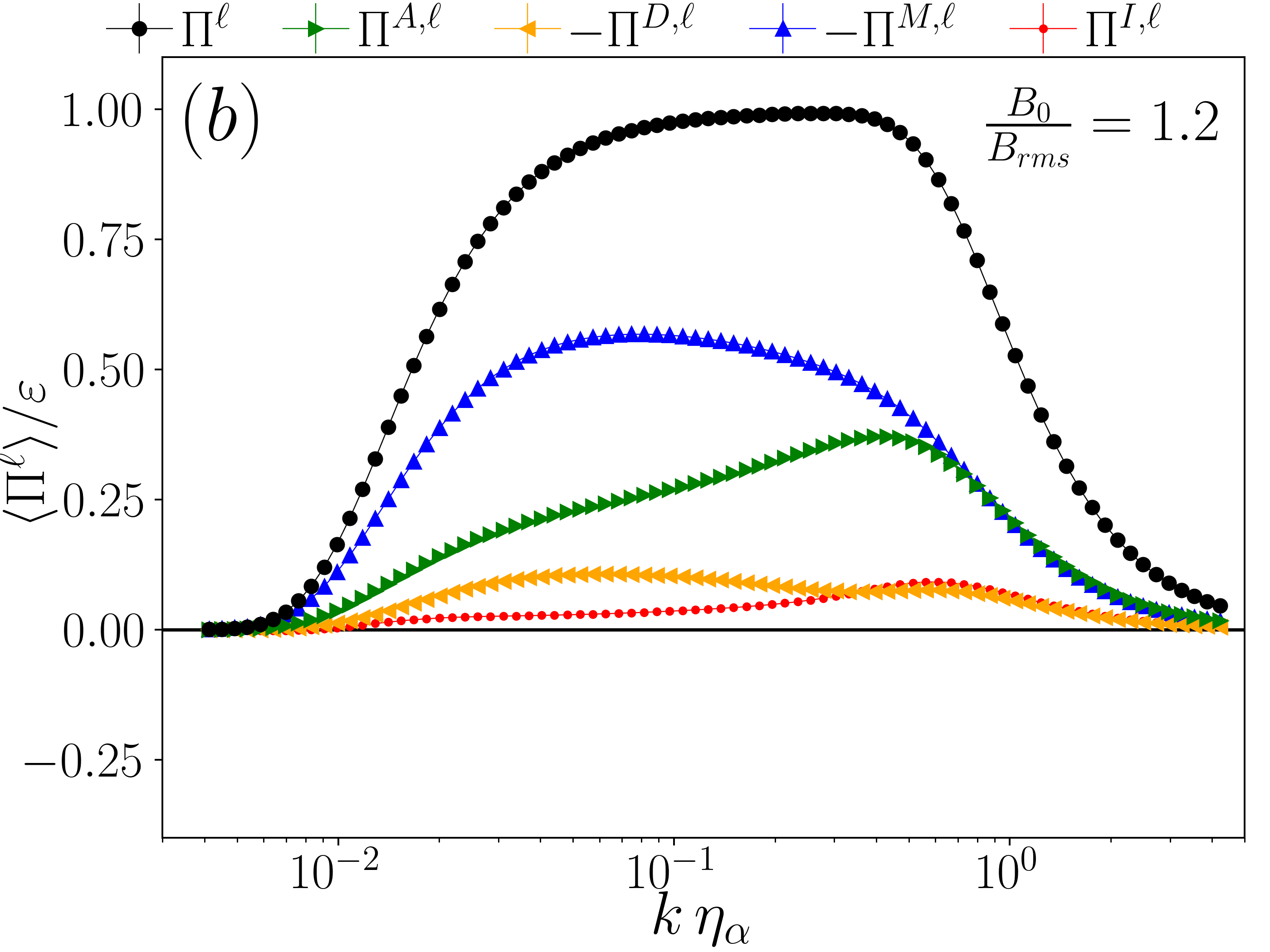}
         \includegraphics[width=.5\columnwidth]{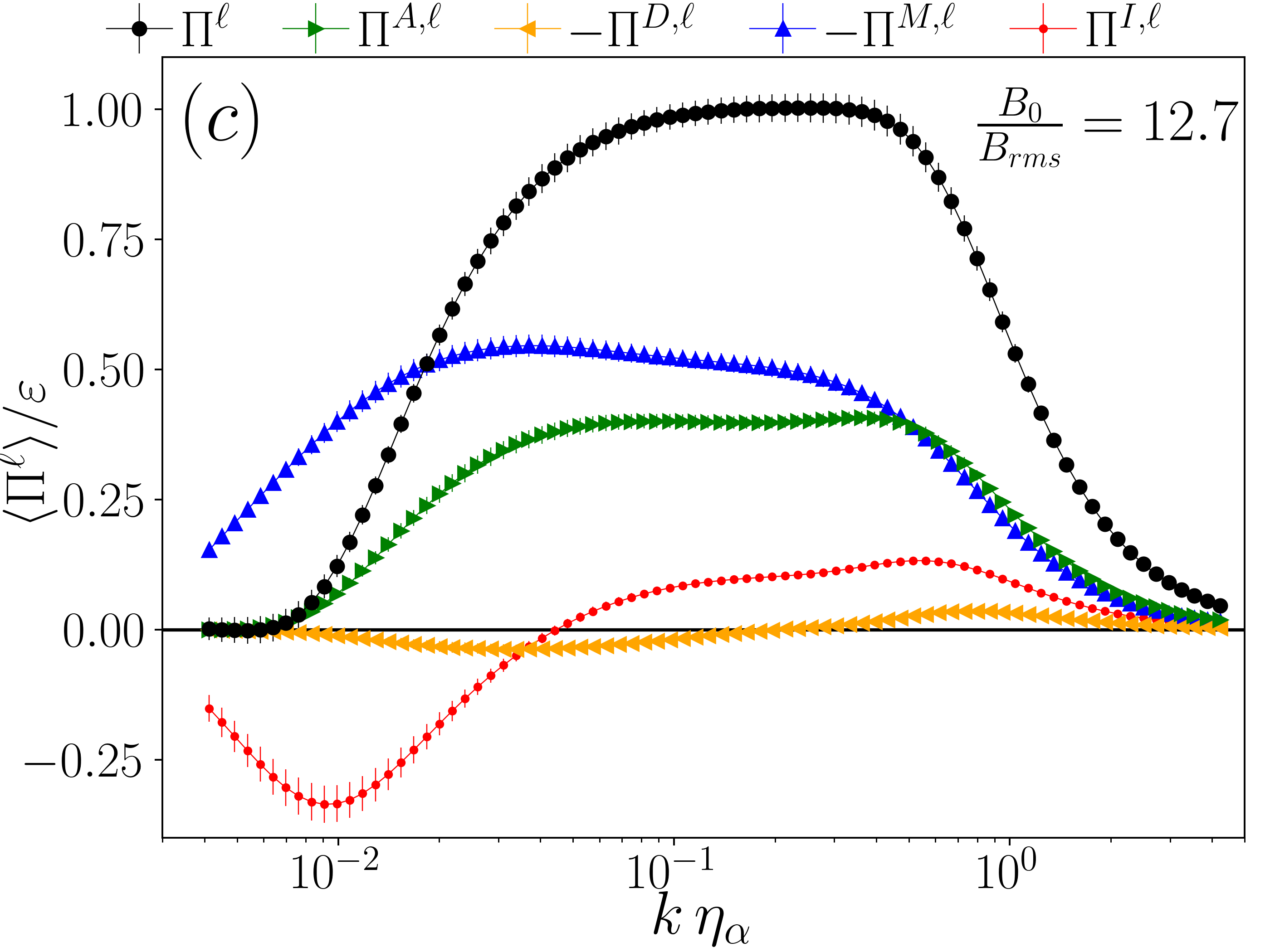} }
    \end{center}
         \caption{ Terms contributing to the MHD filtered energy flux across scale $\ell$ as function of the adimensional parameter $k\eta_\alpha = \pi \eta_\alpha / \ell$ for three different configurations: panel (a) $B_0/B_{rms}=0$, panel (b) $B_0/B_{rms}=1.2$ and panel (c) $B_0/B_{rms}=12.7$ which refers to the late-stage sampling interval. In all the configurations the Gaussian filter has been used. The error bars indicate one standard error. %The three panels share the same legend.
  }
\label{fig:gauss_filter_dynamo_stages}
\end{figure}

% ML: need to rephrase following paragraph
Having discussed both visualisations and spectra, we {now} center our discussion on the mean MHD energy fluxes relative to eqs.~\eqref{eq:Pi-I}--\eqref{eq:Pi-D}. {F}ig.~\ref{fig:gauss_filter_dynamo_stages} shows each subflux, alongside their total sum, normalised by the total energy dissipation rate $\varepsilon= \varepsilon_u + \varepsilon_b$ {and} displayed as function of the non-dimensional parameter $k\,\eta_\alpha = \pi \, \eta /\ell$. Compared to the $B_0/B_{rms}=0$ case, three main effects {emerge} as $B_0$ increases. 

First, the subfluxes become more scale-independent in the inertial range, {here defined as the range where the total energy flux forms a plateau}. {This trend is evident for $B_0/B_{rms}=12.7$,} where all subfluxes are approximately flat at least {over the interval} $3\times 10^{-2} \leq k \eta_\alpha \leq 3 \times 10^{-1}$. {Second,} the Dynamo subflux is increasingly depleted as the strength of $B_0$ increases, {to the point, in the strong BMF case,} that $\Pi^{I,\ell}$ becomes larger in the inertial range. {this behaviour supports} the usefulness of separating Advection and Dynamo terms from the magnetic SGS stress tensor, {which are usually included in the same term \citep[e.g.][]{OffermansEA18,benella2026} }. Finally, the strong $B_0$ case shows a pronounced inverse transfer relative to the Inertial term in the \emph{large} scales. {Since the total energy flux remains positive at each filtering scale, the cascade corresponds to a forward  energy transfer. The negative large-scale contribution to the Inertial flux can, therefore, be interpreted as an enhanced inverse transfer}. This is consistent with the visualisations that display a clear signature of ongoing two-dimensionalisation. Moreover, this behaviour is not an {artefact} of the large-scale forcing scheme, as its fully stochastic nature acts against the formation of a large-scale coherent structures. For $B_0=0$, the negligible {intervals} in which the Inertial flux {becomes negative}, completely disappear with $B_0/B_{rms}=1.2$. This {potentially} suggests that the emergence of the large-scale inverse contribution is not sharp but develops {smoothly} as $B_0$ is increased. %this is still pure Occam's razor though
%TODO, low priority: cite some papers around this, 2D hydro, 2DMHD(Alexakis 2015)

%i just discovered that artefact is BRengl while artifact is USengl.

\section{{Quantification of the} energy flux decomposition}
\label{sec:fluxes}

In this section we {quantify} the flux decompositions derived and interpreted in secs.~\ref{sec:Pi-I_intro}--~\ref{sec:Pi-A_intro} for the cases $B_0/B_{rms}=1.2$, and $B_0/B_{rms}=12.7$, with the aim of assessing which physical processes govern the MHD energy cascade in presence of a background magnetic field.  
{To determine how a non-zero} mean magnetic field modifies the interscale energy transfer, we also analyse the zero BMF dataset A3. This dataset has the same resolution and hyperdiffusivities as for the $B_0/B_{rms}=1.2$ and $B_0/B_{rms}=12.7$ cases, {respectively datasets C1 and C10, allowing for a direct comparison}. For an analysis of the $B_0/B_{rms}=0$ case at higher effective Reynolds number and higher resolution, and for a comparison between simulations using standard diffusivities and hyperdiffusivities, we refer to the analysis by \cite{capocci2025}.
%
%is included for an  \emph{in situ} comparison of MHD without background magnetic field.only the lower $Re$ and lower resolution of the dataset belonging to the earlier investigation \cite{capocci2025}. The purpose of dataset A3 is to have an \emph{in situ} comparison of MHD without background magnetic field.
%
In what follows, we restrict our attention to the mean values of the subfluxes {across the scales}, since their fluctuations do not show significant differences from the $B_0/B_{rms}=0$ case studied in \cite{capocci2025} (not shown). 
%\purple{Consistently with the use of omnidirectional spectra in sec.~\ref{sec:numerics}, the fluxes and their decompositions are evaluated using the same isotropic Gaussian filter for all three configurations. This choice provides a common definition of scale for the zero-, weak-, and strong-BMF datasets, and therefore allows differences in the measured subfluxes to be attributed to the dynamics rather than to a change in the filtering procedure. It also facilitates comparison with Fourier-space analyses based on isotropic Galerkin truncation, such as that of \cite{bian2019}.}
%Moreover, we do not discuss the PDFs of the energy fluxes, as they are not physical observables in consequence of their gauge freedom  \citep{velamartin2022}.

\subsection{Inertial flux}
\label{sec:intertial_flux}
\begin{comment}
The exact decomposition of the Inertial term $\Pi^{I,\ell} $, eq.~\eqref{eq:Pi-I}, yields \citep{capocci2025}:
\begin{equation}
 \label{eq:inertial_dec}
    \Pi^{I,\ell} =  \Pi^{I,\ell}_{s, SSS}  +  \Pi^{I,\ell}_{m,SSS}  +   \Pi^{I,\ell}_{s, S \Omega \Omega}  +  \Pi^{I,\ell}_{m,S \Omega \Omega}  +  \Pi^{I,\ell}_{m, S \Omega S} 
\end{equation}
where the individual expressions of each subflux are reported in Appendix~\ref{app:defintions}. The above expression is the inertial-flux decomposition already obtained for hydrodynamic turbulence \citep{Johnson20,Johnson21}, now applied to the MHD Inertial subflux. The two terms of type $SSS$ are interpreted as strain self-amplification, the two terms of type $S\Omega\Omega$ are associated with vortex stretching, and the term of type $S\Omega S$ is related to vortex-thinning \cite{johnson2021,chen2006,kraichnan1976}. The latter appears only as a multi-scale contribution, since its single-scale counterpart vanishes identically \citep{Johnson20}. 
\end{comment}

\begin{figure}
        \begin{center}
        
         \includegraphics[width=.5\columnwidth]{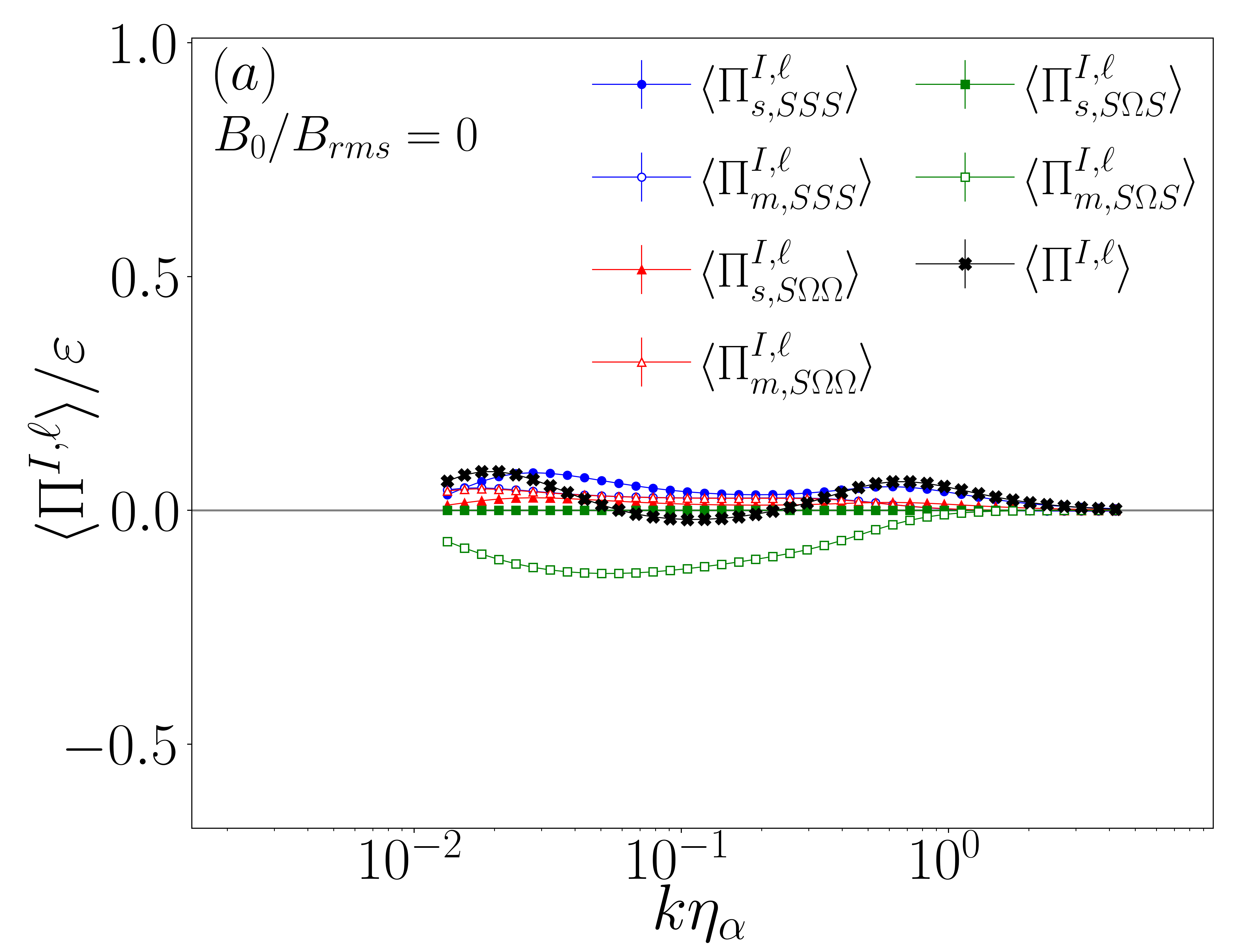}
          \noindent\makebox[\textwidth]{
         \includegraphics[width=.5\columnwidth]
         {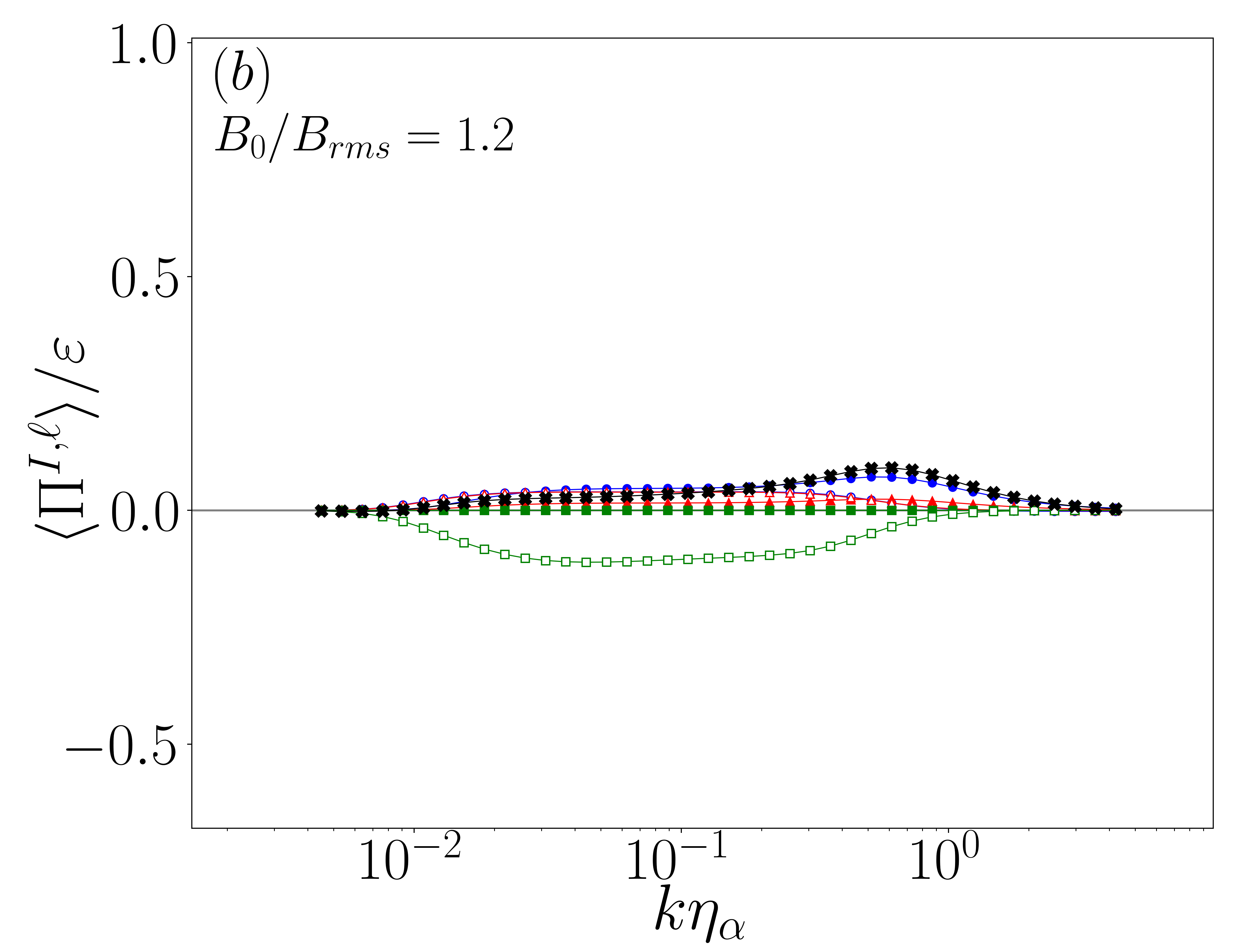}
         \includegraphics[width=.5\columnwidth]{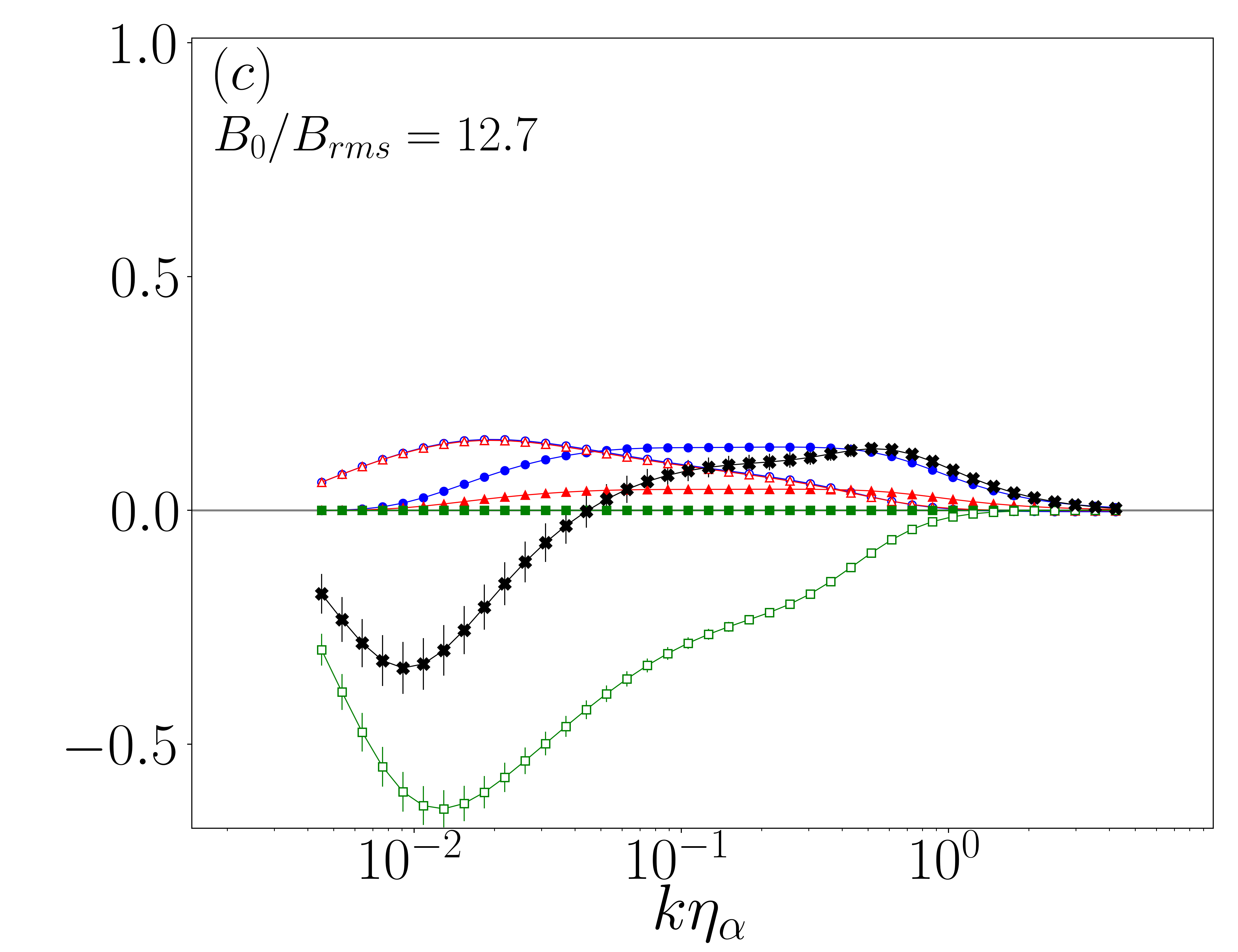} }
    \end{center}
         \caption{Contributions to the Inertial energy flux for (a) the dataset A2 relative to $B_0=0$, (b) the dataset C1 with $B_0=1$ and (c) the dataset C10 with $B_0=10$. Errorbars are for one standard error. }
\label{fig:inertial_bckgr_comp}
\end{figure}

Figure \ref{fig:inertial_bckgr_comp} shows the decomposition {terms} of $\Pi^{I,\ell}$ normalised by the total mean dissipation rate, $\varepsilon$, into subfluxes according to eq.~\eqref{eq:inertial_dec}, with data for $B_0/B_{rms}=0$, $1.2$ and $12.7$ shown in subfigures (a), (b) and (c), respectively, with error bars indicating the standard error of the mean. In all cases, and as reported by \cite{capocci2025} for $B_0 = 0$, the total $\Pi^{I,\ell}$ remains depleted compared to hydrodynamic turbulence. In particular, panel (a), corresponding to $B_0/B_{rms}=0$, does not display any substantial difference with respect to the results shown in fig.~3(b) of \cite{capocci2025}, either in the relative weights of the individual subfluxes or in the total contribution. All subfluxes except the vortex-thinning term $\Pi^{I,\ell}_{m,S \Omega S}$ are individually almost negligible, their combined contribution to forward transfer amounts to about $10\%$ of the total dissipation, and is cancelled by an upscale kinetic energy flux associated with vortex thinning, $\Pi^{I,\ell}_{m,S \Omega S}$. Very similar results hold for $B_0/B_{rms}=1.2$, with results shown in subfigure (b).    

%This is consistent with the behaviour of the mean Inertial term already observed in fig.~\ref{fig:gauss_filter_dynamo_stages}. 
% ML: this is repeating what we say in section 2.
%

%
A different behaviour emerges in the strongly anisotropic case shown in subfigure (c). As observed in earlier in fig.~\ref{fig:gauss_filter_dynamo_stages}, the total Inertial flux is negative at large scales,  indicating a clear inverse transfer at large scales, and a weak residual forward cascade is present for $k \eta_\alpha > 10^{-1}$, accounting for approximately $10\%$ of the total energy flux.  The dominance and sign of the vortex-thinning component, $\lan \Pi^{I,\ell}_{m,S \Omega S} \ran < 0$, is consistent with the pronounced two-dimensionalisation of the flow observed in the visualisations.  Here, the depletion of total energy transfer is due to partial flux cancellation rather than a strong depletion of individual subfluxes, where the forward transfer carried by vortex-stretching and strain self-amplification is counteracted by the inverse transfer from vortex-thinning.
Moreover, we observe scale-independence of $\lan \Pi^{I,\ell}_{s,SSS} \ran$ over an interval in scale-space, and by that, invoking the Betchov relation $\lan \Pi^{I,\ell}_{s,SSS} \ran = 3\lan \Pi^{I,\ell}_{s,S\Omega \Omega} \ran$, also of $\lan \Pi^{I,\ell}_{s,S\Omega \Omega} \ran$. 
The inverse transfer due to partial two-dimensionalisation results in strong flows at scales larger than the forcing scale, and therefore a significantly larger effective Reynolds number (see table \ref{tab:datasets}), and more kinetic energy available for the  cascade to smaller scales. 

%ML: removed, as the discussion is relegated to the RM<HD paper
%\blue{The discussion around the two-dimensional asymptotic regime, for $t \to \infty$, is discussed below in sec.~\ref{sec:2d_sec}.}

\subsection{Maxwell subflux}
\label{sec:Pi-M}

\begin{figure}
        \begin{center}
         \includegraphics[width=.5\columnwidth]{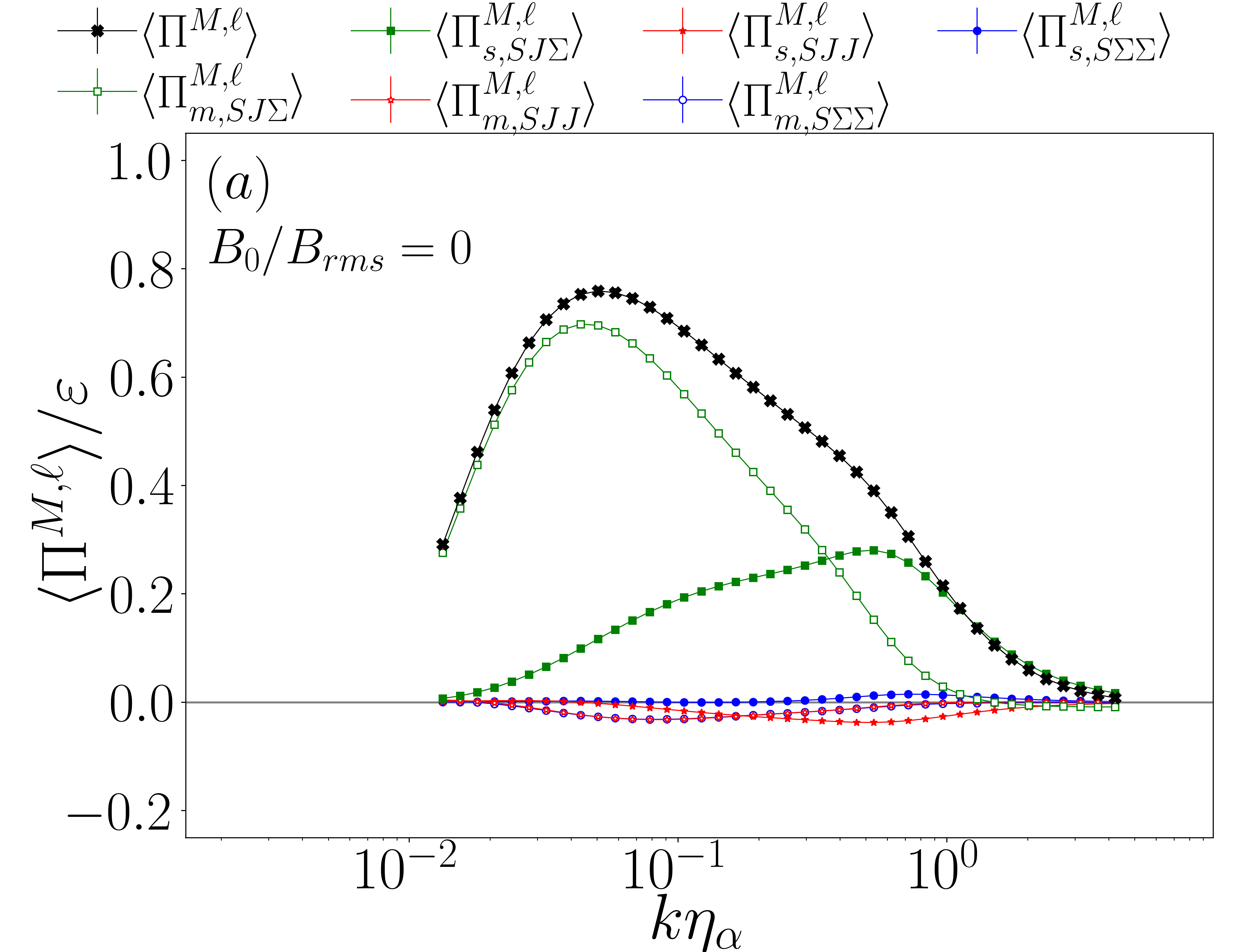}
          \noindent\makebox[\textwidth]{
         \includegraphics[width=.5\columnwidth]
         {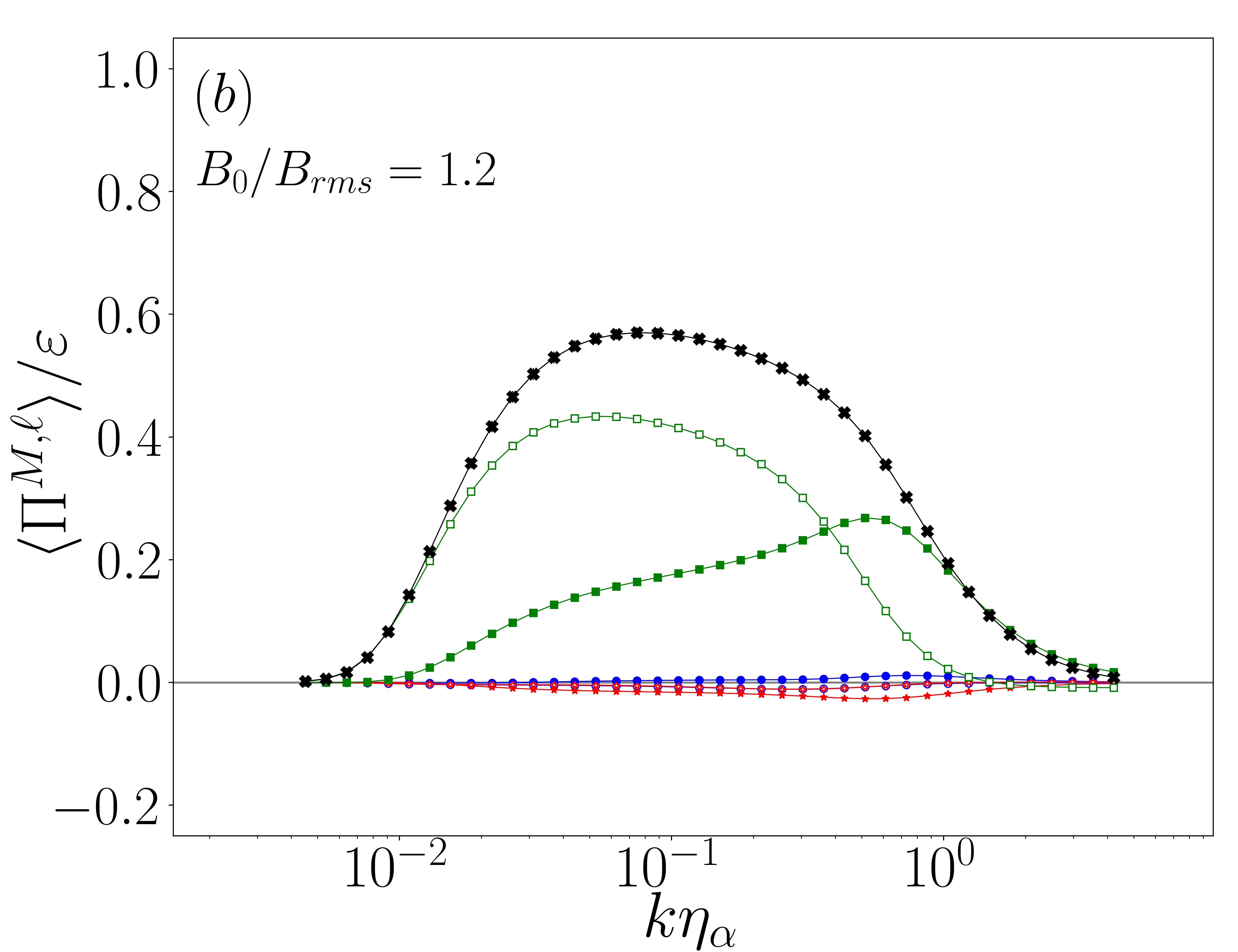}
         \includegraphics[width=.5\columnwidth]{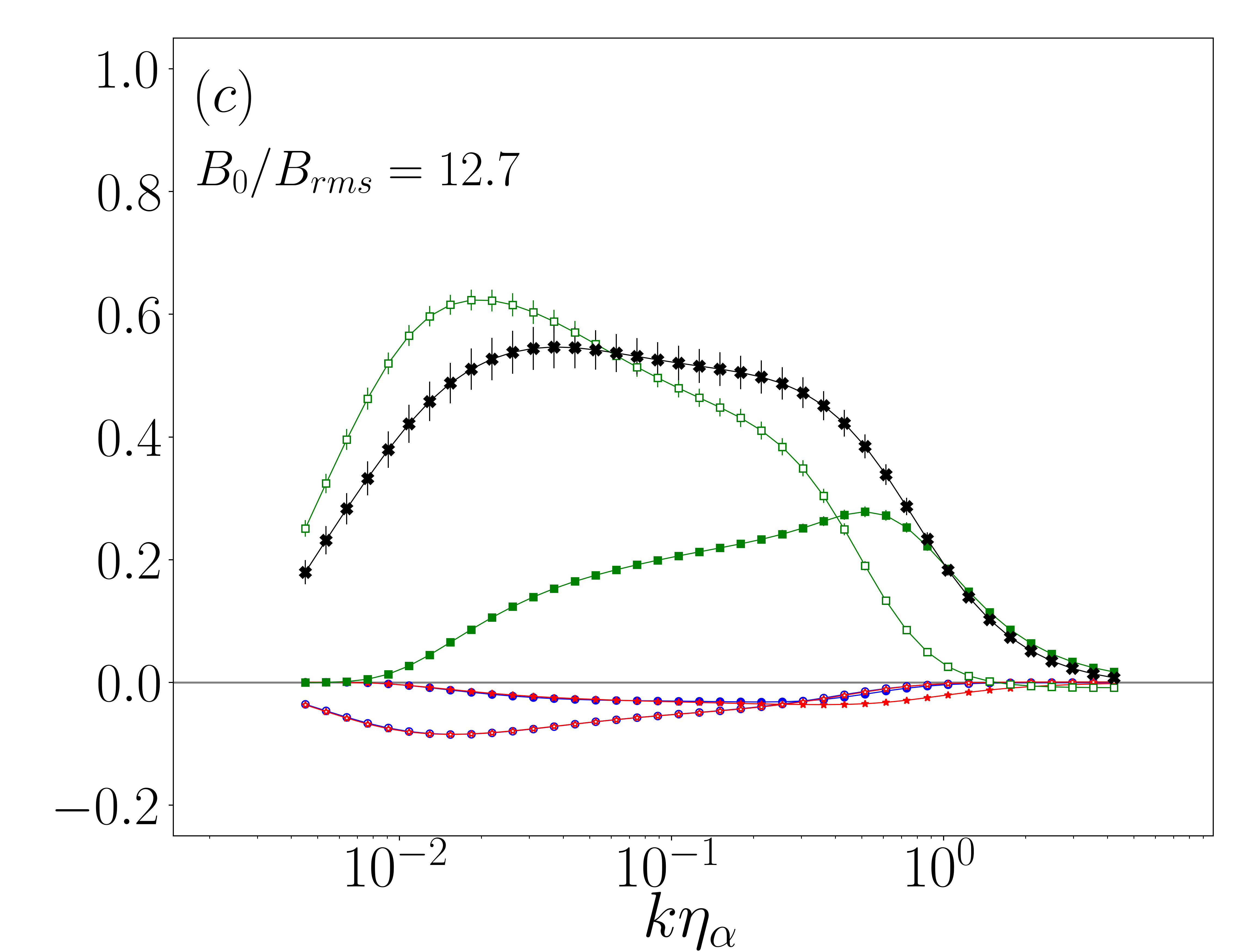} }
    \end{center}
         \caption{Contributions to the Maxwell energy flux for (a) the dataset A2 relative to $B_0=0$, (b) the dataset C1 with $B_0=1$ and (c) the dataset C10 with $B_0=10$. Errorbars are for one standard error. With respect to figure \ref{fig:inertial_bckgr_comp}, the y-axis range has been contracted in order to avoid a further compression of the curves.}
\label{fig:Lorentz_bckgr_comp}
\end{figure}

{In this section we quantify the volume-averaged subfluxes featuring the decomposition of the Maxwell flux in eq.~\eqref{eq:dec_maxwell} across the scales.}
In fig.~\ref{fig:Lorentz_bckgr_comp}, {each term from} the decomposition is displayed for each BMF configuration. The only difference can be found among the configuration with $B_0=0$ and those with $B_0/B_{\rm rms}=1.2$ and $12.7$ that appear quantitatively similar. In fact the leading term $\Pi^{M,\ell}_{m,S J \Sigma}$ is slightly smaller in panels (b) and (c) while its single scale counterpart $\Pi^{M,\ell}_{s, S J \Sigma}$ remains the same. In consequence, unlike the $B_0=0$ case, the total Maxwell subflux $\Pi^{M,\ell}$ in presence of $B_0/B_{\rm rms}=1.2$ and $12.7$ is similar in profile to a hydrodynamic energy flux where the strong $B_0$ case presents a narrow scale independence in the range $0.04 \leq k \eta_\alpha \leq 0.2 $. This effect would be more pronounced in the Fourier-based fluxes 
%shown in fig.~\ref{fig:gauss_filter_dynamo_stages} 
where the corresponding plateau is more extended (not shown). 
In combination with the results of the previous section, we conclude that MHD kinetic energy cascade observed first by \citet{BianAluie19} is mostly due to the back-reaction of the current-sheet thinning process on the flow, with the scale independence likely a consequence of more magnetic energy available at large scale for non-vanishing background magnetic fields. 
%\footnote{Like in the comparison between panel(a) and (b) of fig.~2 of \cite{capocci2025}, the Gaussian filter shrinks the plateaux extension.} 
%this plateau.   

\subsection{Advection and Dynamo fluxes}
\label{sec:Pi-A}
\begin{figure}
        \begin{center}

         \includegraphics[width=.5\columnwidth]{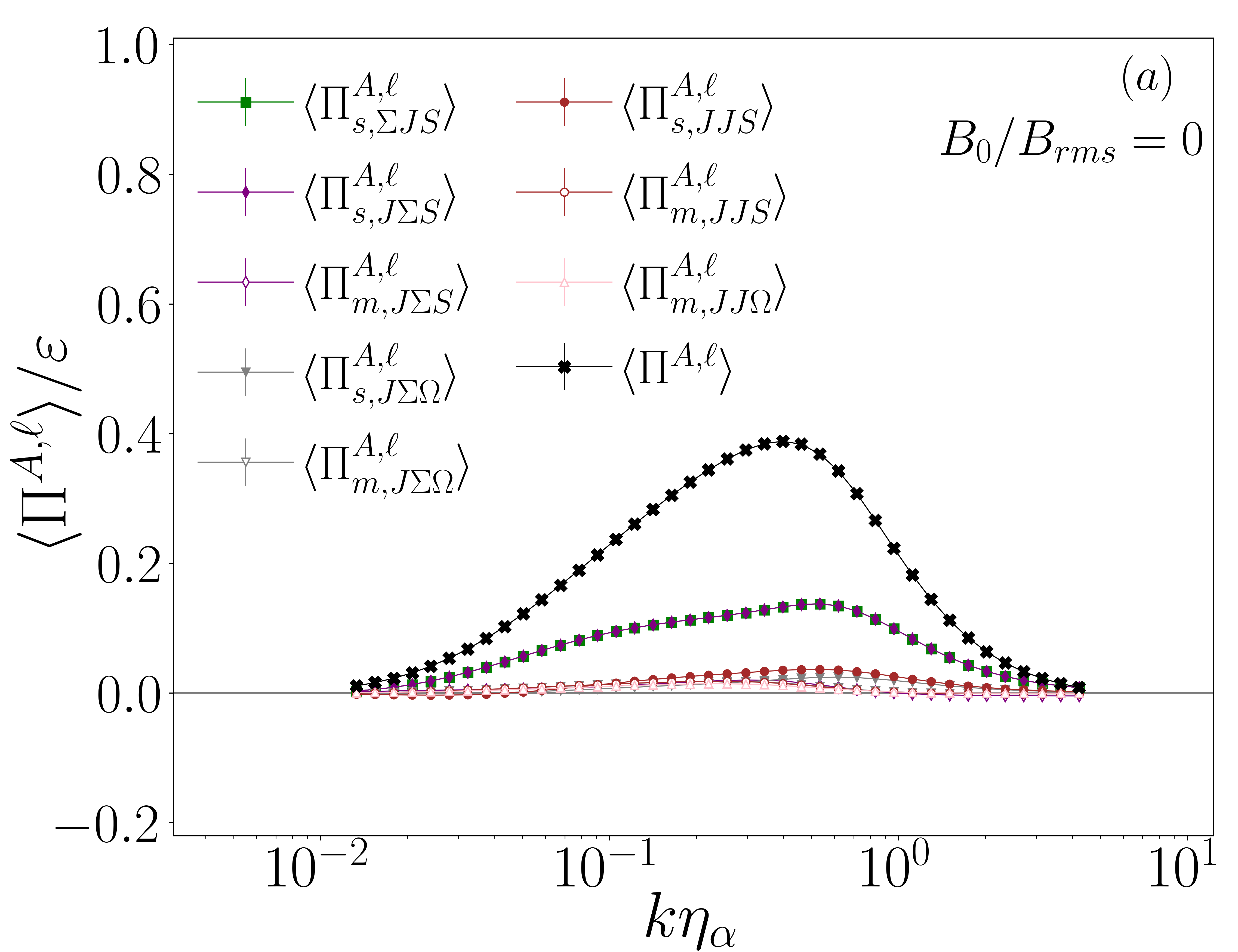}
          \noindent\makebox[\textwidth]{
         \includegraphics[width=.5\columnwidth]
         {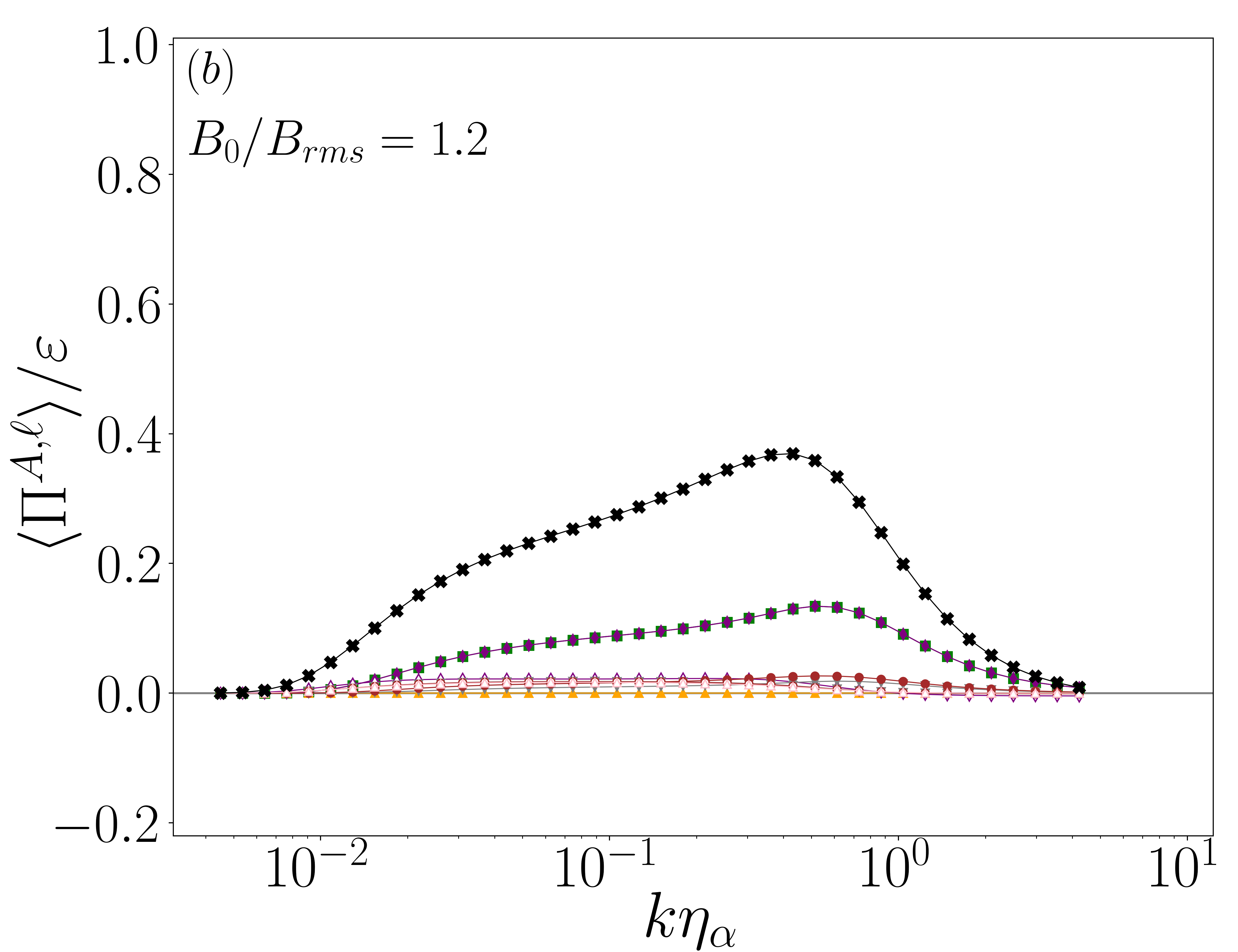}
         \includegraphics[width=.5\columnwidth]{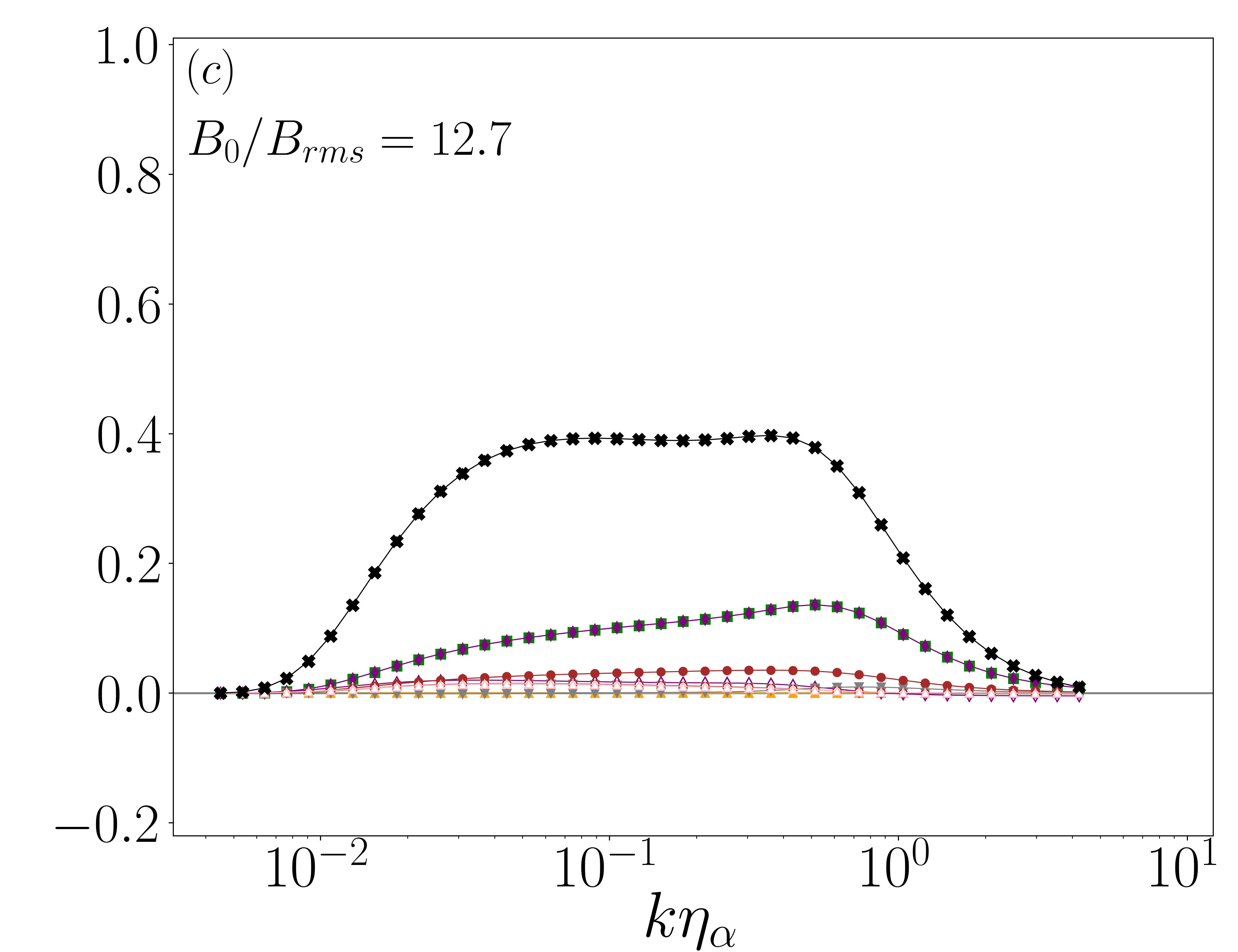} }
    \end{center}
         \caption{Contributions to the Advection energy flux for (a) the dataset A2 relative to $B_0=0$, (b) the dataset C1 with $B_0=1$ and (c) the dataset C10 with $B_0=10$. The errorbars, even though not fully visible, are for one standard error.}
\label{fig:Advection_bckgr_comp}
\end{figure}

{We turn next to the volume-averaged terms in the Advection-flux decomposition of eq.~\eqref{eq:adv_total}. In {the quantification of the Maxwell flux expansion of} sec.~\ref{sec:Pi-M}, we measured the the term corresponding to the back-reaction induced by the current-sheet thinning term} accounts for the bulk of the kinetic-energy transfer towards smaller scales. Here, we show that the same underlying process is also responsible for most of the magnetic-energy transfer from large to small scales. However, unlike the Maxwell flux, which is dominated by a multiscale contribution, the Advection flux is mainly carried by the two single-scale terms $\Pi^{A,\ell}_{s,\Sigma J S}$ and $\Pi^{A,\ell}_{s,J \Sigma S}$, even in the configurations with non-zero BMF. This difference is evident in fig.~\ref{fig:Advection_bckgr_comp}, which presents the exact decomposition of the Advection flux for the datasets. As already noted in fig.~\ref{fig:gauss_filter_dynamo_stages}, the total Advection term $\Pi^{A,\ell}$ becomes increasingly scale-independent as $B_0$ is increased, in close analogy with the Maxwell term. Figure~\ref{fig:Advection_bckgr_comp} indicates, however, that this trend is not driven by the dominant single-scale contributions $\Pi^{A,\ell}_{s,J \Sigma S}$ and $\Pi^{A,\ell}_{s,\Sigma J S}$, whose profiles remain qualitatively similar across the different background-field strengths. Rather, the enhanced scale-independence of the total Advection flux results from the combined contribution of the remaining depleted subfluxes, whose sum is nonetheless slightly larger in the configurations with non-zero BMF. 

For what concerns the decomposition of the Dynamo subflux of eq.~\eqref{eq:dyn_total}, unlike for the other subfluxes, we do not present a detailed quantification of its decomposition here. There are two reasons for this. 
First, as can be seen from the total Dynamo flux shown in fig.~\ref{fig:gauss_filter_dynamo_stages}, it is negligible in the strong-field case. Second, for the other cases, where it accounts for roughly $15\%$ of the total energy flux in the zero BMF case and about $8\%$ in the weak BMF case, the decomposition is qualitatively the same as in the earlier investigation of \cite{capocci2025}: the dominant contributions are multiscale, but they are individually small and numerous, with no clear preference to any particular process. The same applies in the strongly anisotropic case, except that the corresponding contributions are even weaker (not shown).

\section{Scale-resolved energy conversion}
\label{sec:conversion}

In this section we quantify the resolved-scale conversion term introduced in sec.~\ref{sec:theory}. %Our goals are twofold. 
First, we assess how the anisotropy induced by the  background magnetic field modifies the mean conversion at each scale, $\langle \mathcal{W}^\ell \rangle$. Second, we isolate and characterise the contribution that is specific to the guide field, namely the anisotropic part $\mathcal{W}^\ell_{\mathrm{anis}}$. For both, we examine not only the scale dependence of the mean conversion, $\langle \mathcal{W}^\ell\rangle$, but also its fluctuations, as quantified by probability density functions (PDFs). Unlike the previous comparisons, the strongly anisotropic configurations (dataset C10) will be analysed in each of the time intervals where the dynamics appear statistically stationary as indicated in fig.~\ref{fig:B0_10}. 

\begin{figure}
	\begin{center}
         \includegraphics[width=\columnwidth]{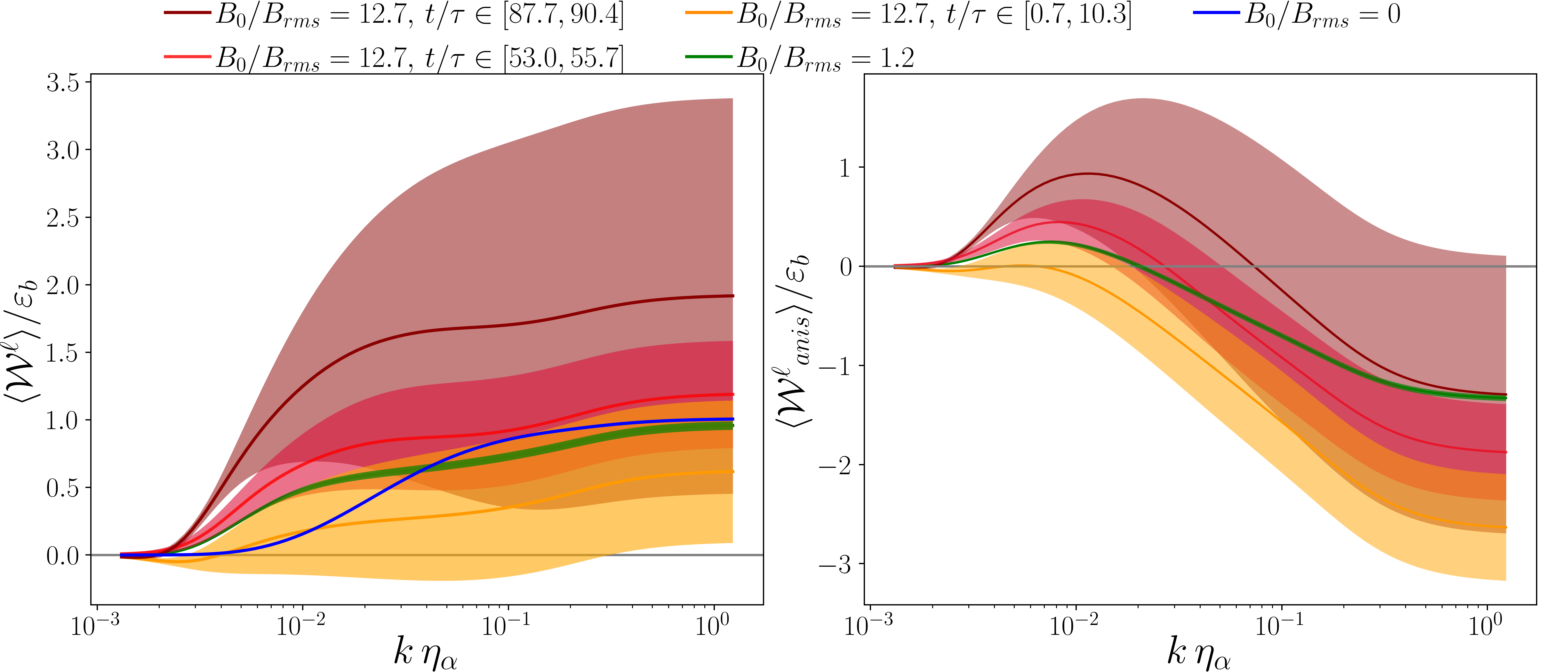}
    \end{center}
    \caption{Mean energy conversion normalised by $\varepsilon_b$ as function of the filtering wave number $k = \pi/\ell$ normalised by the hyperdiffusive Kolmogorov scale $\eta_\alpha$ . Left: total energy conversion term $\langle \mathcal{W}^\ell \rangle$.
    %identified by the LHS of eq.~\eqref{eq:W_def} and normalised by $\varepsilon_b$. 
    Right: anisotropic part of the energy conversion term/energy conversion mediated by the background magnetic field. The shaded regions indicate one standard error. % introduced in the RHS of eq.~\eqref{eq:W_def}. 
    In both the panels, the three curves belonging to $B_0/B_{rms}=12.7$ refer to the three quasi-stationary intervals indicated in fig.~\ref{fig:B0_10} with the same colour palette. 
    %Each axes display non-dimensional quantities.
    %\red{ML: the error bars for some of the curves are not visible. Can you please re-plot using clear markers to be able to distinguish them for all datasets?}
    }
	 \label{fig:conv_term_mean}
\end{figure}

Starting from the mean quantities, the top panel of fig.~\ref{fig:conv_term_mean} shows the averaged RSC term normalised by the mean Joule dissipation rate $\varepsilon_b = \lan \mathcal{D}_b \ran $, with error bars indicating the standard error of the mean. We use the normalisation with $\varepsilon_b$  because, in a statistically stationary state, $\langle \mathcal{W}^\ell\rangle$ has to converge to $\varepsilon_b$ at small scales, as implied by the volume average of eqs.~\eqref{eq:Eu-ls}--\eqref{eq:Eb-ls} since the magnetic-field fluctuations must be driven by the flow on average, see also \citep{BianAluie19}. This is because the filtering approach defines $\mathcal{W}^\ell$ as as the cumulative energy conversion from all scales larger than $\ell$. The zero- and weak-BMF configurations clearly show this convergence, with the weak-field case showing smaller error bars than the strong-background-field counterpart. In contrast, the mean conversion term pertaining to the strong-background-field case has much larger error bars, despite the sampling being restricted to quasi-stationary intervals, reflecting the residual non-stationarity of the flow. We also observe that these error bars increase with time, indicating increasingly strong fluctuations during the time evolution, including instantaneous configurations in which the mean conversion can become negative. At first sight, the large error bars may also suggest values of the averaged RSC term exceeding the mean magnetic dissipation rate $\varepsilon_b$. This is possible over short time intervals in a non-stationary regime. 
%ML: removed \red{Here, we note that the value of $\varepsilon_b$ used in the normalisation has been calculated over the entire trajectory, this is appropriate as the spatial mean of $\mathcal{D}_b$ is statistically stationary \cite{damiano2025data}.} \damiano{this is not true but it is ok.}
% ML we can't do a time average over a the whole trajectory for C10. I comment out as this is not shown in the figure anyway. 
%However, when the average is computed over the full sampled time evolution, we recover the expected small-scale behaviour, $\langle \mathcal{W}^\ell \rangle \approx \varepsilon_b$, in agreement with \cite{BianAluie19}. 
Although the anisotropic contribution was not explicitly isolated in that work, the effect of $B_0$ has been examined in \cite{BianAluie19}, where numerical evidence indicated that $\langle \mathcal{W}^\ell \rangle$ remains bounded for all $\ell$, even in the limit $B_0 \to \infty$. 

%working now on this
%%%%%%%%%%%%%%%%%%%%%%%%%%%%%%%%%%%%%%%%%%%%%%%%%%%%%%%%%%%%%%%%%%%
%ML: let us leave this for now in the interest of time, we can clarify during revisions. 
%\red{ML this is interesting, but I comment out as in quasistatic MHD the magnetic-field fluctuations are not dynamically relevant.}
%\damiano{Quasi-static MHD is the starting point, their prescription is extended to full MHD and applicable to large Re but small Pm.}

%\blue{
%Something more dynamically relevant, however, occurs in presence of \emph{low} $Pm$ configurations, which is characteristics of liquid metals \citep{plunian2013}. To this end, the work \citep{gallet2015} proved rigorously that for sufficiently \emph{large} $B_0$ the dynamics asymptotically converge to a fully two-dimensional hydrodynamic states where the viscous dissipation follows the corresponding two-dimensional non-anomalous behaviour \citep{AlexakisBiferale2018} where Ohmic dissipation, namely $\varepsilon_b$, vanishes. Since, as indicated above, the small-scale limit of the total mean RSC is fixed by the magnetic-energy balance, $\langle \mathcal{W}^{\ell}\rangle \to \varepsilon_b$ ....
%}

%%%%%%%%%%%%%%%%%%%%%%%%%%%%%%%%%%%%%%%%%%%%%%%%%%%%%%%%%%%%%%%%%%%

%\red{ML: can we say more? For $B_0 \to \infty$ there is 2-dimensionalisation, so $u$ and $b$ do not vary in $z$-direction. The question is, does the 2-dimensionalisation occur such that the limit remains finite, and eventually goes to zero? There is a paper by Alexakis and Doering (or Basile Gallet? about this, which could be useful.}

 The anisotropic contribution to the RSC term, $\langle \mathcal{W}^\ell_{anis} \rangle$, which is zero for dataset A3, is presented in the bottom panel of fig.~\ref{fig:conv_term_mean} for dataset C1 and different time intervals from dataset C10. We find that this contribution yields a small kinetic-to-magnetic conversion at large to intermediate scales, whereas, strikingly for a nonlinear dynamo, at small scales it becomes negative and substantially so. As $\langle \mathcal{W}^\ell_{anis} \rangle$ encodes the cumulative energy conversion up to scale $\ell$, this indicates a net magnetic-to-kinetic conversion, which is substantial for all datasets and time intervals.  For the weak-field case $B_0/B_{\rm rms} = 1.2$, $\langle \mathcal{W}^\ell_{anis} \rangle/\varepsilon_b \to -1$ for $\ell \to 0$. In combination with the data shown in the top panel of fig.~\ref{fig:conv_term_mean}, this implies that the magnetic-to-kinetic energy conversion must be compensated for by a dynamo that is twice as strong as for the $B_0 = 0$ case. For the strong-field case, $B_0/B_{\rm rms} = 12$, the magnetic-to-kinetic energy conversion is very strong in the early stages of the time evolution and tends to decrease over time, with again $\langle \mathcal{W}^\ell_{anis} \rangle/\varepsilon_b \to -1$ for $\ell \to 0$ at the late stage. Comparing this again with data for the total conversion term during this time interval shown in the top panel of fig.~\ref{fig:conv_term_mean}, where  $\langle \mathcal{W}^\ell \rangle/\varepsilon_b \to 1.8$ for $\ell \to 0$, indicates that the small-scale magnetic-to-kinetic energy conversion is compensated by a fluctuation dynamo almost three times as strong as for the  $B_0 = 0$ case.  %The former is consistent with the effective magnetic forcing inferred from the spectral peaks (see the right panel of fig.~\ref{fig:spectra}). 
We note that in the late stage of the time evolution, the magnetic-to-kinetic energy conversion has much larger fluctuations of the mean than in the earlier stages, as reflected by the larger error bars. However, part of the variation in error-bar magnitude may be associated with the different numbers of snapshots included in each quasi-stationary sampling interval.

% ML: the below is resolved, as eps_b is stat. stationary, using mean value for all time intervals is fine. 
%\red{ML: is $\varepsilon_b$ constant during time evolution, which values are used in fig. 5 for the normalisation?} \damiano{First of all $\varepsilon_b$ is stationary during the time evolution. In addition, to be very precise, I calculated the (timeseries) average along the time intervals of interest.}

To connect the resolved-scale conversion term, which is a cumulative kinetic-to-magnetic energy exchange associated with all scales larger than the filter scale $\ell$, with the more familiar spectral description, and to quantify energy conversion at each scale,  we introduce the corresponding omni-directional wavenumber $k=\pi/\ell$ and to consider the derivative of the mean resolved-scale conversion term,
\begin{align}
\mathcal{T}_{ub}(k) = \dfrac{\partial \langle \mathcal{W}^{\pi/k} \rangle}{\partial k} 
= -\dfrac{\ell^2}{\pi} \dfrac{\partial \langle \mathcal{W}^{\ell} \rangle}{\partial \ell},
\label{eq:spec_transf}
\end{align}
which represents the contribution to the interfield energy exchange associated with modes in the neighbourhood of the wavenumber $k$. In this sense, differentiation removes the cumulative character inherent to the low-pass filtering, that is, the fact that $\mathcal{W}^\ell$ accounts for the net conversion over all scales larger than $\ell$, and recasts it in a wavenumber-resolved (or scale-resolved) form. We stress that the term \emph{neighbourhood} is due to the use of the Gaussian filter according which $\mathcal{T}_{ub}(k)$ should be interpreted as a spectrally smoothed analogue of the shell-wise conversion term %associated with $k=\sqrt{k_x^2+k_y^2+k_z^2}$ 
that would arise from Galerkin truncation \citep{Verma04,Verma19, AlexakisEA05-shell}.
%ML cite other Alexakis papers, others?

\begin{figure}
	\begin{center}
         \includegraphics[width=\columnwidth]{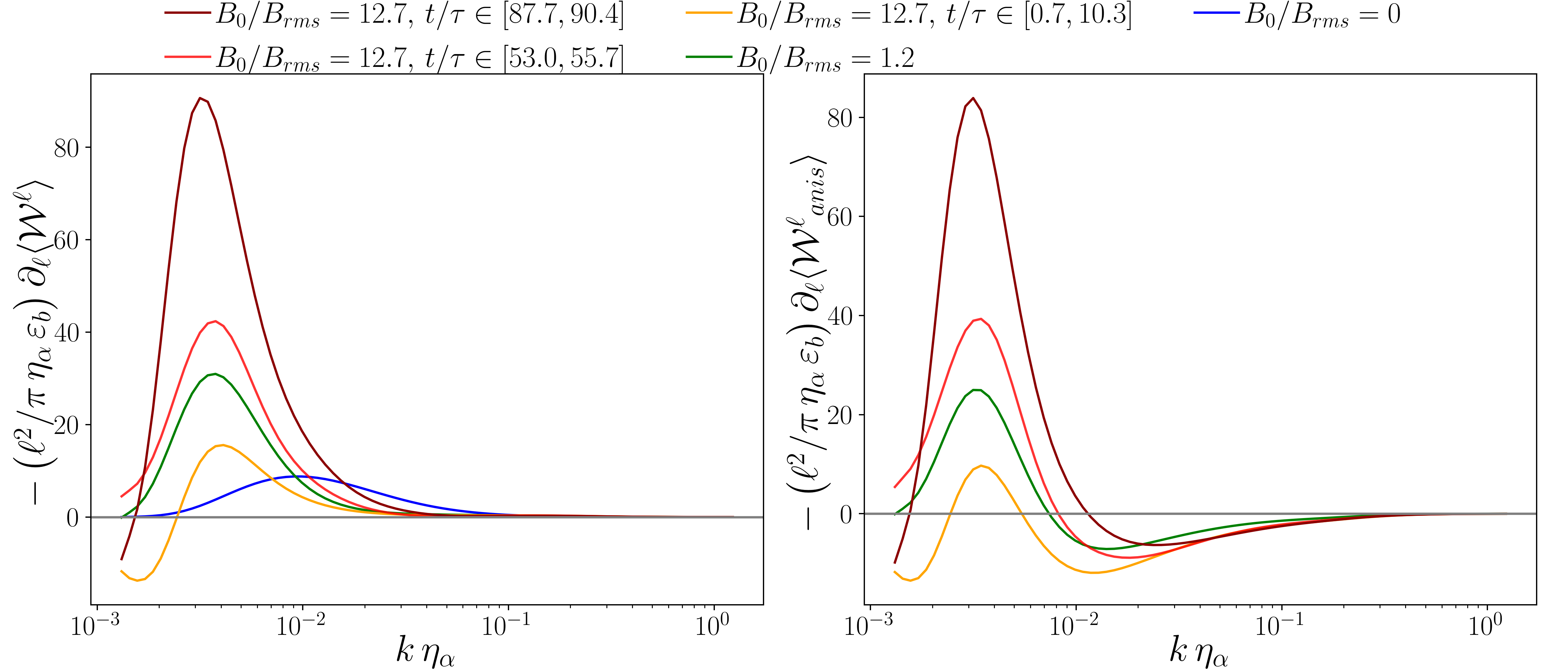}
    \end{center}
    \caption{Normalised scale-resolved energy conversion term $-\dfrac{\ell^2}{\pi \,\eta_\alpha\,\varepsilon_b} \dfrac{\partial \langle \mathcal{W}^{\ell} \rangle}{\partial \ell}$, with total and anisotropic contributions shown in the left and right panels, respectively. In both the panels, the three curves belonging to $B_0/B_{rms}=12.7$ refer to the three quasi-stationary intervals indicated in fig.~\ref{fig:B0_10} with the same colour palette. %Both the panels present non-dimensional values on the axes.
    }
	 \label{fig:conv_term_der}
\end{figure}

We use this representation to highlight the characteristic scales at which the inter-field energy exchange is most active. As such, fig.~\ref{fig:conv_term_der} displays the normalised %spectral 
transfer term defined in eq.~\eqref{eq:spec_transf} on a semi-logarithmic scale, separately for the total energy conversion (left panel) and for the individual contribution associated with the mean magnetic field (right panel) for all datasets. From the data shown in both panels, we observe that, for the configurations with non-zero background magnetic field, the maximum kinetic-to/magnetic energy conversion is localised at large scales, specifically around $k\eta_\alpha \approx 0.003$, and that this conversion is mainly due to the effect of the background magnetic field. This is consistent with the effective large-scale forcing suggested by the form of the magnetic spectra in fig.~\ref{fig:spectra}. We further note that the location of this peak appears to be essentially independent of the background-field strength, which suggests that it may instead be controlled by other non-dimensional parameters of the system, like $Pm$. This peak could already be anticipated from the inflection point of the corresponding curves in fig.~\ref{fig:conv_term_mean}. In contrast, for the $B_0 = 0$ case the maximum kinetic-to-magnetic energy conversion occurs at smaller scales, roughly at $k\eta_\alpha \approx 0.01$, which is also in proximity of the peak of the magnetic energy spectrum shown in fig.~\ref{fig:spectra}. 
For the contribution associated specifically with the background magnetic field shown in the right panel of fig.~\ref{fig:conv_term_der},
%we again observe the same large-scale maximum, followed by 
we observe a negative minimum located approximately in the range $0.01 \lesssim k\eta_\alpha \lesssim 0.024$, depending on the configuration, and a long negative tail with increasing $k$, indicates a magnetic-to-kinetic energy transfer, which is absent in the total $\mathcal{T}_{ub}(k)$ shown in the left panel. This indicates that the small-scale magnetic-to-kinetic energy conversion mediated by the background magnetic field is compensated by a small-scale dynamo, that is stronger than for the $B_0 = 0$ case.  
%As in the case of the total conversion, this behaviour could already be inferred from the inflection point of the anisotropic contribution to the mean resolved-scale conversion shown in fig.~\ref{fig:conv_term_mean}.

\begin{figure}
	\begin{center}
    \noindent\makebox[\textwidth]{
         \includegraphics[width=0.92\columnwidth]{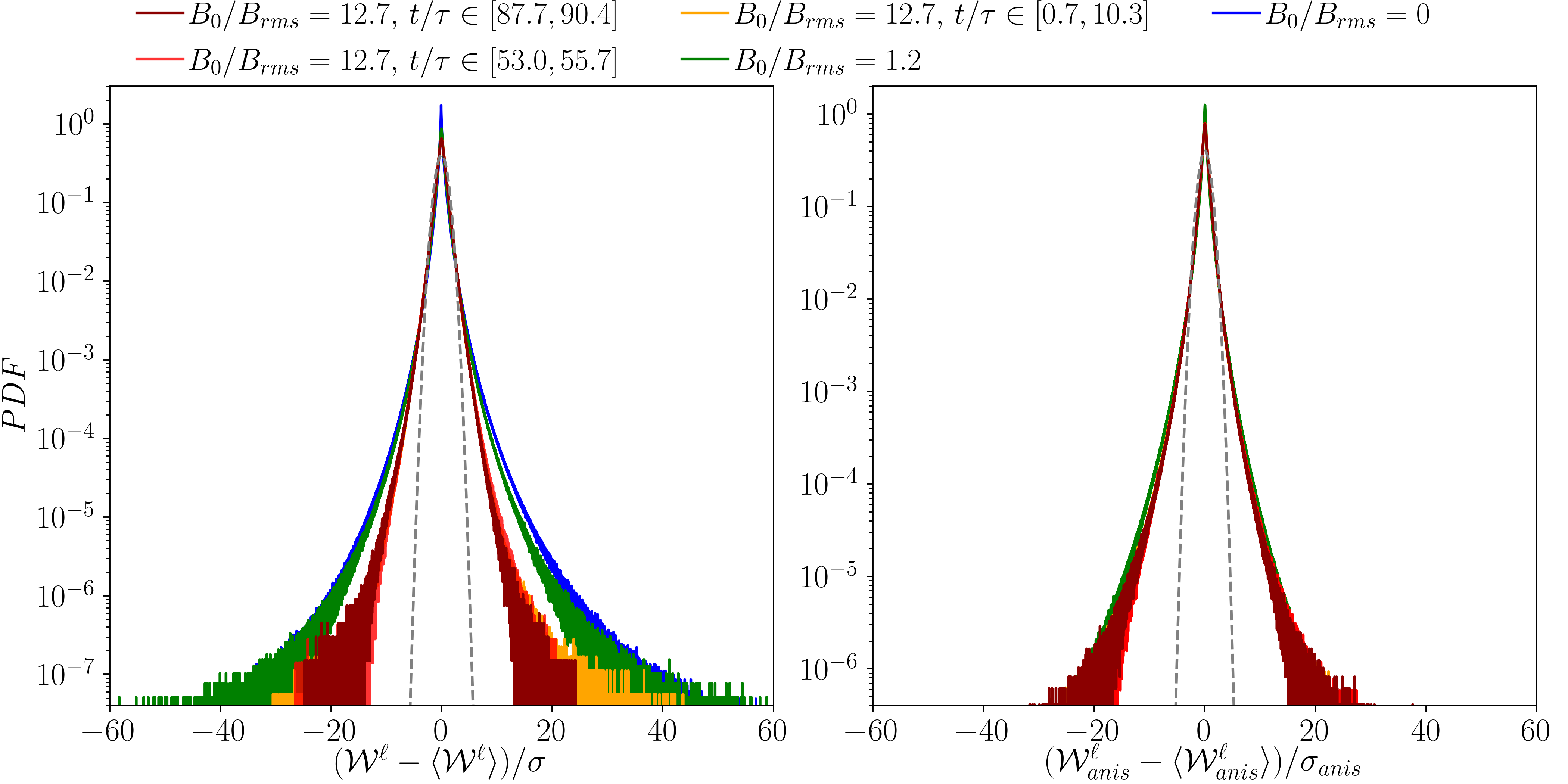}
         }
    \end{center}
    \caption{Standardised PDFs of the RSC terms at $k \eta_\alpha =  0.195$, left $\mathcal{W}^\ell$, right $\mathcal{W}^\ell_{anis}$. %The exponential core of $B_0=10$ in total conversion term cannot be seen without a zoom. 
    }
    %\red{Does this  $\mathcal{W}^\ell_{anis}$ consist of all three terms, or have the spatial transport contributions been subtracted? Fluctuations will be very different if they are included. Same comment applies to the calculation of error bars on the mean of this term. }} point clarified
	 \label{fig:conv_term_pdf}
\end{figure}

Having discussed mean cumulative and mean scale-resolved conversion, we now focus on the statistics of the RSC terms. Fig.~\ref{fig:conv_term_pdf} shows the standardised PDFs of the total RSC term and of its anisotropic contribution (top and bottom panels, respectively), at $k \eta_\alpha = 0.195$, which is the middle of the region where the total energy flux is scale-independent (see fig.~\ref{fig:gauss_filter_dynamo_stages}. For the total RSC term, the PDFs corresponding to the strong-field configuration sampled over different time intervals exhibit noticeably smaller fluctuations than those of the zero- and weak-background-field cases. The latter two appear qualitatively similar, although the $B_0/B_{\rm rms}=0$ case is slightly less skewed than the $B_0/B_{\rm rms}=1.2$ case. 
%
%ML: reinstate if we have a clear point to make, e.g. concerning all scales. 
A quantitative summary of the scale-dependent standard deviation, skewness, and kurtosis for each configuration is provided in fig.~\ref{fig:std_moments} (Appendix~\ref{appendix:moments_conv}).
The standard deviation increases towards smaller scales in all cases, while the skewness is approximately scale-invariant and the kurtosis increases monotonically downscale, except for the $B_0=0$ case.
The standardised PDFs of the anisotropic contribution, $\langle \mathcal{W}^\ell_{anis} \rangle$ shown in the bottom panel of fig.~\ref{fig:conv_term_pdf} are qualitatively indistinguishable. This suggests that, once rescaled by their own variance, the anisotropic contribution to the RSC among datasets C1--C10 presents the same level of fluctuations, even though the corresponding mean values and error bars retain a clearer dependence on the background magnetic field magnitude and dynamo dynamical stage (cf. fig.~\ref{fig:conv_term_mean}). 

%\purple{The statistics of the large-scale RSC have previously been discussed by \cite{bian2019}; however, a direct comparison with that work appears to be difficult. First, the forcing schemes differ, which is relevant because the PDFs are evaluated at large scales, where the details of energy injection may still play a role. Second, \cite{bian2019} quantifies the BMF strength using a ratio based on the maximum of the magnetic-energy spectrum, $B_0/\sqrt{\max_k E_b(k)}$. Since this maximum occurs at large scale, this measure is also sensitive to the type of energy injection. Using their definition, our strong-BMF case would correspond to a value more than three times larger than the strongest case considered in \cite{bian2019}. Further differences arise from the hyperviscosity and hyperdiffusivity values. Finally, their analysis uses a Fourier filter, whereas we employ a Gaussian filter, which can lead to subtle differences in the resulting scale-dependent statistics.}

%ML: I removed as it is not clear.
%Moreover, as concerns the strong anisotropic case(s), the corresponding PDFs of both the panels shows a qualitative similarity on the tails that is robust throughout the Inertial range. 

\section{Conclusions}
\label{sec:conclusions}

We have applied the exact energy flux decomposition to anisotropic MHD turbulence which sees the presence of a superimposed magnetic field. The main result is that the mechanistic picture identified in the case without a mean magnetic field remains unchanged: current-sheet thinning continues to provide the dominant contribution to the forward cascade, while hydrodynamic-like mechanisms such as vortex stretching and strain self-amplification remain depleted, although less so than in the case of zero background field. 

Increasing $B_0$ primarily reorganises the dynamics rather than changing the leading transfer mechanism. In particular, the mean subfluxes become more nearly scale-independent, the Dynamo contribution is progressively suppressed, and the strongly anisotropic case develops a more pronounced inverse-transfer contribution at the largest scales, although the total energy flux remains direct due to subflux cancellations. These features are consistent with the progressive two-dimensionalisation of the flow. 

A second main result concerns the energy conversion term between kinetic and magnetic energy budgets. By separating the total conversion into isotropic and background-field-dependent parts, we have isolated the contribution specific to the background field. This anisotropic term induces an effective kinetic-to-magnetic conversion at large and intermediate scales, but changes sign at smaller scales, where it corresponds to magnetic-to-kinetic conversion. In the strongly anisotropic regime, it also becomes markedly more fluctuating.

Overall, the present results show that a strong guide field does not replace the dominant route by which energy is transferred to small scales in MHD turbulence. Rather, it embeds that transfer in an increasingly anisotropic, quasi two-dimensional state and introduces an additional scale-dependent conversion channel between kinetic and magnetic energies. These findings also suggest that SGS models for anisotropic MHD should retain the same leading cascade physics as in the isotropic case, which is the current-sheet thinning term, while incorporating an antidiffusive term to include an enhanced large-scale accumulation of energy due to the two-dimensionalisation. 
In this context, we point out that current-sheet thinning is the only process involving the magnetic field that survives in the fully 2d case \citep{capocci2025}, and similarly for vortex thinning in the Inertial flux. This suggests that quasi 2D strong-field approximations such as reduced MHD \citep{Strauss76} already capture the most relevant cascade characteristics of the full 3D state. More precisely the dominant physical process, namely the current sheet thinning is retained by the progressive two-dimensionalisation induced by the background magnetic field. The role of the background magnetic field is not to introduce a new leading cascade process, but rather to reduce the full three-dimensional contributions, leaving the mean cascade increasingly controlled by the two-dimensional MHD mechanisms.

\section*{Acknowledgements}
The authors thank Luca Biferale for the financial support
provided by both European Research Council (ERC) under the European Union’s Horizon 2020
research and innovation programme (Grant Agreement No. 882340) and FieldTurb experiment of
the Istituto Nazionale di Fisica Nucleare (INFN). This work used the ARCHER2 UK National
Supercomputing Service (\url{www.archer2.ac.uk.}) with resources provided by the UK Turbulence Consortium (EPSRC
Grant EP/R029326/1). The authors thank %Sean Oughton, 
Perry Johnson and Luca Biferale for helpful discussions, Erin Goldstraw for comments on reduced MHD, and Asif Nawaz for discussions on implications for LES modelling.

\begin{comment}
    
Points from the discussion 29/01
\begin{itemize}

    \item \red{Results are very similar to exact results in 2D / RMHD. That is, RMHD captures most relevant physics of the strongly anisotropic case (reference Erin once that paper si submitted. \damiano{note that the 2D results are basically kinematics. For this reason it might be, that after a long time, the inertial term would become dominant wrt the current sheet thinning.} \\
    \item Same physics of the cascade as without strong background field except for the expected 2d hydro effects. That is, similar LES models can be used, especially as small scale hydro flux is forwards. So depending on LES resolution, no antidiffusive model for the Reynolds stress is required. \\
    \item First characterisation of energy conversion due to $B_0$ specifically. including fluctuations. }

\end{itemize}

\end{comment}
%\newpage   %for now

%\clearpage
\appendix

%\begin{comment}
\section{Moments of RSC}
\label{appendix:moments_conv}

\begin{figure}
        \begin{center}
        
         \includegraphics[width=.6\columnwidth]{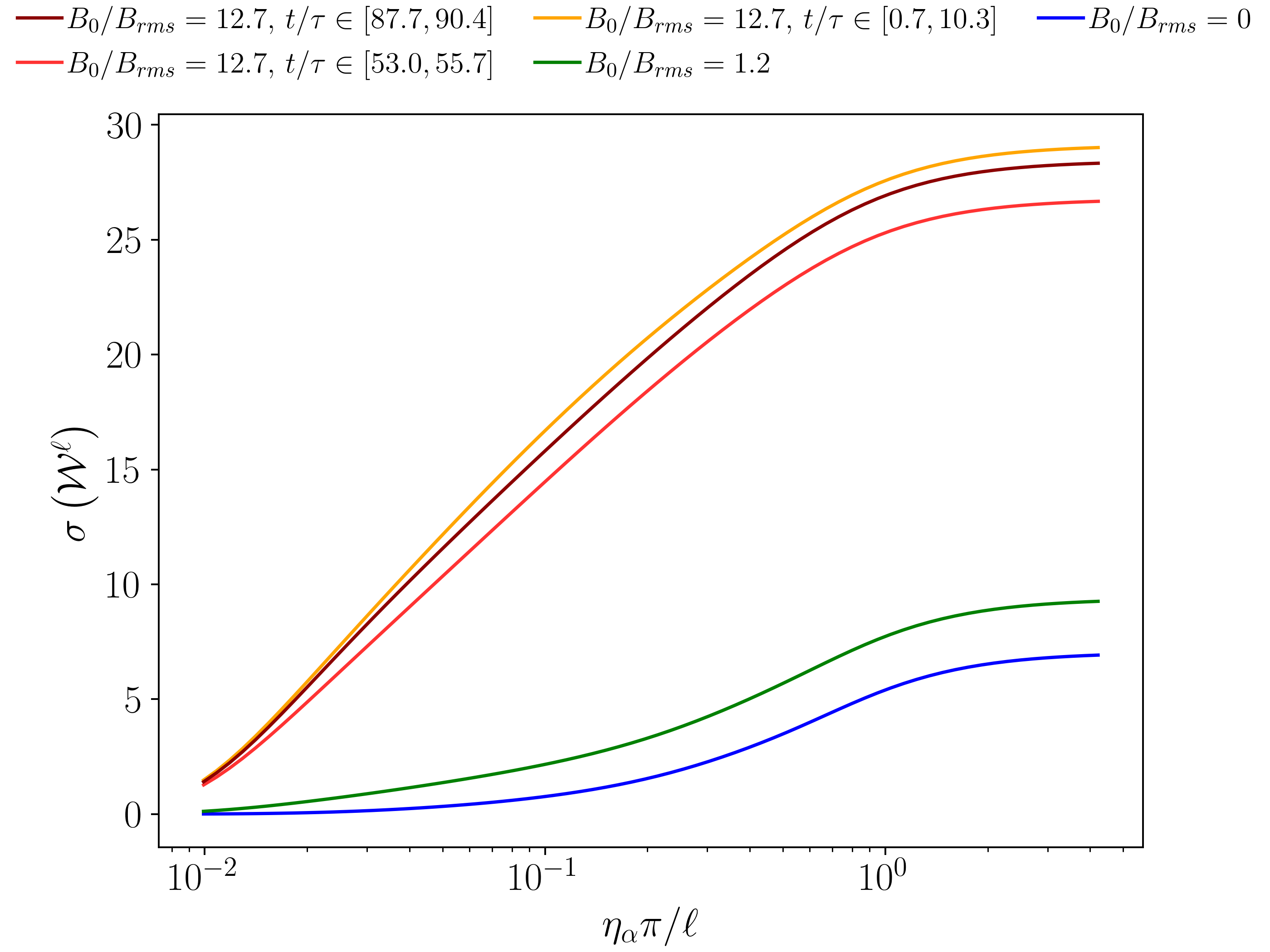}
          \noindent\makebox[\textwidth]{
         \includegraphics[width=.6\columnwidth]{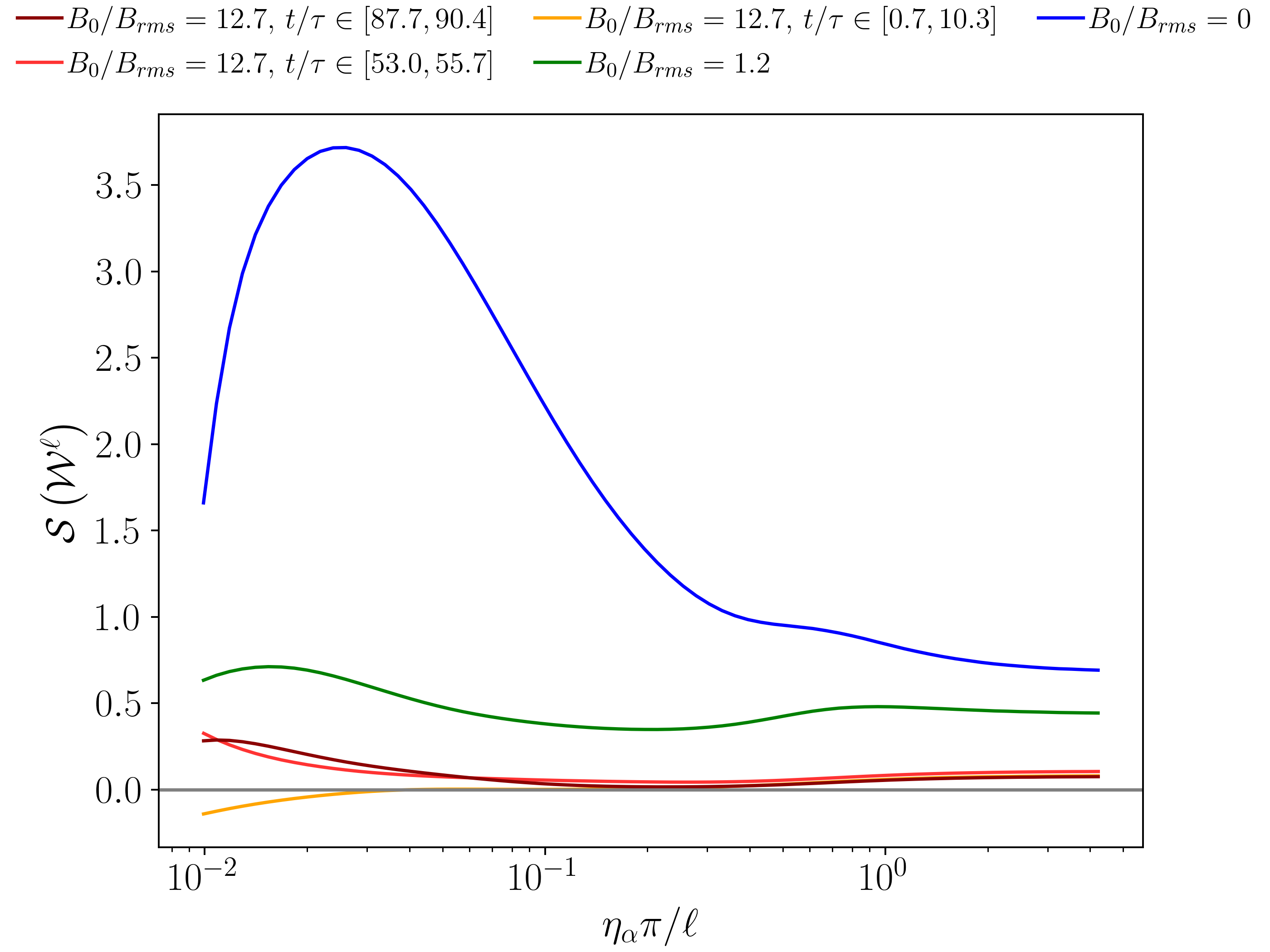}
         \includegraphics[width=.6\columnwidth]{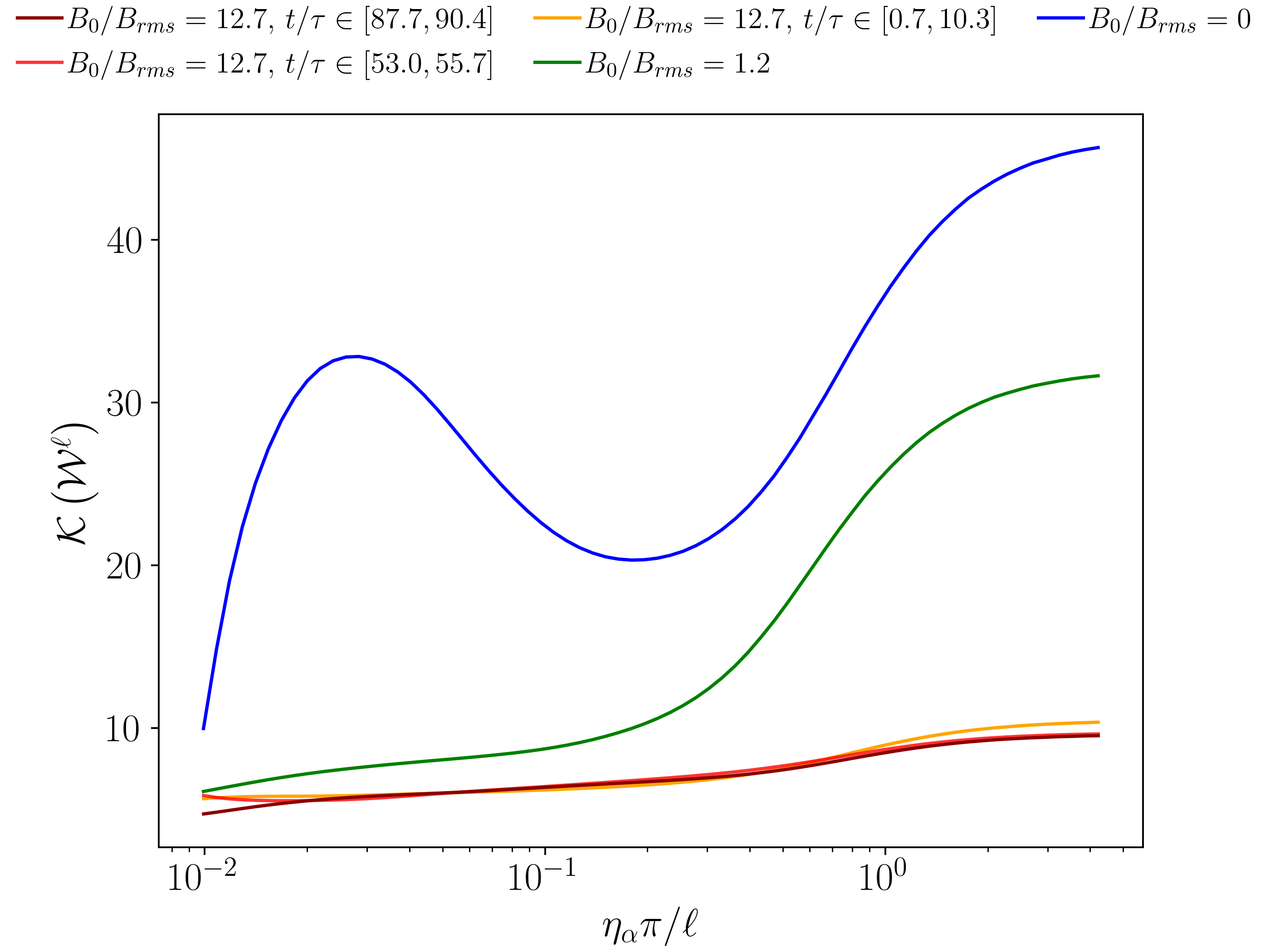} }
    \end{center}
         \caption{Values of standard deviation, skewness and kurtosis of the total $\mathcal{W}^\ell$ as function of $k\,\eta_\alpha=\pi\, \eta_\alpha/\ell$.}
\label{fig:std_moments}
\end{figure}

Figure \ref{fig:std_moments} displays the values of standard deviation (top panel), skewness (bottom-left panel) and kurtosis (bottom-right panel) as function of the wavenumber $k=\eta \,\pi/\ell $ for the datasets A3, C1 and C10. Like the analysis of the main body of the paper, the individual dynamo stages belonging to dataset C10 are displayed individually. Note that the small-scale statistics will be affected by the use  of the hyperviscosity and hyperdissipation in the equations of motion, namely eqs.~\eqref{eq:mom_eq}-\eqref{eq:ind_eq}. More in general, the profile of the standard deviation, on fig.~\ref{fig:std_moments} top panel, presents a gradient-type behaviour as the standard deviation increases as the scale decreases.

%\end{comment}

\section{Subfluxes definitions}
\label{app:defintions}

In this section we provide the definitions of all the subfluxes appearing in the decomposition of the MHD energy fluxes. 
As highlighted in sec.~\ref{sec:exact}, for the  subfluxes 
$\Pi^I_{s, S \Omega S}$, 
$\Pi^I_{m, S \Omega S}$ 
and 
$\Pi^M_{s, SJ \Sigma}$, 
$\Pi^M_{m, SJ \Sigma}$ 
there is an extra factor of two that arises from the symmetry of the corresponding SGS stress tensors. Because in eqs.\eqref{eq:Eu-ls} and \eqref{eq:Eb-ls} 
the fluxes appear with the same leading signs,
both the Maxwell and the Dynamo subfluxes in the definition below acquires an additional minus sign.

\subsection{Inertial}
The following subfluxes are identical to the hydrodynamic counterpart from \cite{Johnson20,Johnson21}.
\begin{align}
   &\Pi^{I,\ell}_{s,S S S}    
     = 
    -  \ell^2 \, \tr{\big( \overline{\vS}^\ell \big)^t \,
       \overline{\vS}^\ell \big(\overline{\vS}^\ell\big)^t } \\
   & \Pi^{I,\ell}_{m,S S S } 
     =
    - \int_0^{\ell^2} \! \! \d\theta \,
       \tr{\big( \overline{\vS}^\ell \big)^t \Big(
       \ol{ \overline{\vS}^{\sqrt{\theta}} 
              \big(\overline{\vS}^{\sqrt{\theta}} \big)^{t} \,}^\phi 
           -
            \ol{\overline{\vS}^{\sqrt{\theta}}\,}^\phi\;
             \ol{ \big( \overline{\vS}^{\sqrt{\theta}} \big)^{\! t} \,}^\phi  
       \Big)} \\
       & \notag  \\
     &\Pi^{I,\ell}_{s,S \Omega \Omega}    
     = 
    -  \ell^2 \, \tr{\big( \overline{\vS}^\ell \big)^t \,
       \overline{\vOmega}^\ell \big(\overline{\vOmega}^\ell\big)^t } \\
   & \Pi^{I,\ell}_{m,S \Omega \Omega } 
     =
    - \int_0^{\ell^2} \! \! \d\theta \,
       \tr{\big( \overline{\vS}^\ell \big)^t \Big(
       \ol{ \overline{\vOmega}^{\sqrt{\theta}} 
              \big(\overline{\vOmega}^{\sqrt{\theta}} \big)^{t} \,}^\phi 
           -
            \ol{\overline{\vOmega}^{\sqrt{\theta}}\,}^\phi\;
             \ol{ \big( \overline{\vOmega}^{\sqrt{\theta}} \big)^{\! t} \,}^\phi  
       \Big)} \\
       &   \notag \\
   &\Pi^{I,\ell}_{s,S \Omega S}    
     = 
    - 2 \ell^2 \, \tr{\big( \overline{\vS}^\ell \big)^t \,
       \overline{\vOmega}^\ell \big(\overline{\vS}^\ell\big)^t } \equiv 0 \\
   & \Pi^{I,\ell}_{m,S \Omega S} 
     =
    - 2\int_0^{\ell^2} \! \! \d\theta \,
       \tr{\big( \overline{\vS}^\ell \big)^t \Big(
       \ol{ \overline{\vOmega}^{\sqrt{\theta}} 
              \big(\overline{\vS}^{\sqrt{\theta}} \big)^{t} \,}^\phi 
           -
            \ol{\overline{\vOmega}^{\sqrt{\theta}}\,}^\phi\;
             \ol{ \big( \overline{\vS}^{\sqrt{\theta}} \big)^{\! t} \,}^\phi  
       \Big)} 
\end{align}

\subsection{Maxwell}
\begin{align}
   &\Pi^{M,\ell}_{s,S \Sigma \Sigma}    
     = 
      \ell^2 \, \tr{\big( \overline{\vS}^\ell \big)^t \,
       \overline{\vSigma}^\ell \big(\overline{\vSigma}^\ell\big)^t } \\
   & \Pi^{M,\ell}_{m,S \Sigma \Sigma} 
     =
     \int_0^{\ell^2} \! \! \d\theta \,
       \tr{\big( \overline{\vS}^\ell \big)^t \Big(
       \ol{ \overline{\vSigma}^{\sqrt{\theta}} 
              \big(\overline{\vSigma}^{\sqrt{\theta}} \big)^{t} \,}^\phi 
           -
            \ol{\overline{\vSigma}^{\sqrt{\theta}}\,}^\phi\;
             \ol{ \big( \overline{\vSigma}^{\sqrt{\theta}} \big)^{\! t} \,}^\phi  
       \Big)} \\
       & \notag  \\
   &\Pi^{M,\ell}_{s,S J J}    
     = 
      \ell^2 \, \tr{\big( \overline{\vS}^\ell \big)^t \,
       \overline{\vJ}^\ell \big(\overline{\vJ}^\ell\big)^t } \\
   & \Pi^{M,\ell}_{m,S J J} 
     =
     \int_0^{\ell^2} \! \! \d\theta \,
       \tr{\big( \overline{\vS}^\ell \big)^t \Big(
       \ol{ \overline{\vJ}^{\sqrt{\theta}} 
              \big(\overline{\vJ}^{\sqrt{\theta}} \big)^{t} \,}^\phi 
           -
            \ol{\overline{\vJ}^{\sqrt{\theta}}\,}^\phi\;
             \ol{ \big( \overline{\vJ}^{\sqrt{\theta}} \big)^{\! t} \,}^\phi  
       \Big)}\\
   &   \notag \\
   &\Pi^{M,\ell}_{s,S J \Sigma}    
     = 
     2 \ell^2 \, \tr{\big( \overline{\vS}^\ell \big)^t \,
       \overline{\vJ}^\ell \big(\overline{\vSigma}^\ell\big)^t } \\
   & \Pi^{M,\ell}_{m,S J \Sigma} 
     =
     2\int_0^{\ell^2} \! \! \d\theta \,
       \tr{\big( \overline{\vS}^\ell \big)^t \Big(
       \ol{ \overline{\vJ}^{\sqrt{\theta}} 
              \big(\overline{\vSigma}^{\sqrt{\theta}} \big)^{t} \,}^\phi 
           -
            \ol{\overline{\vJ}^{\sqrt{\theta}}\,}^\phi\;
             \ol{ \big( \overline{\vSigma}^{\sqrt{\theta}} \big)^{\! t} \,}^\phi  
       \Big)}
\end{align}

\subsection{Advection}
\begin{align}
  &\Pi^{A,\ell}_{s,\Sigma \Sigma S}
     = 
    - \ell^2 \, \tr{\big( \overline{\vSigma}^\ell \big)^t \,
       \overline{\vSigma}^\ell \big(\overline{\vS}^\ell\big)^t } \\
  &\Pi^{A,\ell}_{m,\Sigma \Sigma S}
     =
       \; - \;
       \int_0^{\ell^2} \d\theta \,
       \tr{\big( \overline{\vSigma}^\ell \big)^t \Big(
       \ol{ \overline{\vSigma}^{\sqrt{\theta}} 
              \big(\overline{\vS}^{\sqrt{\theta}} \big)^{t} \,}^\phi 
           -
            \ol{\overline{\vSigma}^{\sqrt{\theta}}\,}^\phi\;
             \ol{ \big( \overline{\vSigma}^{\sqrt{\theta}} \big)^{\! t} \,}^\phi  
       \Big)}\\
  &  \notag \\
  &\Pi^{A,\ell}_{s,\Sigma J S}
     = 
    - \ell^2 \, \tr{\big( \overline{\vSigma}^\ell \big)^t \,
       \overline{\vJ}^\ell \big(\overline{\vS}^\ell\big)^t }\\
  &\Pi^{A,\ell}_{m,\Sigma J S}
  =
       \; - \;
       \int_0^{\ell^2} \d\theta \,
       \tr{\big( \overline{\vSigma}^\ell \big)^t \Big(
       \ol{ \overline{\vJ}^{\sqrt{\theta}} 
              \big(\overline{\vS}^{\sqrt{\theta}} \big)^{t} \,}^\phi 
           -
            \ol{\overline{\vJ}^{\sqrt{\theta}}\,}^\phi\;
             \ol{ \big( \overline{\vS}^{\sqrt{\theta}} \big)^{\! t} \,}^\phi  
       \Big)} \\
  &  \notag \\
  &\Pi^{A,\ell}_{s,\Sigma \Sigma \Omega}
     = - \ell^2 \, \tr{\big( \overline{\vSigma}^\ell \big)^t \,
       \overline{\vSigma}^\ell \big(\overline{\vOmega}^\ell\big)^t } \equiv 0 \\ 
  &\Pi^{A,\ell}_{m,\Sigma \Sigma \Omega} =     
  \; - \;
       \int_0^{\ell^2} \d\theta \,
       \tr{\big( \overline{\vSigma}^\ell \big)^t \Big(
       \ol{ \overline{\vSigma}^{\sqrt{\theta}} 
              \big(\overline{\vOmega}^{\sqrt{\theta}} \big)^{t} \,}^\phi 
           -
            \ol{\overline{\vSigma}^{\sqrt{\theta}}\,}^\phi\;
             \ol{ \big( \overline{\vOmega}^{\sqrt{\theta}} \big)^{\! t} \,}^\phi  
       \Big)}\\
  &  \notag \\
  &\Pi^{A,\ell}_{s,\Sigma J \Omega}   
     = 
    - \ell^2 \, \tr{\big( \overline{\vSigma}^\ell \big)^t \,
       \overline{\vJ}^\ell \big(\overline{\vOmega}^\ell\big)^t } \\
  &\Pi^{A,\ell}_{m,\Sigma J \Omega}
       =
       \; - \;
       \int_0^{\ell^2} \d\theta \,
       \tr{\big( \overline{\vSigma}^\ell \big)^t \Big(
       \ol{ \overline{\vJ}^{\sqrt{\theta}} 
              \big(\overline{\vOmega}^{\sqrt{\theta}} \big)^{t} \,}^\phi 
           -
            \ol{\overline{\vJ}^{\sqrt{\theta}}\,}^\phi\;
             \ol{ \big( \overline{\vOmega}^{\sqrt{\theta}} \big)^{\! t} \,}^\phi  
       \Big)}\\
  & \notag  \\
  &\Pi^{A,\ell}_{s,J \Sigma S}
      = 
    - \ell^2 \, \tr{\big( \overline{\vJ}^\ell \big)^t \,
       \overline{\vSigma}^\ell \big(\overline{\vS}^\ell\big)^t } \\
  &\Pi^{A,\ell}_{m,J \Sigma S}
      =
       \; - \;
       \int_0^{\ell^2} \d\theta \,
       \tr{\big( \overline{\vJ}^\ell \big)^t \Big(
       \ol{ \overline{\vSigma}^{\sqrt{\theta}} 
              \big(\overline{\vS}^{\sqrt{\theta}} \big)^{t} \,}^\phi 
           -
            \ol{\overline{\vSigma}^{\sqrt{\theta}}\,}^\phi\;
             \ol{ \big( \overline{\vS}^{\sqrt{\theta}} \big)^{\! t} \,}^\phi  
       \Big)}\\
       & \notag  \\
  &\Pi^{A,\ell}_{s,J \Sigma \Omega}
      = 
    - \ell^2 \, \tr{\big( \overline{\vJ}^\ell \big)^t \,
       \overline{\vSigma}^\ell \big(\overline{\vOmega}^\ell\big)^t } \\
  &\Pi^{A,\ell}_{m,J \Sigma \Omega}
      =
       \; - \;
       \int_0^{\ell^2} \d\theta \,
       \tr{\big( \overline{\vJ}^\ell \big)^t \Big(
       \ol{ \overline{\vSigma}^{\sqrt{\theta}} 
              \big(\overline{\vOmega}^{\sqrt{\theta}} \big)^{t} \,}^\phi 
           -
            \ol{\overline{\vSigma}^{\sqrt{\theta}}\,}^\phi\;
             \ol{ \big( \overline{\vOmega}^{\sqrt{\theta}} \big)^{\! t} \,}^\phi  
       \Big)}\\
       & \notag   \\
  &\Pi^{A,\ell}_{s,J J S}
     =  
      - \ell^2 \, \tr{\big( \overline{\vJ}^\ell \big)^t \,
       \overline{\vJ}^\ell \big(\overline{\vS}^\ell\big)^t }  \\
  &\Pi^{A,\ell}_{m,J J S}   
    =
    \; - \;
       \int_0^{\ell^2} \d\theta \,
       \tr{\big( \overline{\vJ}^\ell \big)^t \Big(
       \ol{ \overline{\vJ}^{\sqrt{\theta}} 
              \big(\overline{\vS}^{\sqrt{\theta}} \big)^{t} \,}^\phi 
           -
            \ol{\overline{\vJ}^{\sqrt{\theta}}\,}^\phi\;
             \ol{ \big( \overline{\vS}^{\sqrt{\theta}} \big)^{\! t} \,}^\phi  
       \Big)}\\
  & \notag   \\
  &\Pi^{A,\ell}_{s,J J \Omega}
     =  
      - \ell^2 \, \tr{\big( \overline{\vJ}^\ell \big)^t \,
       \overline{\vJ}^\ell \big(\overline{\vOmega}^\ell\big)^t } \equiv 0 \\
  &\Pi^{A,\ell}_{m,J J \Omega}   
    =
    \; - \;
       \int_0^{\ell^2} \d\theta \,
       \tr{\big( \overline{\vJ}^\ell \big)^t \Big(
       \ol{ \overline{\vJ}^{\sqrt{\theta}} 
              \big(\overline{\vOmega}^{\sqrt{\theta}} \big)^{t} \,}^\phi 
           -
            \ol{\overline{\vJ}^{\sqrt{\theta}}\,}^\phi\;
             \ol{ \big( \overline{\vOmega}^{\sqrt{\theta}} \big)^{\! t} \,}^\phi  
       \Big)}
\end{align}

\subsection{Dynamo}
\begin{align}
  &\Pi^{D,\ell}_{s,\Sigma S \Sigma}
     = 
     \ell^2 \, \tr{\big( \overline{\vSigma}^\ell \big)^t \,
       \overline{\vS}^\ell \big(\overline{\vSigma}^\ell\big)^t } \\
  &\Pi^{D,\ell}_{m,\Sigma S \Sigma}
     =
       \;  \;
       \int_0^{\ell^2} \d\theta \,
       \tr{\big( \overline{\vSigma}^\ell \big)^t \Big(
       \ol{ \overline{\vS}^{\sqrt{\theta}} 
              \big(\overline{\vSigma}^{\sqrt{\theta}} \big)^{t} \,}^\phi 
           -
            \ol{\overline{\vS}^{\sqrt{\theta}}\,}^\phi\;
             \ol{ \big( \overline{\vSigma}^{\sqrt{\theta}} \big)^{\! t} \,}^\phi  
       \Big)}\\
  &  \notag \\
  &\Pi^{D,\ell}_{s,\Sigma \Omega \Sigma}
     = 
     \ell^2 \, \tr{\big( \overline{\vSigma}^\ell \big)^t \,
       \overline{\vOmega}^\ell \big(\overline{\vSigma}^\ell\big)^t } \equiv 0 \\
  &\Pi^{D,\ell}_{m,\Sigma \Omega \Sigma}
     =
       \;  \;
       \int_0^{\ell^2} \d\theta \,
       \tr{\big( \overline{\vSigma}^\ell \big)^t \Big(
       \ol{ \overline{\vOmega}^{\sqrt{\theta}} 
              \big(\overline{\vSigma}^{\sqrt{\theta}} \big)^{t} \,}^\phi 
           -
            \ol{\overline{\vOmega}^{\sqrt{\theta}}\,}^\phi\;
             \ol{ \big( \overline{\vSigma}^{\sqrt{\theta}} \big)^{\! t} \,}^\phi  
       \Big)}\\
  &  \notag \\
    &\Pi^{D,\ell}_{s,\Sigma S J}
     = 
     \ell^2 \, \tr{\big( \overline{\vSigma}^\ell \big)^t \,
       \overline{\vS}^\ell \big(\overline{\vJ}^\ell\big)^t } \\
  &\Pi^{D,\ell}_{m,\Sigma S J}
     =
       \;  \;
       \int_0^{\ell^2} \d\theta \,
       \tr{\big( \overline{\vSigma}^\ell \big)^t \Big(
       \ol{ \overline{\vS}^{\sqrt{\theta}} 
              \big(\overline{\vJ}^{\sqrt{\theta}} \big)^{t} \,}^\phi 
           -
            \ol{\overline{\vS}^{\sqrt{\theta}}\,}^\phi\;
             \ol{ \big( \overline{\vJ}^{\sqrt{\theta}} \big)^{\! t} \,}^\phi  
       \Big)}\\
  &  \notag \\
    &\Pi^{D,\ell}_{s,\Sigma \Omega J}
     = 
     \ell^2 \, \tr{\big( \overline{\vSigma}^\ell \big)^t \,
       \overline{\vOmega}^\ell \big(\overline{\vJ}^\ell\big)^t } \\
  &\Pi^{D,\ell}_{m,\Sigma \Omega J}
     =
       \;  \;
       \int_0^{\ell^2} \d\theta \,
       \tr{\big( \overline{\vSigma}^\ell \big)^t \Big(
       \ol{ \overline{\vOmega}^{\sqrt{\theta}} 
              \big(\overline{\vJ}^{\sqrt{\theta}} \big)^{t} \,}^\phi 
           -
            \ol{\overline{\vOmega}^{\sqrt{\theta}}\,}^\phi\;
             \ol{ \big( \overline{\vJ}^{\sqrt{\theta}} \big)^{\! t} \,}^\phi  
       \Big)}\\
  &  \notag \\
      &\Pi^{D,\ell}_{s,J S \Sigma}
     = 
     \ell^2 \, \tr{\big( \overline{\vJ}^\ell \big)^t \,
       \overline{\vS}^\ell \big(\overline{\vSigma}^\ell\big)^t } \\
  &\Pi^{D,\ell}_{m,J S \Sigma}
     =
       \;  \;
       \int_0^{\ell^2} \d\theta \,
       \tr{\big( \overline{\vJ}^\ell \big)^t \Big(
       \ol{ \overline{\vS}^{\sqrt{\theta}} 
              \big(\overline{\vSigma}^{\sqrt{\theta}} \big)^{t} \,}^\phi 
           -
            \ol{\overline{\vS}^{\sqrt{\theta}}\,}^\phi\;
             \ol{ \big( \overline{\vSigma}^{\sqrt{\theta}} \big)^{\! t} \,}^\phi  
       \Big)}\\
  &  \notag \\
      &\Pi^{D,\ell}_{s,J \Omega \Sigma}
     = 
     \ell^2 \, \tr{\big( \overline{\vJ}^\ell \big)^t \,
       \overline{\vOmega}^\ell \big(\overline{\vSigma}^\ell\big)^t } \\
  &\Pi^{D,\ell}_{m,J \Omega \Sigma}
     =
       \;  \;
       \int_0^{\ell^2} \d\theta \,
       \tr{\big( \overline{\vJ}^\ell \big)^t \Big(
       \ol{ \overline{\vOmega}^{\sqrt{\theta}} 
              \big(\overline{\vSigma}^{\sqrt{\theta}} \big)^{t} \,}^\phi 
           -
            \ol{\overline{\vOmega}^{\sqrt{\theta}}\,}^\phi\;
             \ol{ \big( \overline{\vSigma}^{\sqrt{\theta}} \big)^{\! t} \,}^\phi  
       \Big)}\\
  &  \notag \\
      &\Pi^{D,\ell}_{s,J S J}
     = 
     \ell^2 \, \tr{\big( \overline{\vJ}^\ell \big)^t \,
       \overline{\vS}^\ell \big(\overline{\vJ}^\ell\big)^t } \\
  &\Pi^{D,\ell}_{m,J S J}
     =
       \;  \;
       \int_0^{\ell^2} \d\theta \,
       \tr{\big( \overline{\vJ}^\ell \big)^t \Big(
       \ol{ \overline{\vS}^{\sqrt{\theta}} 
              \big(\overline{\vJ}^{\sqrt{\theta}} \big)^{t} \,}^\phi 
           -
            \ol{\overline{\vS}^{\sqrt{\theta}}\,}^\phi\;
             \ol{ \big( \overline{\vJ}^{\sqrt{\theta}} \big)^{\! t} \,}^\phi  
       \Big)}\\
       &  \notag \\
      &\Pi^{D,\ell}_{s,J \Omega J}
     = 
     \ell^2 \, \tr{\big( \overline{\vJ}^\ell \big)^t \,
       \overline{\vOmega}^\ell \big(\overline{\vJ}^\ell\big)^t } \equiv 0  \\
  &\Pi^{D,\ell}_{m,J \Omega J}
     =
       \;  \;
       \int_0^{\ell^2} \d\theta \,
       \tr{\big( \overline{\vJ}^\ell \big)^t \Big(
       \ol{ \overline{\vOmega}^{\sqrt{\theta}} 
              \big(\overline{\vJ}^{\sqrt{\theta}} \big)^{t} \,}^\phi 
           -
            \ol{\overline{\vOmega}^{\sqrt{\theta}}\,}^\phi\;
             \ol{ \big( \overline{\vJ}^{\sqrt{\theta}} \big)^{\! t} \,}^\phi  
       \Big)}
\end{align}

\bibliographystyle{jpp}
\bibliography{ag,hl,mp,qz,extras,bib,thesis}

\end{document}